\documentclass[zpreprint,zbstnp]{zeus_paper}

\usepackage[english]{babel}

\chardef\usc=95
\chardef\til=126
\catcode`\@=11 
\DeclareRobustCommand\xdotspace{\futurelet\@let@token\@xdotspace}
\def\@xdotspace{%
  \ifx\@let@token.\else
  \ifx\@let@token\bgroup.\else
  \ifx\@let@token\egroup.\else
  \ifx\@let@token\/.\else
  \ifx\@let@token\ .\else
  \ifx\@let@token~.\else
  \ifx\@let@token!.\else
  \ifx\@let@token,.\else
  \ifx\@let@token:.\else
  \ifx\@let@token;.\else
  \ifx\@let@token?.\else
  \ifx\@let@token/.\else
  \ifx\@let@token'.\else
  \ifx\@let@token).\else
  \ifx\@let@token-.\else
  \ifx\@let@token\@xobeysp.\else
  \ifx\@let@token\space.\else
  \ifx\@let@token\@sptoken.\else
   .\space
   \fi\fi\fi\fi\fi\fi\fi\fi\fi\fi\fi\fi\fi\fi\fi\fi\fi\fi}
\catcode`\@=12 

\newcommand{\stru}[2]{%
   \relax\ifmmode\hbox{\vrule height#1 depth#2 width0pt}%
   \else\vrule height#1 depth#2 width0pt\fi}

\newcommand{\Ronum}[1]{\uppercase\expandafter{\romannumeral#1}}
\newcommand{\ronum}[1]{\expandafter{\romannumeral#1}}
\DeclareRobustCommand{\LaTeXZ}{%
  \LaTeX\kern-.05em4\kern-.1em
  {\raisebox{-0.2ex}{$\scriptstyle\text{ZEUS}$}}\xspace}

\DeclareMathAlphabet{\mathbf}{OT1}{cmr}{bx}{sl}
\newcommand{\eVdist}{\kern-0.06667em}

\newcommand{\Gev}{{\text{Ge}\eVdist\text{V\/}}}

\newcommand{\pb}{\,\text{pb}}

\newcommand{\ns}{\,\text{ns}}

\newcommand{\Tesla}{\,\text{T}}

\newcommand{\slashfrac}[2]{%
  \raisebox{0.5ex}{\ensuremath #1}\kern-0.12em/\kern-0.08em
  \raisebox{-.8ex}{\ensuremath #2}}

\newcommand{\sqr}[3]{%
    {\vcenter{\hrule height.#3ex\hbox{\vrule width.#2ex height#1ex
     \kern#1ex\vrule width.#3ex}\hrule height.#2ex}}}

\catcode`\@=11 
\newcommand{\parenbar}{\mathpalette\p@renb@r}
\def\p@renb@r#1#2{\vbox{%
  \ifx#1\scriptscriptstyle \dimen@.7em\dimen@ii.2em\else
  \ifx#1\scriptstyle \dimen@.8em\dimen@ii.25em\else
  \dimen@1em\dimen@ii.4em\fi\fi \offinterlineskip
  \ialign{\hfill##\hfill\cr
    \vbox{\hrule width\dimen@ii}\cr
    \noalign{\vskip-.3ex}%
    \hbox to\dimen@{$\mathchar300\hfil\mathchar301$}\cr
    \noalign{\vskip-.3ex}%
    $#1#2$\cr}}}
\catcode`\@=12 

\newcommand{\IP}{{\rm I$\kern-0.01667em$P}\xspace}

\mathchardef\qsm=63
\mathchardef\pls=43
\mathchardef\mns=512
\mathchardef\plm=518
\mathchardef\eql=61
\mathchardef\smallleft=300
\mathchardef\smallright=301
\mathchardef\les=316
\mathchardef\gre=318
\mathchardef\leq=532
\mathchardef\grq=533

\catcode`\@=11 
\newcounter{pict@width}
\newcounter{pict@height}
\newlength{\pict@scale}
\newcommand{\psfigadd}[4]{%
\setcounter{pict@width}{1*\ratio{#2+\pict@scale/2}{\pict@scale}}
\setcounter{pict@height}{1*\ratio{#3+\pict@scale/2}{\pict@scale}}
\setlength{\unitlength}{\pict@scale}
\hbox to #2{\hspace{-\fill}\begin{picture}(\thepict@width,\thepict@height)
\put(0,0){\psfig{figure=#1,width=#2,height=#3,clip=}}
\SetScale{0.283466457}
\SetWidth{1.763889}
{#4}
\end{picture}}
}
\newcounter{pict@widthfst}
\newcounter{pict@widthscd}
\newcounter{pict@widthtot}
\newcommand{\psfigaddtwo}[7]{%
\setcounter{pict@widthfst}{1*\ratio{#2+\pict@scale/2}{\pict@scale}}
\setcounter{pict@widthscd}{1*\ratio{#2+#4+\pict@scale/2}{\pict@scale}}
\setcounter{pict@widthtot}{1*\ratio{#2+#4+#6+\pict@scale/2}{\pict@scale}}
\setcounter{pict@height}{1*\ratio{#3+\pict@scale/2}{\pict@scale}}
\setlength{\unitlength}{\pict@scale}
\hbox{\hspace{-\fill}\begin{picture}(\thepict@widthtot,\thepict@height)
\put(0,0){\psfig{figure=#1,width=#2,height=#3,clip=}}
\put(\thepict@widthscd,0){\psfig{figure=#5,width=#6,height=#3,clip=}}
\SetScale{0.283466457}
\SetWidth{1.763889}
{#7}
\end{picture}}
}
\newcommand{\psfigror}[4]{%
\setcounter{pict@width}{1*\ratio{#2+\pict@scale/2}{\pict@scale}}
\setcounter{pict@height}{1*\ratio{#3+\pict@scale/2}{\pict@scale}}
\setlength{\unitlength}{\pict@scale}
\hbox{\begin{picture}(\thepict@width,\thepict@height)
\put(0,\thepict@height){\psfig{figure=#1,width=#3,height=#2,clip=,angle=270}}
\SetScale{0.283466457}
\SetWidth{1.763889}
{#4}
\end{picture}}
}
\newcommand{\psfigrol}[4]{%
\setcounter{pict@width}{1*\ratio{#2+\pict@scale/2}{\pict@scale}}
\setcounter{pict@height}{1*\ratio{#3+\pict@scale/2}{\pict@scale}}
\setlength{\unitlength}{\pict@scale}
\hbox{\begin{picture}(\thepict@width,\thepict@height)
\put(0,0){\psfig{figure=#1,width=#3,height=#2,clip=,angle=90}}
\SetScale{0.283466457}
\SetWidth{1.763889}
{#4}
\end{picture}}
}
\catcode`\@=12 
\newlength\listtextwidth

\catcode`\@=11 
\newlength{\@tabfninsert}
\newlength{\@tabfnwidth}
\newcommand{\tabfootnote}[2]{%
  \setlength{\@tabfninsert}{0.8em}
  \setlength{\@tabfnwidth}{\textwidth}
  \addtolength{\@tabfnwidth}{-\@tabfninsert}
  \addtolength{\@tabfnwidth}{-0.4em}
  \noindent\makebox[\@tabfninsert][r]{\footnotesize$^{#1}$\hfil}\hfill%
  \parbox[t]{\@tabfnwidth}{\footnotesize #2\hfill}}
\catcode`\@=12 

\def\JHEP{JHEP}

\def\etjet{E_T^{\rm jet}}
\def\etajet{\eta^{\rm jet}}
\def\phijet{\varphi^{\rm jet}}

\def\etaj{\eta^{\rm jet1}}
\def\etajj{\eta^{\rm jet2}}
\def\etj{E_T^{\rm jet1}}
\def\etjj{E_T^{\rm jet2}}

\def\etcal{E_{T,{\rm cal}}^{\rm jet}}
\def\etacal{\eta_{\rm cal}^{\rm jet}}
\def\phical{\varphi_{\rm cal}^{\rm jet}}

\def\etaphi{\eta-\varphi}

\def\etar{-1<\etajet<2.5}
\def\etacr{-1<\etacal<2.5}

\def\costr{\cost<0.8}

\def\seta{d\sigma/d\etajet}
\def\set{d\sigma/d\etjet}
 
\def\q2{Q^2}
\def\pb1{pb$^{-1}$}
\def\gp{\gamma p}

\def\g2{GeV$^2$}

\def\mj{M^{\rm jj}}

\def\cost{\vert\cos\theta^*\vert}

\def\smj{d\sigma/d\mj}
\def\scost{d\sigma/d\cost}
\def\sccos{d\sigma/d\ccos}

\def\kt{k_T}

\def\xo{x_{\gamma}^{\rm obs}}

\def\qq{q\bar q}

\def\colab#1{#1 Coll.}

\def\ccos{\cos\theta^*}

\def\ptmis{p_T^{\rm miss}}

\def\ns{n_{\rm subjet}}

\def\yc{y_{\rm cut}}

\def\sxo{d\sigma/d\xo}

\def\z0{Z^0}
\def\mz{M_Z}

\def\as{\alpha_s}
\def\oalphas2{{\cal O}(\alpha\as^2)}

\def\asz{\as(\mz)}
\def\asmz#1#2#3#4#5#6{\asz = #1\pm #2\ {\rm (stat.)}\ ^{+#4}_{-#3}\ {\rm (exp.)}\ ^{+#6}_{-#5}\ {\rm (th.)}}

\def\mr2{\mu_R^2}
\def\mf2{\mu_F^2}

\def\etal{et al.}

\def\wrn{$142<\wgp<293$ GeV}
\def\wgp{W_{\gp}}

\begin{document}

\prepnum{{DESY--04--072}}

\title{
Substructure dependence of jet cross sections at HERA and
determination of {\boldmath $\as$}
}                                                       

\author{ZEUS Collaboration}
\date{May 2004}

\abstract{
Jet substructure and differential cross sections for jets produced in
the photoproduction and deep inelastic $ep$ scattering regimes
have been measured with the ZEUS detector at HERA using an integrated
luminosity of $82.2$ \pb1. The substructure of jets has been studied
in terms of the jet shape and subjet multiplicity for jets with
transverse energies $\etjet>17$ GeV. The data are well described by
the QCD calculations. The jet shape and subjet multiplicity are used
to tag gluon- and quark-initiated jets. Jet cross sections as
functions of $\etjet$, jet pseudorapidity, the jet-jet scattering
angle, dijet invariant mass and the fraction of the photon energy
carried by the dijet system are presented for gluon- and quark-tagged
jets. The data exhibit the behaviour expected from the underlying parton
dynamics. A value of $\as(\mz)$ of 
$\asmz{0.1176}{0.0009}{0.0026}{0.0009}{0.0072}{0.0091}$
was extracted from the measurements of jet shapes in deep inelastic
scattering.
}

\makezeustitle

\def\3{\ss}

\pagenumbering{Roman}

\begin{center}
{                      \Large  The ZEUS Collaboration              }
\end{center}

  S.~Chekanov,
  M.~Derrick,
  J.H.~Loizides$^{   1}$,
  S.~Magill,
  S.~Miglioranzi$^{   1}$,
  B.~Musgrave,
  \mbox{J.~Repond},
  R.~Yoshida\\
 {\it Argonne National Laboratory, Argonne, Illinois 60439-4815},
 USA~$^{n}$
\par \filbreak

  M.C.K.~Mattingly \\
 {\it Andrews University, Berrien Springs, Michigan 49104-0380}, USA
\par \filbreak

  N.~Pavel \\
  {\it Institut f\"ur Physik der Humboldt-Universit\"at zu Berlin,
    Berlin, Germany}
% ...

\newpage

$^{\    1}$ also affiliated with University College London, UK \\
% ...

\newpage

\begin{tabular}[h]{rp{14cm}}
$^{a}$ &  supported by the Natural Sciences and Engineering Research
Council of Canada (NSERC) \\
$^{b}$ &  supported by the German Federal Ministry for Education and
Research (BMBF), under contract numbers HZ1GUA 2, HZ1GUB 0, HZ1PDA 5,
HZ1VFA 5\\
$^{c}$ &  supported in part by the MINERVA Gesellschaft f\"ur
Forschung GmbH, the Israel Science Foundation (grant no. 293/02-11.2),
the U.S.-Israel Binational Science Foundation and the Benozyio Center
for High Energy Physics\\
$^{d}$ &  supported by the German-Israeli Foundation and the Israel
Science Foundation\\
$^{e}$ &  supported by the Italian National Institute for Nuclear
Physics (INFN) \\
$^{f}$ &  supported by the Japanese Ministry of Education, Culture,
Sports, Science and Technology (MEXT) and its grants for Scientific
Research\\
$^{g}$ &  supported by the Korean Ministry of Education and Korea
Science and Engineering Foundation\\
$^{h}$ &  supported by the Netherlands Foundation for Research on
Matter (FOM)\\
$^{i}$ &  supported by the Polish State Committee for Scientific
Research, grant no. 620/E-77/SPB/DESY/P-03/DZ 117/2003-2005\\
$^{j}$ &  partially supported by the German Federal Ministry for
Education and Research (BMBF)\\
$^{k}$ &  supported by RF President grant N 1685.2003.2 for the
leading scientific schools and by the Russian Ministry of Industry,
Science and Technology through its grant for Scientific Research on
High Energy Physics\\
$^{l}$ &  supported by the Spanish Ministry of Education and Science
through funds provided by CICYT\\
$^{m}$ &  supported by the Particle Physics and Astronomy Research
Council, UK\\
$^{n}$ &  supported by the US Department of Energy\\
$^{o}$ &  supported by the US National Science Foundation\\
$^{p}$ &  supported by the Polish Ministry of Scientific Research and
Information Technology, grant no. 112/E-356/SPUB/DESY/P-03/DZ
116/2003-2005\\
$^{q}$ &  supported by the Polish State Committee for Scientific
Research, grant no. 115/E-343/SPUB-M/DESY/P-03/DZ 121/2001-2002, 2
P03B 07022\\
\end{tabular}

\newpage

\pagenumbering{arabic} 
\pagestyle{plain}

\section{Introduction}
\label{intro}
Jet production in $ep$ collisions provides a fruitful testing ground
of perturbative QCD (pQCD). Measurements of differential cross
sections for jet 
production~\cite{pl:b507:70,pl:b531:9,epj:c23:615,pl:b547:164,pl:b560:7,epj:c31:149,epj:c23:13,*pl:b443:394,epj:c19:289,epj:c25:13,*pl:b542:193,*epj:c29:497,*epj:c19:429,*pl:b515:17}
have allowed detailed studies of
parton dynamics, tests of the proton and photon parton distribution
functions (PDFs) as well as precise determinations of the strong coupling
constant, $\as$. Most of these measurements refer to the production
of jets irrespective of their partonic origin -- quarks or gluons --
and, therefore, have only provided general tests of the partonic
structure of the short-distance process and of combinations of the
proton and/or photon PDFs. The identification of quark- and
gluon-initiated jets would allow more stringent tests of the QCD
predictions. Such measurements of the production of jets containing a
heavy quark have been made by means of tagging specific decay 
channels~\cite{pl:b565:87,*hep-ex-0312057}. In the present
study, quark- and gluon-initiated jets are identified on a statistical 
basis by utilising their internal structure.

Two kinematic regimes have been studied: photoproduction ($\gp$) and
neutral current (NC) deep inelastic $ep$ scattering
(DIS). Photoproduction at HERA is studied by means of $ep$ scattering
at low four-momentum transfers ($\q2\approx 0$, where $\q2$ is the
virtuality of the exchanged photon). In photoproduction, two
types of QCD processes contribute to jet production at leading order
(LO)~\cite{pl:b79:83,*np:b166:413,*pr:d21:54,*zfp:c6:241,proc:hera:1987:331,*prl:61:275,*prl:61:682,*pr:d39:169,*zfp:c42:657,*pr:d40:2844}:
either the photon interacts directly with a parton in the proton (the
direct process) or the photon acts as a source of partons which
scatter off those in the proton (the resolved process). Jet production
in NC DIS up to LO in $\as$ proceeds as in the quark-parton model
($Vq\rightarrow q$, where $V=\gamma$ or $\z0$) or via the boson-gluon
fusion ($Vg\rightarrow \qq$) and QCD-Compton ($Vq\rightarrow qg$)
processes.

This paper is organised as follows. Section~\ref{thexp} gives the
theoretical expectations for the measurements presented. The
experimental set-up and data selection are described in
Sections~\ref{expcon} and \ref{datsel},
respectively. Section~\ref{qcdcal} explains the QCD calculations used
in this analysis. The corrections applied to the data and systematic
uncertainties are given in Section~\ref{corsys}.
The results on the mean integrated jet shape and subjet multiplicity
in photoproduction and NC DIS are presented in Section~\ref{jetsub}. 
The measurements of differential inclusive jet cross sections as a
function of the jet pseudorapidity\footnote{The ZEUS coordinate system
  is a right-handed Cartesian system, with the $Z$ axis pointing in
  the proton beam direction, referred to as the ``forward direction'',
  and the $X$ axis pointing left towards the centre of HERA. The
  coordinate origin is at the nominal interaction point.}, $\etajet$,
and jet transverse energy, $\etjet$, for samples of jets in the
photoproduction and NC DIS regimes, separated according to their shape
and subjet multiplicity, are presented in Section~\ref{xsections}. In
addition, measurements of differential dijet cross sections in
photoproduction as a function of $\ccos$, where $\theta^*$ is the
angle between the jet-jet axis and the beam direction in the dijet
centre-of-mass system, the dijet invariant mass, $\mj$, and the
fraction of the photon momentum participating in the production of the
two jets with highest $\etjet$, $\xo$, are also presented for a
variety of tagged-jet configurations. The results are compared to
leading-logarithm parton-shower calculations and used to investigate
the dynamics underlying the production of specific tagged-jet final
states. Finally, in Section~\ref{asset}, the measurements of the mean
integrated jet shape in NC DIS are compared to next-to-leading-order
(NLO) QCD predictions and used to extract $\as$.

\section{Theoretical expectations}
\label{thexp}

The internal structure of a jet depends mainly on the type of primary
parton -- quark or gluon -- from which it originated and to a lesser
extent on the particular hard scattering process. At sufficiently high
jet transverse energy, where the influence of fragmentation effects
becomes negligible, the internal structure of a jet is calculable in
pQCD. Such calculations predict that gluon-initiated jets are broader
than quark-initiated jets due to the larger colour charge of the
gluon. The jet shape~\cite{prl:69:3615} and subjet
multiplicity~\cite{np:b383:419,*np:b421:545,*pl:b378:279,*jhep:9909:009}
can be used to study the internal structure of the jets and to
classify them: a ``broad''-jet sample is enriched in gluon-initiated
jets, whereas a ``narrow''-jet sample is enriched in quark-initiated
jets. Thus, measurements of cross sections for broad- and narrow-jet
samples allow the contributing hard-scattering subprocesses to be
disentangled.

The dominant partonic subprocesses responsible for jet photoproduction
in the kinematic region presented in this paper are 
$\gamma g\rightarrow \qq$ and $q_{\gamma}g_p\rightarrow qg$, where
$q_{\gamma}$ ($g_p$) denotes a quark (gluon) coming from the photon
(proton). The kinematics of these two-to-two subprocesses are such
that the majority of the jets in the region $\etajet<0$ originate from
outgoing quarks, whereas the fraction of gluon-initiated jets increases
as $\etajet$ increases.

The distribution in $\theta^*$ reflects the underlying parton
dynamics and is sensitive to the spin of the exchanged particle. In the
case of direct-photon interactions, the contributing subprocesses at LO 
QCD are (i) $\gamma q(\bar q)\rightarrow gq(\bar q)$ and (ii)
$\gamma g\rightarrow \qq$, which involve quark exchange. The behaviour
of the dijet angular distribution as $\cost\rightarrow 1$ is the same
for all direct subprocesses and proportional to $(1-\cost)^{-1}$. In
the case of resolved-photon interactions, the contributing
subprocesses are $qg\rightarrow qg$, 
$qq^{\prime}\rightarrow qq^{\prime}$, $gg\rightarrow gg$,~...~. The
dominant subprocesses are those that involve gluon exchange and the
behaviour of the dijet angular distribution as $\cost\rightarrow 1$ is
proportional to $(1-\cost)^{-2}$. The different behaviour of the dijet
angular distribution for resolved and direct processes has been
measured in photoproduction at
HERA~\cite{pl:b384:401,epj:c23:615}. The study of the angular
distribution for dijet events with tagged quark- and/or
gluon-initiated jets in the final state, provides then a handle to
investigate the underlying parton dynamics further.

Measurements of jet substructure in NC DIS allow a determination of 
$\as$. In zeroth-order pQCD, a jet consists of only one parton and the
jets have no substructure. The first non-trivial contribution to the jet
substructure is given by ${\cal O}(\as)$ processes in which, e.g., a
quark radiates a gluon at a small angle; these are
proportional to the rate of parton emission and thus to $\as$. For DIS
in the laboratory frame, all necessary QCD corrections to the jet
cross sections for the determination of $\as$ from the jet
substructure are available.

\subsection{Jet-shape and subjet-multiplicity definitions}
\label{jssmdef}
The $\kt$ cluster algorithm~\cite{np:b406:187} was used in the
longitudinally invariant inclusive mode~\cite{pr:d48:3160} to define
jets in the hadronic final state.
The integrated jet shape, $\psi(r)$, is defined using only
those particles belonging to the jet as the fraction of the jet
transverse energy that lies inside a cone in the $\etaphi$ plane of
radius $r$ concentric with the jet axis:

$$\psi(r)=\frac{E_T(r)}{\etjet},$$
where $E_T(r)$ is the transverse energy within the given cone of
radius $r$. The mean integrated jet shape, $\langle\psi(r)\rangle$, is
defined as the averaged fraction of the jet transverse energy inside
the cone $r$:

$$\langle\psi(r)\rangle=\frac{1}{N_{\rm jets}}\sum_{\rm jets}\frac{E_T(r)}{\etjet},$$
where $N_{\rm jets}$ is the total number of jets in the sample.

The integrated jet shape is calculated at LO in pQCD as the fraction of
the jet transverse energy, due to parton emission, that lies in the
cone segment between $r$ and $R=1$:
$$\langle 1-\psi(r)\rangle ={\int dE_T\  (E_T/\etjet)\ [d\sigma (ep\rightarrow 2\ {\rm partons})/dE_T]\over\sigma_{\rm jet}(\etjet)},$$
where $\sigma_{\rm jet}(\etjet)$ is the cross section for inclusive jet
production. In the NLO QCD predictions of the integrated jet shape,
the numerator in the above formula is calculated to 
${\cal O}(\alpha\as^2)$ and the denominator to ${\cal O}(\alpha\as)$.

Studies of QCD using jet production in NC DIS at HERA are usually
performed in the Breit frame. The analysis of jet shapes presented
here was performed in the laboratory frame, since calculations of this
observable in the Breit frame can, at present, only be performed to
${\cal O}(\as)$, precluding a reliable determination of
$\as$. However, calculations of the jet shape can be performed up to
${\cal O}(\as^2)$ in the laboratory frame. Furthermore, the analysis
was performed in the kinematic region defined by $\q2>125$ \g2\ since,
at lower values of $\q2$, the sample of events with at least one jet
with $\etjet>17$ GeV is dominated by dijet events. The calculation of
the integrated jet shape for dijet events can be performed only
up to ${\cal O}(\as)$, which would severely restrict the accuracy of
the predictions.

Subjets were resolved within a jet by considering all particles
associated with the jet and repeating the application of the
$\kt$ cluster algorithm until, for every pair of particles $i$ and
$j$ the quantity 
$d_{ij}={\rm min}(E_{T,i},E_{T,j})^2\cdot((\eta_i-\eta_j)^2+(\varphi_i-\varphi_j)^2)$, 
where $E_{T,i}$, $\eta_i$ and $\varphi_i$ are the transverse energy,
pseudorapidity and azimuth of particle $i$, respectively,
was greater than $d_{\rm cut}=\yc(\etjet)^2$. All remaining
clusters were called subjets. The subjet multiplicity, $\ns$, depends
upon the value chosen for the resolution parameter $\yc$. The mean
subjet multiplicity, $\langle\ns\rangle$, is defined as the average
number of subjets contained in a jet at a given value of $\yc$:

$$\langle\ns(\yc)\rangle=\frac{1}{N_{\rm jets}}\sum_{i=1}^{N_{\rm jets}} \ns^i(\yc),$$
where $\ns^i(\yc)$ is the number of subjets in jet $i$.

\section{Experimental set-up}
\label{expcon}
The data used in this analysis were collected during the 1998-2000
running period, when HERA operated with protons of energy
$E_p=920$~GeV and electrons or positrons\footnote{Here and in the
  following, the term ``electron'' denotes generically both the
  electron ($e^-$) and the positron ($e^+$).}
of energy $E_e=27.5$~GeV, and correspond to an integrated luminosity
of $82.2\pm 1.9$~\pb1.

A detailed description of the ZEUS detector can be found
elsewhere~\cite{pl:b293:465,zeus:1993:bluebook}. A brief outline of
the components that are most relevant for this analysis is given below.
Charged particles are tracked in the central tracking detector 
(CTD)~\cite{nim:a279:290,*npps:b32:181,*nim:a338:254}, which operates
in a magnetic field of $1.43\Tesla$ provided by a narrow
superconducting solenoid. The CTD consists of 72~cylindrical
drift-chamber layers, organized in nine superlayers covering the
polar-angle region 
\mbox{$15^\circ<\theta<164^\circ$}. The transverse-momentum resolution
for full-length tracks can be parameterised as 
$\sigma(p_T)/p_T=0.0058p_T\oplus0.0065\oplus0.0014/p_T$, with $p_T$ in
$\Gev$. The tracking system was used to measure the interaction vertex
with a typical resolution along (transverse to) the beam direction of
0.4~(0.1)~cm and to cross-check the energy scale of the calorimeter.

The high-resolution uranium--scintillator calorimeter
(CAL)~\cite{nim:a309:77,*nim:a309:101,*nim:a321:356,*nim:a336:23} covers 
$99.7\%$ of the total solid angle and consists 
of three parts: the forward (FCAL), the barrel (BCAL) and the rear (RCAL)
calorimeters. Each part is subdivided transversely into towers and
longitudinally into one electromagnetic section (EMC) and either one
(in RCAL) or two (in BCAL and FCAL) hadronic sections (HAC). The
smallest subdivision of the calorimeter is called a cell. Under
test-beam conditions, the CAL single-particle relative energy
resolutions were $\sigma(E)/E=0.18/\sqrt E$ for electrons and
$\sigma(E)/E=0.35/\sqrt E$ for hadrons, with $E$ in GeV.

The luminosity was measured from the rate of the bremsstrahlung
process $ep\rightarrow e\gamma p$. The resulting small-angle
energetic photons were measured by the luminosity 
monitor~\cite{desy-92-066,*zfp:c63:391,*acpp:b32:2025}, a
lead-scintillator calorimeter placed in the HERA tunnel at $Z=-107$ m.

\section{Data selection and jet search}
\label{datsel}
A three-level trigger system was used to select events
online~\cite{zeus:1993:bluebook,proc:chep:1992:222}. At the first
level, events were triggered by a coincidence of a regional or
transverse energy sum in the CAL and at least one track from the
interaction point measured in the CTD. At the second level, a total
transverse energy of at least $8$~GeV, excluding the energy in the
eight CAL towers immediately surrounding the forward beampipe, was
required, and cuts on CAL energies and timing were used to suppress
events caused by interactions between the proton beam and residual gas
in the beampipe. At the third level, a jet algorithm was applied to
the CAL cells and jets were reconstructed using the energies and
positions of these cells. Events with at least one (two) jet(s) with
$E_T>10\ (6)$~GeV and $\eta<2.5$ were accepted for the inclusive jet
(dijet) samples. For systematic trigger studies, all events with a
total transverse energy of at least $25$~GeV, excluding the energy in
the eight CAL towers immediately surrounding the forward beampipe,
were accepted. No jet algorithm was applied in this case.

In the offline selection, a reconstructed event vertex consistent with
the nominal interaction position was required and cuts based on the
tracking information were applied to reduce beam-induced interactions
and cosmic-ray events. The main steps of the selection of
photoproduction and DIS events are briefly explained below.

\subsection{Selection of the photoproduction sample}
\label{gpsel}
Events from collisions between quasi-real photons and protons were
selected offline using similar criteria to those reported in a previous 
publication~\cite{pl:b531:9,pl:b560:7}. Charged current DIS events
were rejected by requiring the total missing transverse momentum,
$\ptmis$, to be small compared to the total transverse energy, 
$E^{\rm tot}_T$, \mbox{$\ptmis/\sqrt{E^{\rm tot}_T}<2\ \sqrt{\rm GeV}$}. 
Any NC DIS events with an identified scattered-electron candidate in
the CAL~\cite{nim:a365:508,*nim:a391:360} were removed from the sample
using the method described previously~\cite{pl:b322:287}. The
remaining background from NC DIS events was estimated by Monte Carlo
(MC) techniques to be below $0.3\%$ and was neglected.

The selected sample consisted of events from $ep$ interactions with
$\q2\lesssim 1$ \g2\ and a median $\q2\approx 10^{-3}$~\g2. The $\gp$
centre-of-mass energy is given by $\wgp=\sqrt{sy}$, where $y$ is the
inelasticity variable and $\sqrt{s}$ is the $ep$ centre-of-mass energy,
$s=4E_eE_p$. The inelasticity variable was reconstructed using the
method of Jacquet-Blondel~\cite{proc:epfacility:1979:391}, 
$y_{\rm JB}=(E-p_Z)/2E_e$, where $E$ is the total CAL energy and $p_Z$
is the $Z$ component of the energy measured in the CAL cells. The
value of $y$ was systematically underestimated by $\sim 20\%$
with an r.m.s. of $\sim 10\%$. This effect, which was due to energy
lost in the inactive material in front of the CAL and to particles
lost in the rear beampipe, was satisfactorily reproduced by the MC
simulation of the detector. The MC event samples were therefore used
to correct for this underestimation. The photoproduction sample was
restricted to \mbox{\wrn}~\cite{pl:b531:9,pl:b560:7}.

\subsection{Selection of the NC DIS sample}
\label{dissel}
Events from NC DIS interactions were selected offline using similar
criteria to those reported in a previous
publication~\cite{pl:b558:41}. The scattered-electron candidate was
identified using the pattern of energy deposits in the
CAL~\cite{nim:a365:508,*nim:a391:360}. The energy, $E_e^{\prime}$, and
polar angle, $\theta_e$, of the electron candidate were
also determined from the CAL measurements. The double-angle
method~\cite{proc:hera:1991:23,*proc:hera:1991:43}, which uses
$\theta_e$ and an angle $\gamma$ that corresponds, in the quark-parton
model, to the direction of the scattered quark, was used to
reconstruct $\q2$, $\q2_{\rm DA}$. The angle $\gamma$ was
reconstructed using the CAL measurements of the hadronic final
state.

An electron candidate of energy $E_e^{\prime}>10$~GeV was required to
ensure a high and well understood electron-finding efficiency and to
suppress background from photoproduction. The inelasticity variable as
reconstructed from the electron, $y_e$, was required to be below
$0.95$. This condition removed events in which fake electron
candidates from photoproduction background were found in the FCAL. The
requirements $38<(E-p_Z)<65$~GeV, to remove events with large
initial-state radiation and to reduce further the photoproduction
background, and 
\mbox{$\ptmis/\sqrt{E^{\rm tot}_T}<3 \ \sqrt{\rm GeV}$}, to remove
cosmic rays and beam-related background, were applied. The kinematic
range was restricted to $\q2_{\rm DA}>125$ \g2.

\subsection{Jet search}
\label{jets}
The $\kt$ cluster algorithm was used in the longitudinally invariant
inclusive mode to reconstruct jets in the hadronic final state from
the energy deposits in the CAL cells. For DIS events, the jet
algorithm was applied after excluding those cells associated with the
scattered-electron candidate. The jet search was performed in the
$\etaphi$ plane of the laboratory frame. The jet variables were
defined according to the Snowmass
convention~\cite{proc:snowmass:1990:134}. The jets reconstructed from
the CAL cell energies are called calorimetric jets and the variables
associated with them are denoted by $\etcal$, $\etacal$ and
$\phical$. A total of $199\;\! 237\ (98\;\! 240)$~events with at least
one jet satisfying $\etcal>13$~GeV and $\etacr$ were selected in the
photoproduction (DIS) sample.

\section{QCD calculations}
\label{qcdcal}
\subsection{Leading-logarithm parton-shower Monte Carlo models}
\label{mc}
The programs {\sc Pythia}~6.1~\cite{cpc:82:74,*cpc:135:238} and
{\sc Herwig}~6.1~\cite{cpc:67:465,*jhep:0101:010} were used to
generate photoproduction events for resolved and direct
processes. Events were generated using
GRV-HO~\cite{pr:d45:3986,*pr:d46:1973} for the photon and
CTEQ4M~\cite{pr:d55:1280} for the proton PDFs. In both generators, the
partonic processes are simulated using LO matrix elements, with the
inclusion of initial- and final-state parton showers. Fragmentation
into hadrons is performed using the Lund string
model~\cite{prep:97:31} as implemented in 
{\sc Jetset}~\cite{cpc:82:74,*cpc:135:238,cpc:39:347,*cpc:43:367} in
the case of {\sc Pythia}, and a cluster model~\cite{np:b238:492} in
the case of {\sc Herwig}. Samples of {\sc Pythia} including
multiparton interactions (MI)~\cite{pr:d36:2019} with a minimum
transverse momentum for the secondary scatter of
1~GeV~\cite{epj:c2:61} were used to study the effects of a possible
``underlying event''.

Neutral current DIS events including radiative effects were simulated
using the {\sc Heracles}~4.6.1~\cite{cpc:69:155,*spi:www:heracles}
program with the {\sc
  Djangoh}~1.1~\cite{cpc:81:381,*spi:www:djangoh11} interface to the
hadronisation programs. {\sc Heracles} includes corrections for
initial- and final-state radiation, vertex and propagator terms, and
two-boson exchange. The QCD cascade is simulated using the
colour-dipole
model (CDM)~\cite{pl:b165:147,*pl:b175:453,*np:b306:746,*zfp:c43:625}
including the LO QCD diagrams as implemented in
{\sc Ariadne}~4.08~\cite{cpc:71:15,*zfp:c65:285} and, as a systematic
check of the final results, with the MEPS model of {\sc Lepto}
6.5~\cite{cpc:101:108}. Both MC programs use the Lund string model for
the hadronisation. The CTEQ5D~\cite{epj:c12:375} proton PDFs were used
for these simulations.

These MC samples were used to correct the data to
the hadron level, defined as those hadrons with lifetime 
$\tau\geq 10$~ps. For this purpose, the generated events were passed
through the ZEUS detector- and trigger-simulation programs based on
{\sc Geant}~3.13~\cite{tech:cern-dd-ee-84-1}. They were reconstructed and
analysed by the same program chain as the data. The jet search was
performed using the energy measured in the CAL cells in the
same way as for the data. The same jet algorithm was also applied to
the final-state particles and to the partons available after the
parton shower; the jets found in this way are referred to
as hadronic and partonic jets, respectively.

Electroweak-radiative, hadronisation and $\z0$-exchange effects are
not at present included in the NLO QCD programs described in
Section~\ref{nlo}. Therefore, samples of MC events were generated
without electroweak-radiative effects so that the data could be
corrected for these effects when comparing with the NLO QCD
predictions. Additional samples of MC events without $\z0$-exchange
effects were generated to correct the NLO QCD calculations for these
effects and for hadronisation.

\subsection{NLO QCD calculations}
\label{nlo}

The NLO QCD calculations of the mean integrated jet shapes
in DIS\footnote{Only QCD calculations of jet shapes in DIS are compared
  to the data because there is no NLO program available for similar
  calculations in photoproduction.} are based on the program
{\sc Disent}~\cite{np:b485:291}. The calculations use a
generalised version of the subtraction method~\cite{np:b178:421} and
are performed in the massless $\overline{\rm MS}$ renormalisation and
factorisation schemes. The number of flavours was set to five; the
renormalisation ($\mu_R$) and factorisation ($\mu_F$) scales were both
set to $\mu_R=\mu_F=Q$; $\as$ was calculated at two loops using 
$\Lambda^{(5)}_{\overline{\rm MS}}=220$~MeV, which corresponds to
$\asz=0.1175$. The MRST99~\cite{epj:c4:463,*epj:c14:133}
parameterisations of the proton PDFs were used as defaults for the
comparisons with the data. The calculations obtained with {\sc Disent}
were cross-checked by using the program
{\sc Disaster}++~\cite{disaster1,*disaster2}. The differences were
smaller than $0.5\%$ for $r\geq 0.3$.

Since the measurements refer to jets of hadrons, whereas the QCD
calculations refer to partons, the predictions were corrected to the
hadron level using the MC samples described in Section~\ref{mc}. The
multiplicative correction factor, $C_{\rm had}$, defined as 
$\langle\psi(r)\rangle_{\rm had}/\langle\psi(r)\rangle_{\rm par}$,
where $\langle\psi(r)\rangle_{\rm par\ (had)}$ is the mean integrated
jet shape before (after) the hadronisation process, was estimated with
both the CDM and MEPS models. The procedure for applying hadronisation
corrections to the NLO QCD calculations was verified by checking that
the MC predictions for the integrated jet shape at the parton level
reproduced the NLO QCD calculations. The agreement was well within
$0.2\%$ after adjusting the contributions of the different subprocesses
($eq\rightarrow eq$, $eg\rightarrow e \qq$ and $eq\rightarrow eqg$) in
the MC to reproduce the $\etajet$ cross section and jet shape of the
NLO calculations. The values of $C_{\rm had}$ obtained with the CDM
model were taken as the defaults; the predictions from the two models
were in good agreement. The value of $C_{\rm had}$ was $0.95$ at
$r=0.5$ for $\etjet=21$ GeV and approached unity as $\etjet$ increased.

\section{Corrections and systematic uncertainties}
\label{corsys}
\subsection{Jet energy corrections}
\label{jenecor}
The comparison of the reconstructed jet variables for the hadronic and 
the calorimetric jets in simulated events showed that no correction
was needed for $\etajet$ and $\phijet$ ($\etajet\simeq\etacal$ and
$\phijet \simeq \phical$). However, the transverse energy of the
calorimetric jet was an underestimate of the corresponding
hadronic jet energy by an average of $\sim 15\%$, with an r.m.s. of
$\sim 10\%$. This underestimation was mainly due to the energy lost by
the particles in the inactive material in front of the CAL. The
transverse-energy corrections to calorimetric jets, as a function of
$\etacal$ and $\etcal$ and averaged over $\phical$, were determined
for the photoproduction and DIS samples using the corresponding
MC-generated
events~\cite{epj:c23:615,pl:b531:9,pl:b558:41}. Henceforth, jet
variables without subscript refer to the corrected values. After these
corrections to the jet transverse energy, events with at least one jet
satisfying $\etjet>17$~GeV and $\etar$ were retained for the studies
of inclusive jet observables and events with at least two jets with
$\etj>17$ GeV, $\etjj>14$ GeV and $\etar$, where the jets are labelled
in decreasing $\etjet$ order, were retained for the dijet studies.

\subsection{Acceptance corrections}
\label{corfac}
Using the selected data sample of inclusive jets with $\etjet>17$~GeV
and $-1<\etajet<2.5$, in the kinematic region defined by
$\q2< 1$~\g2\ and \wrn\ in photoproduction and $\q2>125$~\g2\ in DIS,
the mean integrated jet shape and mean subjet multiplicity were
reconstructed using the CAL cells and corrected to the hadron level by
MC techniques. The corrected values were determined bin-by-bin 
using the MC samples separately for each region of
$\etajet$ and $\etjet$ studied. For this approach to be valid, the
distributions of the uncorrected integrated jet shape and 
subjet multiplicity in the data must be well
described by the MC simulations at the detector level. This condition
was in general satisfied by the MC photoproduction models. 
In DIS, to obtain the best description of the uncorrected jet shape in
the data by the MC simulations, the contributions of the different
subprocesses ($eq\rightarrow eq$, $eg\rightarrow e \qq$ and 
$eq\rightarrow eqg$) were reweighted so as to reproduce the mean
integrated jet shape and $\etajet$ distributions in the data. This
procedure was applied to the simulations of CDM and MEPS and
for each region in $\etajet$ and $\etjet$. The correction factors
were evaluated using these tuned versions of CDM and MEPS in DIS and
the default mixture of resolved and direct processes in {\sc Pythia},
{\sc Herwig} and {\sc Pythia} MI.

Differential jet cross sections were measured using the selected data
sample of inclusive jet events. Dijet differential cross sections in
photoproduction were measured in the same kinematic region as above
and refer to the two highest-$\etjet$ jets of hadrons in the event
with $\etj>17$~GeV, $\etjj>14$~GeV and $\etar$. The {\sc Pythia} (CDM)
MC samples of photoproduction (DIS) events were used to compute the
acceptance corrections to the jet distributions. These correction
factors took into account the efficiency of the trigger, the selection
criteria and the purity and efficiency of the jet reconstruction. The
inclusive jet cross sections were obtained by applying bin-by-bin
corrections to the measured distributions. The samples of {\sc Herwig}
and MEPS were used to compute the systematic uncertainties coming from
the fragmentation and parton-shower models in photoproduction and DIS,
respectively (see Section~\ref{syst}).

\subsection{Systematic uncertainties}
\label{syst}
A study of the main sources contributing to the systematic
uncertainties of the measurements was performed. The sources
considered in the photoproduction measurements are:
\begin{itemize}
 \item the effect of the treatment of the parton shower and hadronisation
   was estimated by using the {\sc Herwig} generator to evaluate the
   correction factors;
 \item the effect of the simulation of the trigger was evaluated by
   using an alternative trigger configuration, as explained in
   Section~\ref{datsel}, in both data and MC events;
 \item the effect of the uncertainty on $\wgp$ was
   estimated by varying $y_{\rm JB}$ by its uncertainty of $\pm 1\%$
   in simulated events;
 \item the effect of the uncertainty on the parameterisations of the
   proton and photon PDFs was estimated by using alternative sets of PDFs
   in the MC simulation to calculate the correction factors;
 \item the effect of the uncertainty on the absolute energy scale of
   the calorimetric jets was estimated by varying $\etjet$ by its
   uncertainty of $\pm 1\%$ in simulated  events. The method used was
   the same as in earlier
   publications~\cite{epj:c23:615,pl:b531:9,proc:calor:2002:767} and
   verified with the 98-00 data sample~\cite{pl:b560:7}.
\end{itemize}

In the DIS regime, the main sources contributing to the systematic
uncertainties of the measurements are:
\begin{itemize}
 \item the effect of the treatment of the parton shower was estimated
   by using the MEPS model to evaluate the correction factors;
\item the effect of the simulation of the trigger was evaluated by
   using an alternative trigger configuration, as explained in
   Section~\ref{datsel}, in both data and MC events;
\item the effect of the uncertainty on the scattered-electron
  identification was estimated by using an alternative
  technique~\cite{epj:c11:427} to select the candidates, in both data
  and MC events;
\item the effect of the uncertainty of $\pm 1\%$ in the absolute
  energy scale of the jets was applied to the simulated events;
\item the effect of the uncertainty of $\pm 1\%$ in the absolute
  energy scale of the scattered-electron candidate was applied to the
  simulated events.
\end{itemize}

These uncertainties, for each regime, were added in quadrature to the
statistical uncertainty of the data and are shown as error bars in the
figures showing the substructure measurements. For the cross-section
measurements, the uncertainty arising from that on the absolute energy
scale of the jets is shown separately. The uncertainty in the
luminosity determination of $2.25\%$ was not included.

\section{Measurements of jet substructure}
\label{jetsub}

\subsection{Jet-shape measurements}
\label{resjsgp}
The measured mean integrated jet shape as a function of $r$,
$\langle\psi(r)\rangle$, for different regions in $\etajet$ is shown
in Fig.~\ref{fig1} and Table~\ref{tabthree} for the photoproduction
regime. The jets broaden as $\etajet$ increases. Leading-logarithm
parton-shower predictions from {\sc Pythia} for resolved plus direct
processes and gluon- and quark-initiated jets are compared to the data
in Fig.~\ref{fig1}. The measured $\langle\psi(r)\rangle$ is reasonably
well described by the MC calculations of {\sc Pythia} for resolved and
direct processes for $-1<\etajet<1.5$, whereas for $1.5<\etajet<2.5$,
the measured jets are slightly broader than the predictions. From the
comparison with the predictions for gluon- and quark-initiated jets,
it is seen that the measured jets are quark-like for $-1<\etajet<0$
and become increasingly more gluon-like as $\etajet$ increases.

Figure~\ref{fig2} shows the same measurements as Fig.~\ref{fig1},
compared to the predictions of {\sc Pythia} including MI. This model
gives rise to jets that are much broader than those observed. The
predictions using {\sc Herwig}, also shown in Fig.~\ref{fig2}, describe 
the data well for $-1<\etajet<1$, are slightly narrower than the data
for $1<\etajet<1.5$ and fail to describe the data for
$1.5<\etajet<2.5$. These results and those presented below are
consistent with the previous ZEUS study of jet shapes in
photoproduction~\cite{epj:c2:61} which was performed using an
iterative cone algorithm and at lower $\etjet$.

Figure~\ref{fig3} and Tables~\ref{tabfour} and \ref{tabfive} show the
$\langle\psi(r)\rangle$ in different regions of $\etjet$ for the
photoproduction regime. The jets become narrower as $\etjet$
increases. The predictions of {\sc Pythia} for resolved plus direct
processes reproduce the data reasonably well. For $17<\etjet<29$ GeV,
the predictions for resolved processes alone also describe the data,
consistent with the dominance of resolved processes in this $\etjet$
region.

Figure~\ref{fig4}a shows the measured mean integrated jet shape at a
fixed value of $r=0.5$, $\langle\psi(r=0.5)\rangle$, as a function of
$\etajet$ in photoproduction. The predictions of {\sc Pythia} for
quark-initiated jets lie above the data, while those for
gluon-initiated jets lie below the data. The prediction of 
{\sc Pythia}, including resolved and direct processes, also shown in
Fig.~\ref{fig4}a, fails to describe the relatively strong broadening
of the measured jet shape for $\etajet>1.5$. This might be because the
fraction of gluon-initiated jets in the region $\etajet>1.5$ is
underestimated or that the effects of a possible underlying event in
the data have not been properly taken into account~\cite{epj:c2:61};
however, the prediction of {\sc Pythia} MI fails to describe the data
over the whole $\etajet$ range. Since $\langle\psi(r=0.5)\rangle$
changes from a value close to the upper curve (quark-initiated jets)
to a value near the lower curve (gluon-initiated jets) as $\etajet$
increases, the broadening of $\langle\psi(r=0.5)\rangle$ as $\etajet$
increases is consistent with an increase of the fraction of
gluon-initiated jets. The $\langle\psi(r=0.5)\rangle$ shows an
increase with $\etjet$ (see Fig.~\ref{fig4}b). The predictions of 
{\sc Pythia} for the dependence of $\langle\psi(r=0.5)\rangle$ on
$\etjet$ in resolved plus direct processes reproduce the data well for
$\etjet>21$ GeV. Therefore, the discrepancies between data and MC are
concentrated at low $\etjet$ and high $\etajet$ values.

The measured $\langle\psi(r)\rangle$ for different regions of
$\etajet$ and $\etjet$ is shown in Figs.~\ref{fig5} and \ref{fig6} and
Tables~\ref{tabsix} to \ref{tabeight} for DIS
events. Figure~\ref{fig7} shows the measured
$\langle\psi(r=0.5)\rangle$ as a function of $\etajet$ and $\etjet$.
There is no significant variation of $\langle\psi(r=0.5)\rangle$ with
$\etajet$ in DIS, whereas $\langle\psi(r=0.5)\rangle$ increases as
$\etjet$ increases, as observed in a previous study~\cite{epj:c8:367}
using an iterative cone algorithm. These conclusions are also in
agreement with those of a previous publication~\cite{pl:b558:41}, in
which the internal structure of jets in NC DIS was studied using the
mean subjet multiplicity.

The NLO QCD calculations of $\langle\psi(r)\rangle$, corrected for
hadronisation and $\z0$-exchange effects, are compared to the
data in Figs.~\ref{fig5} to \ref{fig7}. The NLO QCD calculations give
a good description of $\langle\psi(r)\rangle$ for $r\geq 0.2$; the
fractional differences between the measurements and the predictions,
also shown in Figs.~\ref{fig5} and \ref{fig6}, are less than $0.2\%$
for $r=0.5$. The sensitivity of the measurements to the value of
$\as(\mz)$ is illustrated in Fig.~\ref{fig7}b by comparing the
measured $\langle \psi(r=0.5)\rangle$ to NLO QCD calculations using
three different values of $\as(\mz)$. The NLO QCD calculations provide
a good description of the measured $\langle\psi(r=0.5)\rangle$ and
thus this observable was used to determine $\as(\mz)$, as explained in
Section~\ref{asset}.

Figure~\ref{fig8} shows the $\etajet$ and $\etjet$ dependence of
$\langle\psi(r=0.5)\rangle$ for photoproduction and DIS events; the
MC predictions of CDM and {\sc Pythia} for quark- and gluon-initiated
jets are compared to the data. Figure~\ref{fig8}a shows that the
sample of jets in DIS is consistent with being dominated by
quark-initiated jets with an approximately constant fraction over the
$\etajet$ region measured. Photoproduced jets in the backward region
are similar to jets in DIS and this agreement confirms, independently
of the comparison to MC predictions, that they are dominated by
quark-initiated jets. The increasing deviation in the integrated jet
shape for photoproduced jets from that of jets in DIS as $\etajet$
($\etjet$) increases (decreases) can be attributed to the increasing
fraction of gluon-initiated jets arising from resolved processes.

\subsection{Subjet-multiplicity measurements}
\label{smgp}
The measured mean subjet multiplicity as a function of $\yc$
for different regions of $\etajet$ and $\etjet$ for photoproduction is
shown in Figs.~\ref{fig9} and \ref{fig10} and Tables~\ref{tabnine} to
\ref{tabeleven}. Figure~\ref{fig11} shows the measured mean subjet
multiplicity at a fixed value of $\yc=10^{-2}$,
$\langle\ns(\yc=10^{-2})\rangle$, as a function of $\etajet$ and
$\etjet$. The measured mean subjet multiplicity increases as $\etajet$
increases and decreases as $\etjet$ increases. 

The predictions of {\sc Pythia} for quark-initiated, gluon-initiated
and all jets at the hadron level are compared to the measurements in
Figs.~\ref{fig9} to \ref{fig11}. The predicted
$\langle\ns(\yc)\rangle$ is larger for gluon-initiated jets than for
quark-initiated jets in each region of $\etajet$. For quark- or
gluon-initiated jets alone, $\langle\ns(\yc=10^{-2})\rangle$ exhibits
only a small dependence on $\etajet$ (see Fig.~\ref{fig11}a). The
$\etajet$-dependence of $\langle\ns(\yc=10^{-2})\rangle$ in the
calculation for all jets is dictated by the $\etajet$ variation of the
fractions of quark- and gluon-initiated jets. This variation, in turn,
originates from the different dominant two-body subprocess. The
calculations using {\sc Pythia} based on the predicted admixture of
quark- and gluon-initiated jets give a good description of the
measured $\langle\ns(\yc)\rangle$ as a function of $\yc$, $\etajet$
and $\etjet$. These results are in agreement with those from the
mean integrated jet shape (see Section~\ref{resjsgp}).

\section{Study of quark and gluon dynamics}
\label{xsections}
The predictions of the MC for the jet shape and subjet multiplicity
generally reproduce the data well and show the expected differences
for quark- and gluon-initiated jets. These differences are used now to
select samples enriched in quark- and gluon-initiated jets to study the
dynamics of the hard subprocesses in more detail.

\subsection{Selection of quark- and gluon-initiated jets}
\label{qgsel}
Quark- and gluon-initiated jets were selected on a statistical
basis based on their substructure. The integrated jet shape at
$r=0.3$, $\psi(r=0.3)$, and the subjet multiplicity at 
$\yc=5\cdot 10^{-4}$, $\ns(\yc=5\cdot 10^{-4})$, were used to select
quark- and gluon-initiated jets in the photoproduction and DIS
samples. The values $r=0.3$ and $\yc=5\cdot 10^{-4}$ were chosen to be
as small as possible to be sensitive to the differences between quark
and gluons, but large enough to avoid uncertainties due to the
detector resolution. The different behaviour of these distributions
for gluon- and quark-initiated jets is shown in Fig.~\ref{fig12} for
samples of {\sc Pythia} and {\sc Herwig} generated events. These
observables were used to classify the jets into:
\begin{itemize}
\item a gluon-enriched sample (broad jets), defined as those
  jets with $\psi(r=0.3)<0.6$ and/or $\ns(\yc=5\cdot 10^{-4})\geq 6$,
  and  
\item a quark-enriched sample (narrow jets), defined as those
  jets with $\psi(r=0.3)>0.8$ and/or $\ns(\yc=5\cdot 10^{-4})<4$. 
\end{itemize}
Non-overlapping ranges were chosen to suppress migration effects. The
values for the cuts in $\psi(r=0.3)$ and $\ns(\yc=5\cdot 10^{-4})$
chosen were a compromise between purity and statistics. The purity of
the gluon-initiated sample is around $50\%$, whereas for the
quark-enriched sample it is around $90\%$. Table~\ref{tabone} shows the
purities and efficiencies for the different MC and selected samples.

The remaining number of jets after applying the jet-shape,
subjet-multiplicity and the combination of both selection cuts in the
broad and narrow inclusive jet photoproduction data sample and in the 
broad-broad, narrow-narrow and broad-narrow dijet data samples
selected using the jet-shape method are shown in
Table~\ref{tabtwo}. The same table shows also the number of jets in
the broad and narrow inclusive jet NC DIS data sample selected
according to the jet shape. In the next sections, measurements of
cross sections are presented for samples of jets separated according
to their shape and/or subjet multiplicity.

\subsection{Measurements of $\seta$}
\label{seta}
The differential inclusive-jet cross-section $\seta$ for
photoproduction is shown in Fig.~\ref{fig13}a and
Table~\ref{tabtwelve} for samples of broad and narrow jets, separated
according to the jet-shape selection. The measured cross sections
exhibit different behaviours: the $\etajet$ distribution for broad
jets increases up to the highest $\etajet$ value measured,
whereas the distribution for narrow jets peaks at
$\etajet\approx 0.7$. Measurements of $\seta$ in photoproduction for
samples of broad and narrow jets separated according to either
the subjet-multiplicity selection or a combination of jet shape and
subjet multiplicity are shown in Figs.~\ref{fig13}c and \ref{fig13}d
and Tables~\ref{tabthirteen} and \ref{tabfourteen},
respectively. These measurements also exhibit a difference in shape
for the samples of broad and narrow jets. The same conclusions as
in the case of using the integrated-jet-shape selection method can be
drawn.

Leading-logarithm parton-shower MC calculations using {\sc Pythia},
{\sc Herwig} and {\sc Pythia} MI for resolved plus direct photon
processes are compared to the measurements in Figs.~\ref{fig13}a and
\ref{fig13}b. The same selection method was applied to the jets
of hadrons in the MC event samples and the calculations have been
normalised to the total measured cross section of each sample. The MC
predictions provide a good description of the shape of the
narrow-jet distribution in the data. The shape of the
broad-jet distribution in the data is reasonably well described by
{\sc Pythia} or {\sc Pythia} MI, but the prediction of {\sc Herwig}
fails to describe this distribution.  From the calculation of
{\sc Pythia} ({\sc Herwig}), the sample of broad jets selected
according to the jet shape is predicted to contain $15(12)\%$ of $gg$
subprocesses in the final state and $50(47)\%$ of $gq$, and a
contamination from processes with only quarks in the final state of
$35(41)\%$. There is a large contribution from $gq$ final states in
the broad-jet sample because the partonic cross section for the
resolved subprocess $q_{\gamma}g_p\rightarrow qg$ is much larger than
the cross section for the subprocesses $\qq\rightarrow gg$ plus
$gg\rightarrow gg$. The sample of narrow jets contains $62(61)\%$
of $qq$ subprocesses and $34(36)\%$ of $qg$, with a contamination of
$4(3)\%$ from $gg$ subprocesses. The measured cross section for the
broad-jet sample is $(32.1\pm 0.1)\%$ of the total cross section,
whereas the narrow-jet sample is $(40.6\pm 0.1)\%$ using the jet-shape
selection. {\sc Pythia} ({\sc Herwig}) predicts $31.5 (27.1)\%$ for
the broad-jet sample and $37.4 (44.0)\%$ for the narrow-jet
sample. Similar conclusions can be drawn from the subjet-multiplicity
selection and the combined jet-shape and subjet-multiplicity selection.

Figure~\ref{fig13}a also shows the predictions of {\sc Pythia} for
jets of quarks and gluons separately. These predictions have been
obtained without any jet-shape selection and are normalised to the
data cross sections. The calculation that includes only
quark-initiated jets gives a good description of the narrow-jet
cross section, whereas the calculation for gluon-initiated jets
provides a reasonable description of the broad-jet cross
section. This result supports the expectation that the broad
(narrow)-jet sample is dominated by gluon (quark)-initiated jets.

Figure~\ref{fig14}a and Table~\ref{tabfifteen} show $\seta$ in DIS for
samples of broad and narrow jets separated according to the jet
shape. The two cross sections have the same variation with $\etajet$,
as can be seen from the ratio of the narrow- to the broad-jet cross
sections. However, the narrow-jet cross section is about five times
larger than the broad-jet cross section, which shows that the DIS
sample is enriched in quark-initiated jets. Since the ratio of the
cross sections is approximately constant, the quark and gluon content
of the final state in DIS does not change with $\etajet$, as also was
concluded from Fig.~\ref{fig7}a. The predictions of the CDM model are
compared to the data in Fig.~\ref{fig14}a and give a good description
of the data. The predictions of MEPS give a poorer
description. Figure~\ref{fig14}b shows the same measured cross
sections together with the calculations of CDM for gluon- and
quark-initiated jets; no jet-shape selection has been applied in this
case. These predictions describe well the shapes of the broad- and
narrow-jet samples, respectively.

\subsection{Measurements of $\set$}
\label{set}
The differential inclusive jet cross-section $\set$, measured in
the range $17<\etjet<95$ GeV, is presented in Figs.~\ref{fig15}a
and \ref{fig15}b and Tables~\ref{tabsixteen} and \ref{tabseventeen}
for samples of broad and narrow jets, separated according to the jet
shape, for photoproduction and DIS events, respectively. The cross
sections for the narrow-jet samples have a harder spectrum than that
for the broad-jet sample. Figures~\ref{fig15}c and \ref{fig15}d and
Tables~\ref{tabeighteen} and \ref{tabnineteen} show
the $\set$ cross section for samples of broad and narrow jets
separated according to the subjet-multiplicity selection and the
combined integrated-jet-shape and subjet-multiplicity selection in the
photoproduction regime. These measurements exhibit the same behaviour
as in Fig.~\ref{fig15}a, but the cross-over between the broad- and
narrow-jet distributions takes place at slightly higher $\etjet$. The
MC calculations using {\sc Pythia}, which have been obtained using the
same selection method as for the data, are compared to the
measurements in Figs.~\ref{fig15}a, c and d. The MC predictions provide
a good description of the shapes of the data distributions. The
predictions of the CDM MC are compared to the measurements in
Fig.~\ref{fig15}b and give a good description of the data. In
photoproduction, the different $\etjet$ spectra exhibited by the
narrow- and broad-jet samples are understood in terms of the
increasing fraction of gluon-initiated jets as $\etjet$ decreases.

\subsection{Measurements of $\sccos$}
\label{sccos}
For samples of broad-broad or narrow-narrow dijet events, only
the absolute value of $\ccos$ can be determined because the outgoing
jets are indistinguishable. The differential dijet cross section as a
function of $\cost$ has been measured in the range $\costr$ for dijet
invariant masses $\mj>52$ GeV for photoproduction. The region of phase
space in the $(\mj,\cost)$ plane was chosen in order to minimise the
bias introduced by selecting jets with $\etj>17$~GeV and
$\etjj>14$~GeV. The measured $\scost$ for the samples of
broad-broad dijet events and narrow-narrow dijet events are
presented in Fig.~\ref{fig16}a and Table~\ref{tabtwenty}. The measured
and predicted cross sections were normalised to unity at
$\cost=0.1$. The $\cost$ distribution for the two samples of dijet
events increases as $\cost$ increases; however they exhibit a
different slope. The cross section at $\cost=0.7$ for broad-broad
dijet events is more than seven times larger than the measured value at
$\cost=0.1$, whereas for narrow-narrow dijet events, the cross section
at $\cost=0.7$ is only twice as large as at $\cost=0.1$.

Calculations using {\sc Pythia} for broad-broad and
narrow-narrow dijet events are compared to the data in
Fig.~\ref{fig16}a. The predictions from {\sc Pythia} give a good
description of the shape of the measured $\scost$. {\sc Pythia}
predicts $16\%$ of $gg$-final-state subprocesses, $52\%$ of $gq$ and
$32\%$ of $qq$ for the broad-broad dijet sample in the kinematic
region of this measurement. For the narrow-narrow dijet sample,
the predictions are: $71\%$ of $qq$ subprocesses, $28\%$ of $qg$ and
$1\%$ of $gg$. The differences observed in the measured $\scost$ for
the two samples are adequately reproduced by the calculations and
understood in terms of the dominant two-body processes: the resolved
subprocess $q_{\gamma}g_p\rightarrow qg$, mediated by gluon exchange
for the broad-broad dijet sample and the direct subprocess 
$\gamma g\rightarrow \qq$, mediated by quark exchange for the 
narrow-narrow dijet sample.

The sample of photoproduced dijet events with one broad jet and
one narrow jet allows a measurement of the unfolded 
$\sccos_{\rm broad}$ cross section. Since in this case the two jets
can be distinguished, $\theta^*_{\rm broad}$ refers to the
scattering angle measured with respect to the broad jet. 
Figure~\ref{fig16}b and Table~\ref{tabtwentyone} show the measured dijet
cross section as a function of $\ccos_{\rm broad}$. The measured and
predicted cross sections were normalised to unity at $\ccos_{\rm
  broad}=0.1$. The dijet angular distribution shows a different
behaviour on the negative and positive sides; the measured cross
section at $\ccos_{\rm broad}=0.7$ is approximately twice as large as
at $\ccos_{\rm broad}=-0.7$. The calculation from {\sc Pythia} gives a
good description of the shape of the measured 
$\sccos_{\rm broad}$. The predictions of {\sc Pythia} for the partonic
content are: $52\%$ of $qg$ subprocesses, $4\%$ of $gg$ and $44\%$ of
$qq$. The observed asymmetry is adequately reproduced by the
calculation and is understood in terms of the dominant resolved
subprocess $q_{\gamma}g_p \rightarrow qg$. The $\ccos_{\rm broad}$
distribution for this subprocess is asymmetric due to the different
dominant diagrams in the regions $\ccos_{\rm broad}\rightarrow\pm 1$:
$t$-channel gluon exchange ($\ccos_{\rm broad} \rightarrow +1$) and
$u$-channel quark exchange ($\ccos_{\rm broad} \rightarrow -1$).

\subsection{Measurements of $\smj$ and $\sxo$}
\label{smjsxo}
The photoproduction differential dijet cross section as a function of
$\mj$ has been measured in the range $52<\mj<123$ GeV for
$\cost<0.8$. The measured $\smj$ for the samples of broad-broad and
narrow-narrow dijet events are presented in Fig.~\ref{fig16}c and
Table~\ref{tabtwentytwo}. The measured $\smj$ cross sections decrease as
$\mj$ increases, but the distribution for the narrow-narrow dijet
sample exhibits a harder spectrum, as was also seen for the inclusive
jet cross section as a function of $\etjet$. The MC calculations from
{\sc Pythia} are compared to the data in Fig.~\ref{fig16}c and give a
good description of the shape of the measured $\smj$. The different
shape in both cross sections is understood in terms of the dominant
two-body processes: the broad-broad dijet sample is dominated by the
resolved subprocess $q_{\gamma}g_p\rightarrow qg$ and the
narrow-narrow dijet sample is dominated by the direct subprocess
$\gamma g\rightarrow\qq$. Direct processes reach larger values of
$\mj$ than those of resolved since the full incoming-photon energy is
available at the hard interaction.

Resolved and direct processes can be separated by using the $\xo$
variable, which is defined as
$$ \xo = \frac{1}{2 y E_e}(\etj e^{-\etaj} + \etjj e^{-\etajj}).$$
Resolved and direct processes populate different regions in $\xo$,
with the direct processes concentrated at high values. The dijet cross
section as a function of $\xo$ is presented in Fig.~\ref{fig16}d and 
Table~\ref{tabtwentythree} and is reasonably well described by the MC
predictions of {\sc Pythia}. The cross section for the broad-broad
dijet sample is approximately constant as a function of $\xo$ whereas
the cross section for the narrow-narrow dijet sample peaks at high
values. The shape of the distribution for the broad-broad
(narrow-narrow) dijet events is consistent with the dominance of
resolved (direct) processes.

\section{Determination of $\as$}
\label{asset}
The measured $\langle\psi(r=0.5)\rangle$ for $\etjet>21$~GeV in DIS
was used to determine $\asz$ using a method similar to one presented
previously~\cite{pl:b558:41}. The NLO QCD calculations were
performed using the program {\sc Disent} with three different MRST99
sets of proton PDFs, central, MRST99$\downarrow\downarrow$ and
MRST99$\uparrow\uparrow$; the value of $\asz$ used in each partonic
cross-section calculation was that associated with the corresponding
set of PDFs. The $\asz$ dependence of the predicted
$\langle\psi(r=0.5)\rangle$ in each bin $i$ of $\etjet$ was
parameterised according to

$$\left [ \langle\psi(r=0.5)\rangle(\asz) \right ]_i=C_1^i+C_2^i\asz,$$
where $C_1^i$ and $C_2^i$ were determined from a $\chi^2$ fit by using
the NLO QCD calculations corrected for hadronisation and $\z0$-exchange
effects. Finally, a value of $\asz$ was determined in each $\etjet$
region as well as from all the data points by a $\chi^2$ fit.

The values of $\asz$ as determined from the measured
$\langle\psi(r=0.5)\rangle$ in each region of $\etjet$ are shown in
Fig.~\ref{fig17} and Table~\ref{tabtwentyfour}. Taking into account only
the statistical uncertainties, the value of $\asz$ obtained by
combining all the $\etjet$ regions is 
$\asz=0.1176 \pm 0.0009\ {\rm (stat.)}$. 

The uncertainties on the extracted value of $\asz$ due to the
experimental systematic uncertainties were evaluated by repeating the
analysis above for each systematic check described in
Section~\ref{syst}. The total experimental systematic uncertainty on
the value of $\asz$ is $\Delta\asz/\asz={}^{+0.8}_{-2.2}\%$. The main
contribution to the positive (negative) systematic uncertainty comes
from the uncertainty in the jet energy scale (scattered-electron
identification).

The following sources of theoretical uncertainties on the extracted
value of $\asz$ were considered:
\begin{itemize}
\item terms beyond NLO were estimated by varying $\mu_R$ between $Q/2$
  and $2Q$ and keeping $\mu_F$ fixed at $Q$; this results in a
  variation of $\Delta\asz={}^{+0.0089}_{-0.0070}$;
\item the uncertainty on the modelling of the parton shower was
  estimated by using the MEPS model to calculate the parton-to-hadron
  correction factors; this results in a variation of $\Delta\asz=0.0018$;
\item the uncertainty in the choice of $\mu_R$ was estimated by using
  $\mu_R=\etjet$ instead of $Q$ and $\mu_F$ was set to $Q$; this
  results in a variation of $\Delta\asz=0.0003$;
\item the uncertainty in the NLO QCD calculations due to the
  uncertainties in the proton PDFs was estimated by repeating the
  calculations using 40 additional sets from
  CTEQ6~\cite{jhep:0207:012,*jhep:0310:046}; this results in a
  variation of $\Delta\asz=0.0002$;
\item the uncertainty of the calculations in the value of $\mu_F$ was
  estimated by repeating the calculations with $\mu_F=Q/2$ and $2Q$;
  this results in a variation of $\Delta\asz=0.0001$.
\end{itemize}

These uncertainties were added in quadrature and give a total
theoretical uncertainty of $\Delta\asz/\asz={}^{+7.7}_{-6.1}\%$. As a
cross-check of the extracted value of $\asz$, the fit procedure was
repeated by using the five sets of the CTEQ4 ``A-series'', resulting
in $\asz=0.1178\pm 0.0009$, in very good agreement with the central
value determined above. As a consistency check, the whole procedure
was repeated for $\langle\psi(r=0.4)\rangle$ and
$\langle\psi(r=0.6)\rangle$, giving values of 
$\asz=0.1158\pm 0.0008$ and $\asz=0.1193\pm 0.0010$, respectively,
which are compatible with the value determined from
$\langle\psi(r=0.5)\rangle$. The determination of $\asz$ was also
repeated using the calculations from the {\sc Disaster}++ program; this
gave $\asz=0.1166\pm 0.0009$, which is compatible with the value
quoted above.

The value of $\asz$ as determined from the measured
$\langle\psi(r=0.5)\rangle$ is therefore
\begin{center}
$\asmz{0.1176}{0.0009}{0.0026}{0.0009}{0.0072}{0.0091}$.
\end{center}
This result is in agreement with recent determinations using
measurements of jet production in
DIS~\cite{pl:b558:41,pl:b547:164,pl:b507:70,epj:c31:149,epj:c19:289} and
photoproduction~\cite{pl:b560:7} and with the current world
average of $0.1183\pm 0.0027$~\cite{jp:g26:r27}. This determination of
$\as$ has experimental uncertainties as small as those based on previous
measurements. However, the theoretical uncertainty is large and
dominated by terms beyond NLO. Further theoretical work on
higher-order contributions would allow an improved determination of
$\as$ from the integrated jet shape in DIS.

\section{Summary and conclusions}
\label{sumcon}
Measurements of the mean integrated jet shape and mean subjet
multiplicities for inclusive jet photoproduction and DIS at a
centre-of-mass energy of $318$~GeV using the data collected by ZEUS in
1998 to 2000, which correspond to an integrated luminosity of
$82.2$~\pb1, have been presented. The measurements refer to jets
identified with the $\kt$ cluster algorithm in the longitudinally
invariant inclusive mode in the laboratory frame and selected
according to $\etjet>17$~GeV and $\etar$. The measurements are given
in the kinematic region defined by $\q2< 1$~\g2\ and \wrn\ for
photoproduction and $\q2>125$~\g2\ for DIS. The jet shape broadens
(narrows) and the mean subjet multiplicity increases (decreases) as
$\etajet$ ($\etjet$) increases in photoproduction. The observed
broadening of the jet shape and the increase of the mean subjet
multiplicity as $\etajet$ increases are consistent with an increase of
the fraction of gluon-initiated jets. In DIS, the data show no
significant dependence with $\etajet$ and a moderate dependence with
$\etjet$. Leading-logarithm parton-shower MC models for
photoproduction and NLO QCD calculations for DIS give a good
description of the data.

Measurements of differential inclusive jet and dijet cross sections in
the photoproduction and DIS regimes separated into broad and
narrow jets according to their internal structure have been
presented. Leading-logarithm parton-shower MC models give a
good description of the data. The inclusive jet cross-sections $\seta$
and $\set$ for broad- and narrow-jet samples show the expected
behaviour for samples enriched in gluon- and quark-initiated jets,
respectively. The dijet cross section as a function of $\cost$,
measured in the range $\costr$ and integrated over $\mj>52$ GeV,
displays for broad-broad dijets a behaviour consistent with that
expected for a sample enriched in processes mediated by gluon
exchange. Narrow-narrow dijets, however, show a behaviour consistent
with a sample enriched in processes mediated by quark exchange. The
dijet cross-section $\sccos_{\rm broad}$, measured in the region
$-0.8<\ccos_{\rm broad} <0.8$ and integrated over $\mj>52$ GeV, for a
sample of events with broad-narrow dijets, exhibits a large asymmetry
consistent with the expected dominance of gluon (quark) exchange as
$\ccos_{\rm broad}\rightarrow +1$ ($\ccos_{\rm broad}\rightarrow-1$). 
The dijet cross section as a function of $\mj$ and $\xo$ for the
sample of broad-broad dijets shows a behaviour consistent with the
dominance of the resolved $q_{\gamma}g_p\rightarrow qg$ subprocess,
whereas the sample with narrow-narrow dijets is consistent with the
dominance of the direct subprocess $\gamma g\rightarrow\qq$.

The measurements of the mean integrated jet shape in DIS have been
used to extract a value of $\asz$ by comparing to the predictions
of NLO QCD as a function of $\etjet$. The calculations reproduce the
measured observables well, demonstrating the validity of the
description of the internal structure of jets by pQCD.
The value of $\asz$ as determined by fitting the NLO QCD calculations
to the measured mean integrated jet shape $\langle\psi(r=0.5)\rangle$
for $\etjet>21$~GeV is 
\begin{center}
$\asmz{0.1176}{0.0009}{0.0026}{0.0009}{0.0072}{0.0091}$.
\end{center}
This value is in good agreement with the current world
average. 

\newpage
\vspace{0.5cm}
\noindent {\Large\bf Acknowledgements}
\vspace{0.3cm}

We thank the DESY Directorate for their strong support and
encouragement. The remarkable achievements of the HERA machine group
were essential for the successful completion of this work and are
greatly appreciated. We would like to thank M. Seymour for useful
discussions.

\vfill\eject

\providecommand{\etal}{et al.\xspace}
\providecommand{\coll}{Coll.\xspace}
\catcode`\@=11
\def\@bibitem#1{%
\ifmc@bstsupport
  \mc@iftail{#1}%
    {;\newline\ignorespaces}%
    {\ifmc@first\else.\fi\orig@bibitem{#1}}
  \mc@firstfalse
\else
  \mc@iftail{#1}%
    {\ignorespaces}%
    {\orig@bibitem{#1}}%
\fi}%
\catcode`\@=12

\newpage
\clearpage
\begin{table}[p]
\begin{center}
\begin{tabular}{|c||c||c|}                \hline
$r$ & \multicolumn{2}{c|}{$\langle\psi(r)\rangle \pm\ {\rm stat.}\ \pm\ {\rm syst.}$} \\ \cline{2-3}
 &  $-1<\etajet<0$&  $0<\etajet<1$ \\ 
\hline\hline
$ 0.1$ & $  0.3135 \pm   0.0025_{-  0.0048}^{+  0.0033}$ & $  0.3040 \pm   0.0012_{-  0.0032}^{+  0.0021}$ \\ 
$ 0.2$ & $  0.5870 \pm   0.0025_{-  0.0138}^{+  0.0038}$ & $  0.5730 \pm   0.0012_{-  0.0100}^{+  0.0025}$ \\ 
$ 0.3$ & $  0.7322 \pm   0.0020_{-  0.0106}^{+  0.0029}$ & $  0.7189 \pm   0.0010_{-  0.0071}^{+  0.0019}$ \\ 
$ 0.4$ & $  0.8165 \pm   0.0015_{-  0.0059}^{+  0.0021}$ & $  0.8075 \pm   0.0007_{-  0.0047}^{+  0.0015}$ \\ 
$ 0.5$ & $  0.8720 \pm   0.0011_{-  0.0041}^{+  0.0014}$ & $  0.8649 \pm   0.0006_{-  0.0024}^{+  0.0010}$ \\ 
$ 0.6$ & $  0.9106 \pm   0.0008_{-  0.0025}^{+  0.0010}$ & $  0.9048 \pm   0.0004_{-  0.0015}^{+  0.0008}$ \\
\hline
\end{tabular}
\caption{
Measured mean integrated jet shape corrected to the hadron level
for jets in photoproduction with
$\etjet>17$~GeV in different $\etajet$ regions.
The statistical and systematic uncertainties are also indicated.}
\label{tabthree}
\end{center}
\end{table}

\newpage
\clearpage
\begin{table}[p]
\begin{center}
\begin{tabular}{|c||c||c|}                \hline
$r$ & \multicolumn{2}{c|}{$\langle\psi(r)\rangle \pm\ {\rm stat.}\ \pm\ {\rm syst.}$} \\ \cline{2-3}
 &  $17<\etjet<21$ GeV&  $21<\etjet<25$ GeV \\ 
\hline\hline
$ 0.1$ & $  0.2406 \pm   0.0009_{-  0.0019}^{+  0.0054}$ & $  0.2915 \pm   0.0016_{-  0.0027}^{+  0.0033}$ \\ 
$ 0.2$ & $  0.4934 \pm   0.0010_{-  0.0025}^{+  0.0021}$ & $  0.5470 \pm   0.0017_{-  0.0037}^{+  0.0030}$ \\ 
$ 0.3$ & $  0.6477 \pm   0.0009_{-  0.0020}^{+  0.0019}$ & $  0.6938 \pm   0.0014_{-  0.0026}^{+  0.0026}$ \\ 
$ 0.4$ & $  0.7489 \pm   0.0007_{-  0.0017}^{+  0.0023}$ & $  0.7857 \pm   0.0011_{-  0.0018}^{+  0.0019}$ \\ 
$ 0.5$ & $  0.8188 \pm   0.0006_{-  0.0013}^{+  0.0025}$ & $  0.8469 \pm   0.0009_{-  0.0013}^{+  0.0014}$ \\ 
$ 0.6$ & $  0.8691 \pm   0.0004_{-  0.0010}^{+  0.0024}$ & $  0.8901 \pm   0.0007_{-  0.0010}^{+  0.0012}$ \\
\hline
\end{tabular}
\caption{
Measured mean integrated jet shape corrected to the hadron level
for jets in photoproduction with
$\etar$ in different $\etjet$ regions. 
The statistical and systematic uncertainties are also indicated.}
\label{tabfour}
\end{center}
\end{table}

\newpage
\clearpage
\begin{table}[p]
\begin{center}
\begin{tabular}{|c||c||c|}                \hline
$r$ & \multicolumn{2}{c|}{$\langle\psi(r)\rangle \pm\ {\rm stat.}\ \pm\ {\rm syst.}$} \\ \cline{2-3}
 &  $35<\etjet<41$ GeV&  $41<\etjet<47$ GeV \\ 
\hline\hline
$ 0.1$ & $  0.4045 \pm   0.0068_{-  0.0068}^{+  0.0069}$ & $  0.431 \pm   0.012_{-  0.014}^{+  0.013}$ \\ 
$ 0.2$ & $  0.6538 \pm   0.0061_{-  0.0067}^{+  0.0063}$ & $  0.681 \pm   0.010_{-  0.010}^{+  0.012}$ \\ 
$ 0.3$ & $  0.7792 \pm   0.0047_{-  0.0060}^{+  0.0049}$ & $  0.7960 \pm   0.0078_{-  0.0080}^{+  0.0091}$ \\ 
$ 0.4$ & $  0.8491 \pm   0.0035_{-  0.0041}^{+  0.0037}$ & $  0.8607 \pm   0.0058_{-  0.0059}^{+  0.0080}$ \\ 
$ 0.5$ & $  0.8959 \pm   0.0026_{-  0.0026}^{+  0.0029}$ & $  0.9019 \pm   0.0042_{-  0.0043}^{+  0.0052}$ \\ 
$ 0.6$ & $  0.9279 \pm   0.0019_{-  0.0020}^{+  0.0021}$ & $  0.9326 \pm   0.0030_{-  0.0030}^{+  0.0050}$ \\
\hline
\end{tabular}
\caption{
Measured mean integrated jet shape corrected to the hadron level
for jets in photoproduction with
$\etar$ in different $\etjet$ regions. 
The statistical and systematic uncertainties are also indicated.}
\label{tabfive}
\end{center}
\end{table}

\newpage
\clearpage
\begin{table}[p]
\begin{center}
\begin{tabular}{|c||c||c|}                \hline
$r$ & \multicolumn{2}{c|}{$\langle\psi(r)\rangle \pm\ {\rm stat.}\ \pm\ {\rm syst.}$} \\ \cline{2-3}
 &  $-1<\etajet<0$&  $0<\etajet<1$ \\ 
\hline\hline
$ 0.1$ & $  0.3434 \pm   0.0034_{-  0.0037}^{+  0.0199}$ & $  0.3759 \pm   0.0019_{-  0.0070}^{+  0.0028}$ \\ 
$ 0.2$ & $  0.6332 \pm   0.0030_{-  0.0033}^{+  0.0128}$ & $  0.6487 \pm   0.0016_{-  0.0073}^{+  0.0027}$ \\ 
$ 0.3$ & $  0.7720 \pm   0.0022_{-  0.0024}^{+  0.0075}$ & $  0.7811 \pm   0.0012_{-  0.0071}^{+  0.0019}$ \\ 
$ 0.4$ & $  0.8474 \pm   0.0016_{-  0.0017}^{+  0.0041}$ & $  0.8542 \pm   0.0009_{-  0.0048}^{+  0.0013}$ \\ 
$ 0.5$ & $  0.8950 \pm   0.0011_{-  0.0012}^{+  0.0031}$ & $  0.8999 \pm   0.0006_{-  0.0031}^{+  0.0009}$ \\ 
$ 0.6$ & $  0.9271 \pm   0.0008_{-  0.0008}^{+  0.0023}$ & $  0.9309 \pm   0.0005_{-  0.0029}^{+  0.0006}$ \\
\hline
\end{tabular}
\caption{
Measured mean integrated jet shape corrected to the hadron level
and for electroweak radiative effects for jets in DIS with
$\etjet>17$~GeV in different $\etajet$ regions. 
The statistical and systematic uncertainties are also indicated.}
\label{tabsix}
\end{center}
\end{table}

\newpage
\clearpage
\begin{table}[p]
\begin{center}
\begin{tabular}{|c||c||c|}                \hline
$r$ & \multicolumn{2}{c|}{$\langle\psi(r)\rangle \pm\ {\rm stat.}\ \pm\ {\rm syst.}$} \\ \cline{2-3}
 &  $17<\etjet<21$ GeV&  $21<\etjet<25$ GeV \\ 
\hline\hline
$ 0.1$ & $  0.3211 \pm   0.0016_{-  0.0067}^{+  0.0034}$ & $  0.3817 \pm   0.0025_{-  0.0070}^{+  0.0054}$ \\ 
$ 0.2$ & $  0.6093 \pm   0.0015_{-  0.0044}^{+  0.0031}$ & $  0.6588 \pm   0.0022_{-  0.0028}^{+  0.0047}$ \\ 
$ 0.3$ & $  0.7547 \pm   0.0012_{-  0.0031}^{+  0.0024}$ & $  0.7881 \pm   0.0016_{-  0.0018}^{+  0.0028}$ \\ 
$ 0.4$ & $  0.8351 \pm   0.0009_{-  0.0022}^{+  0.0017}$ & $  0.8587 \pm   0.0012_{-  0.0013}^{+  0.0028}$ \\ 
$ 0.5$ & $  0.8866 \pm   0.0007_{-  0.0012}^{+  0.0013}$ & $  0.9030 \pm   0.0009_{-  0.0009}^{+  0.0020}$ \\ 
$ 0.6$ & $  0.9216 \pm   0.0005_{-  0.0014}^{+  0.0011}$ & $  0.9330 \pm   0.0007_{-  0.0008}^{+  0.0014}$ \\
\hline
\end{tabular}
\caption{
Measured mean integrated jet shape corrected to the hadron level
and for electroweak radiative effects for jets in DIS with
$\etar$ in different $\etjet$ regions. 
The statistical and systematic uncertainties are also indicated.}
\label{tabseven}
\end{center}
\end{table}

\newpage
\clearpage
\begin{table}[p]
\begin{center}
\begin{tabular}{|c||c||c|}                \hline
$r$ & \multicolumn{2}{c|}{$\langle\psi(r)\rangle \pm\ {\rm stat.}\ \pm\ {\rm syst.}$} \\ \cline{2-3}
 &  $35<\etjet<41$ GeV&  $41<\etjet<47$ GeV \\ 
\hline\hline
$ 0.1$ & $  0.5194 \pm   0.0067_{-  0.0075}^{+  0.0128}$ & $  0.5630 \pm   0.0095_{-  0.0100}^{+  0.0144}$ \\ 
$ 0.2$ & $  0.7503 \pm   0.0049_{-  0.0058}^{+  0.0095}$ & $  0.7755 \pm   0.0066_{-  0.0071}^{+  0.0087}$ \\ 
$ 0.3$ & $  0.8464 \pm   0.0035_{-  0.0038}^{+  0.0068}$ & $  0.8629 \pm   0.0047_{-  0.0049}^{+  0.0058}$ \\ 
$ 0.4$ & $  0.8990 \pm   0.0025_{-  0.0027}^{+  0.0038}$ & $  0.9110 \pm   0.0033_{-  0.0048}^{+  0.0046}$ \\ 
$ 0.5$ & $  0.9305 \pm   0.0018_{-  0.0019}^{+  0.0030}$ & $  0.9395 \pm   0.0023_{-  0.0042}^{+  0.0033}$ \\ 
$ 0.6$ & $  0.9530 \pm   0.0013_{-  0.0013}^{+  0.0019}$ & $  0.9584 \pm   0.0017_{-  0.0018}^{+  0.0023}$ \\
\hline
\end{tabular}
\caption{
Measured mean integrated jet shape corrected to the hadron level
and for electroweak radiative effects for jets in DIS with
$\etar$ in different $\etjet$ regions. 
The statistical and systematic uncertainties are also indicated.}
\label{tabeight}
\end{center}
\end{table}

\newpage
\clearpage
\begin{table}[p]
\begin{center}
\begin{tabular}{|c||c||c|}                \hline
$\yc$ & \multicolumn{2}{c|}{$\langle\ns(\yc)\rangle \pm\ {\rm stat.}\ \pm\ {\rm syst.}$} \\ \cline{2-3}
 &  $-1<\etajet<0$&  $0<\etajet<1$ \\ 
\hline\hline
$0.0005$ & $   4.627 \pm    0.011_{-   0.070}^{+   0.013}$ & $   4.723 \pm    0.006_{-   0.088}^{+   0.008}$ \\ 
$ 0.001$ & $   3.760 \pm    0.009_{-   0.028}^{+   0.013}$ & $   3.836 \pm    0.005_{-   0.055}^{+   0.008}$ \\ 
$ 0.003$ & $   2.638 \pm    0.007_{-   0.010}^{+   0.009}$ & $   2.687 \pm    0.004_{-   0.016}^{+   0.007}$ \\ 
$ 0.005$ & $   2.218 \pm    0.006_{-   0.009}^{+   0.010}$ & $   2.265 \pm    0.003_{-   0.007}^{+   0.006}$ \\ 
$  0.01$ & $   1.754 \pm    0.005_{-   0.007}^{+   0.016}$ & $   1.797 \pm    0.003_{-   0.005}^{+   0.012}$ \\ 
$  0.03$ & $   1.271 \pm    0.004_{-   0.005}^{+   0.029}$ & $   1.301 \pm    0.002_{-   0.003}^{+   0.017}$ \\
\hline
\end{tabular}
\caption{
Measured mean subjet multiplicity corrected to the hadron level
for jets in photoproduction with
$\etjet>17$~GeV in different $\etajet$ regions. 
The statistical and systematic uncertainties are also indicated.}
\label{tabnine}
\end{center}
\end{table}

\newpage
\clearpage
\begin{table}[p]
\begin{center}
\begin{tabular}{|c||c||c|}                \hline
$\yc$ & \multicolumn{2}{c|}{$\langle\ns(\yc)\rangle \pm\ {\rm stat.}\ \pm\ {\rm syst.}$} \\ \cline{2-3}
 &  $17<\etjet<21$ GeV&  $21<\etjet<25$ GeV \\ 
\hline\hline
$0.0005$ & $   5.273 \pm    0.005_{-   0.161}^{+   0.008}$ & $   4.966 \pm    0.008_{-   0.129}^{+   0.011}$ \\ 
$ 0.001$ & $   4.304 \pm    0.004_{-   0.114}^{+   0.008}$ & $   4.011 \pm    0.007_{-   0.099}^{+   0.010}$ \\ 
$ 0.003$ & $   3.027 \pm    0.003_{-   0.060}^{+   0.008}$ & $   2.799 \pm    0.005_{-   0.051}^{+   0.007}$ \\ 
$ 0.005$ & $   2.549 \pm    0.003_{-   0.044}^{+   0.007}$ & $   2.344 \pm    0.004_{-   0.029}^{+   0.007}$ \\ 
$  0.01$ & $   2.020 \pm    0.002_{-   0.023}^{+   0.005}$ & $   1.856 \pm    0.004_{-   0.008}^{+   0.006}$ \\ 
$  0.03$ & $   1.429 \pm    0.002_{-   0.003}^{+   0.006}$ & $   1.348 \pm    0.003_{-   0.005}^{+   0.008}$ \\
\hline
\end{tabular}
\caption{
Measured mean subjet multiplicity corrected to the hadron level
for jets in photoproduction with
$\etar$ in different $\etjet$ regions. 
The statistical and systematic uncertainties are also indicated.}
\label{tabten}
\end{center}
\end{table}

\newpage
\clearpage
\begin{table}[p]
\begin{center}
\begin{tabular}{|c||c||c|}                \hline
$\yc$ & \multicolumn{2}{c|}{$\langle\ns(\yc)\rangle \pm\ {\rm stat.}\ \pm\ {\rm syst.}$} \\ \cline{2-3}
 &  $35<\etjet<41$ GeV&  $41<\etjet<47$ GeV \\ 
\hline\hline
$0.0005$ & $   4.152 \pm    0.027_{-   0.090}^{+   0.030}$ & $   4.000 \pm    0.043_{-   0.235}^{+   0.044}$ \\ 
$ 0.001$ & $   3.264 \pm    0.022_{-   0.076}^{+   0.022}$ & $   3.128 \pm    0.035_{-   0.123}^{+   0.037}$ \\ 
$ 0.003$ & $   2.272 \pm    0.016_{-   0.063}^{+   0.017}$ & $   2.148 \pm    0.026_{-   0.062}^{+   0.027}$ \\ 
$ 0.005$ & $   1.885 \pm    0.014_{-   0.025}^{+   0.015}$ & $   1.805 \pm    0.022_{-   0.036}^{+   0.023}$ \\ 
$  0.01$ & $   1.514 \pm    0.012_{-   0.012}^{+   0.028}$ & $   1.484 \pm    0.020_{-   0.022}^{+   0.026}$ \\ 
$  0.03$ & $   1.226 \pm    0.009_{-   0.011}^{+   0.014}$ & $   1.202 \pm    0.015_{-   0.015}^{+   0.022}$ \\
\hline
\end{tabular}
\caption{
Measured mean subjet multiplicity corrected to the hadron level
for jets in photoproduction with
$\etar$ in different $\etjet$ regions. 
The statistical and systematic uncertainties are also indicated.}
\label{tabeleven}
\end{center}
\end{table}

\newpage
\clearpage
\begin{table}[p]
\begin{center}
\begin{tabular}{|c||c|c|c|c|} \hline
 & {\sc Pythia} & {\sc Herwig} & CDM & MEPS \\ \hline\hline
 \multicolumn{5}{|c|}{jet shape ($\%$)} \\ \hline\hline
gluons & 57/48 & 50/43 & 43/47 & 33/65 \\
quarks & 84/51 & 85/55 & 93/65 & 97/57 \\
\hline\hline
 \multicolumn{5}{|c|}{subjet multiplicity ($\%$)} \\ \hline\hline
gluons & 56/63 & 53/56 & \multicolumn{2}{c|}{ } \\
quarks & 90/23 & 92/32 & \multicolumn{2}{c|}{ } \\
\hline\hline
  \multicolumn{5}{|c|}{combination ($\%$)} \\ \hline\hline
gluons & 61/43 & 55/37 & \multicolumn{2}{c|}{ } \\
quarks & 90/22 & 93/30 & \multicolumn{2}{c|}{ } \\
\hline
\end{tabular}
\caption{
Purity/efficiency of gluon-initiated jets in the broad-jet sample and of
quark-initiated jets in the narrow-jet sample from the 
{\sc Pythia} and {\sc Herwig} photoproduction Monte Carlo generated
samples and from the CDM and MEPS NC DIS Monte Carlo generated samples.}
\label{tabone}
\end{center}
\end{table}

\begin{table}[p]
\begin{center}
\begin{tabular}{|c||c|c|} \hline
 \multicolumn{3}{c}{jet shape} \\ \hline
 & photoproduction & NC DIS \\ \hline\hline
broad  & 50737 & 12032 \\
narrow & 65038 & 47549 \\
broad-broad  & 5744 & \\
narrow-narrow & 11826 & \\
broad-narrow  & 13813 & \\
\hline
 \multicolumn{3}{c}{subjet multiplicity} \\ \hline
 & \multicolumn{2}{c|}{photoproduction} \\ \hline\hline
broad  & \multicolumn{2}{c|}{55750} \\
narrow & \multicolumn{2}{c|}{54654} \\
\hline
 \multicolumn{3}{c}{combination} \\ \hline
 & \multicolumn{2}{c|}{photoproduction} \\ \hline\hline
broad  & \multicolumn{2}{c|}{37213} \\
narrow & \multicolumn{2}{c|}{43928} \\
\hline
\end{tabular}
\caption{
Number of jets (events) in the inclusive-jet (dijet) photoproduction and
NC DIS samples selected according to various methods.}
\label{tabtwo}
\end{center}
\end{table}

\newpage
\clearpage
\begin{table}[p]
\begin{center}
\begin{tabular}{|c||c||c|}                \hline
 & \multicolumn{2}{c|}{$\seta \pm\ {\rm stat.}\ \pm\ {\rm
    syst.}\ \pm\ \etjet-{\rm scale}$ (pb)} \\ \cline{2-3}
$\etajet$ & broad jets & narrow jets  \\ \hline
$ -1.0 -  -0.5$ & $     7.1 \pm      0.3{}_{-     2.6}^{+     2.6} {}_{-     0.9}^{+     0.8}$ & $    20.9 \pm      0.7{}_{-     1.9}^{+     1.9} {}_{-     1.9}^{+     1.9}$ \\ 
$ -0.5 -   0.0$ & $    60.3 \pm      1.1{}_{-    17.6}^{+    17.6} {}_{-     5.0}^{+     4.1}$ & $   137.8 \pm      1.7{}_{-    22.2}^{+    22.2} {}_{-     8.5}^{+     6.7}$ \\ 
$  0.0 -   0.5$ & $   130.0 \pm      1.6{}_{-    27.6}^{+    27.6} {}_{-     8.0}^{+     7.6}$ & $   274.3 \pm      2.5{}_{-    26.4}^{+    26.5} {}_{-    11.1}^{+    10.8}$ \\ 
$  0.5 -   1.0$ & $   190.8 \pm      2.1{}_{-    20.3}^{+    20.3} {}_{-    10.5}^{+     9.5}$ & $   349.0 \pm      2.8{}_{-    32.4}^{+    32.4} {}_{-    11.7}^{+    11.9}$ \\ 
$  1.0 -   1.5$ & $   240.1 \pm      2.5{}_{-     4.6}^{+     4.6} {}_{-    11.6}^{+    11.7}$ & $   299.3 \pm      2.7{}_{-     7.0}^{+     7.0} {}_{-     8.5}^{+     8.7}$ \\ 
$  1.5 -   2.0$ & $   259.1 \pm      2.5{}_{-    11.7}^{+    11.7} {}_{-    12.2}^{+    12.8}$ & $   227.9 \pm      2.2{}_{-     3.4}^{+     3.4} {}_{-     6.7}^{+     6.6}$ \\ 
$  2.0 -   2.5$ & $   270.8 \pm      2.5{}_{-    17.1}^{+    17.2} {}_{-    14.4}^{+    12.7}$ & $   151.7 \pm      1.8{}_{-     4.1}^{+     4.1} {}_{-     5.4}^{+     4.9}$ \\ 
\hline
\end{tabular}
\caption{
Measured differential $ep$ cross-section $\seta$ for inclusive jet
photoproduction with $\etjet>17$ GeV. The jets have been selected
according to their shape into broad and narrow jets. The statistical
and systematic uncertainties $-$not associated with the absolute
energy scale of the jets$-$ are also indicated. The systematic
uncertainties associated to the absolute energy scale of the jets are
quoted separately.}
\label{tabtwelve}
\end{center}
\end{table}

\begin{table}[p]
\begin{center}
\begin{tabular}{|c||c||c|}                \hline
 & \multicolumn{2}{c|}{$\seta \pm\ {\rm stat.}\ \pm\ {\rm
    syst.}\ \pm\ \etjet-{\rm scale}$ (pb)} \\ \cline{2-3}
$\etajet$ & broad jets & narrow jets  \\ \hline
$ -1.0 -  -0.5$ & $     8.1 \pm      0.3{}_{-     2.0}^{+     2.0} {}_{-     0.9}^{+     0.9}$ & $     8.9 \pm      0.3{}_{-     3.6}^{+     3.6} {}_{-     0.8}^{+     0.8}$ \\ 
$ -0.5 -   0.0$ & $    72.1 \pm      1.1{}_{-    14.8}^{+    14.8} {}_{-     5.3}^{+     4.4}$ & $    63.8 \pm      0.9{}_{-    26.5}^{+    26.5} {}_{-     4.2}^{+     3.1}$ \\ 
$  0.0 -   0.5$ & $   153.1 \pm      1.6{}_{-    23.6}^{+    23.6} {}_{-     7.9}^{+     7.6}$ & $   132.8 \pm      1.4{}_{-    36.1}^{+    36.1} {}_{-     5.7}^{+     5.4}$ \\ 
$  0.5 -   1.0$ & $   218.7 \pm      2.3{}_{-    13.0}^{+    13.0} {}_{-    10.8}^{+     9.6}$ & $   171.9 \pm      1.6{}_{-    43.7}^{+    43.7} {}_{-     6.3}^{+     6.0}$ \\ 
$  1.0 -   1.5$ & $   261.4 \pm      2.8{}_{-     8.1}^{+     8.1} {}_{-    11.8}^{+    12.0}$ & $   137.4 \pm      1.2{}_{-    33.1}^{+    33.1} {}_{-     4.3}^{+     4.6}$ \\ 
$  1.5 -   2.0$ & $   300.7 \pm      2.9{}_{-    22.3}^{+    22.3} {}_{-    12.6}^{+    12.7}$ & $   103.1 \pm      1.1{}_{-    18.8}^{+    18.8} {}_{-     3.6}^{+     3.3}$ \\ 
$  2.0 -   2.5$ & $   315.5 \pm      2.6{}_{-    25.1}^{+    25.1} {}_{-    15.7}^{+    13.5}$ & $    62.4 \pm      0.8{}_{-    17.5}^{+    17.5} {}_{-     2.4}^{+     2.2}$ \\ 
\hline
\end{tabular}
\caption{
Measured differential $ep$ cross-section $\seta$ for inclusive jet
photoproduction with $\etjet>17$ GeV. The jets have been selected
according to their subjet multiplicity into broad and narrow
jets. The statistical
and systematic uncertainties $-$not associated with the absolute
energy scale of the jets$-$ are also indicated. The systematic
uncertainties associated to the absolute energy scale of the jets are
quoted separately.}
\label{tabthirteen}
\end{center}
\end{table}

\newpage
\clearpage
\begin{table}[p]
\begin{center}
\begin{tabular}{|c||c||c|}                \hline
 & \multicolumn{2}{c|}{$\seta \pm\ {\rm stat.}\ \pm\ {\rm
    syst.}\ \pm\ \etjet-{\rm scale}$ (pb)} \\ \cline{2-3}
$\etajet$ & broad jets & narrow jets  \\ \hline
$ -1.0 -  -0.5$ & $     4.7 \pm      0.3{}_{-     1.7}^{+     1.7} {}_{-     0.6}^{+     0.5}$ & $     8.5 \pm      0.3{}_{-     2.7}^{+     2.7} {}_{-     0.7}^{+     0.7}$ \\ 
$ -0.5 -   0.0$ & $    44.0 \pm      0.9{}_{-    10.1}^{+    10.1} {}_{-     3.7}^{+     3.0}$ & $    58.8 \pm      0.9{}_{-    22.1}^{+    22.1} {}_{-     3.7}^{+     2.8}$ \\ 
$  0.0 -   0.5$ & $    96.7 \pm      1.4{}_{-    16.3}^{+    16.4} {}_{-     5.7}^{+     5.6}$ & $   121.7 \pm      1.4{}_{-    30.7}^{+    30.7} {}_{-     5.0}^{+     4.8}$ \\ 
$  0.5 -   1.0$ & $   139.9 \pm      1.9{}_{-    14.5}^{+    14.5} {}_{-     7.2}^{+     6.8}$ & $   160.4 \pm      1.6{}_{-    36.7}^{+    36.7} {}_{-     5.4}^{+     5.2}$ \\ 
$  1.0 -   1.5$ & $   182.0 \pm      2.3{}_{-     4.4}^{+     4.4} {}_{-     8.6}^{+     9.2}$ & $   126.5 \pm      1.3{}_{-    26.9}^{+    26.9} {}_{-     3.7}^{+     3.8}$ \\ 
$  1.5 -   2.0$ & $   213.5 \pm      2.4{}_{-    11.9}^{+    11.9} {}_{-     9.9}^{+    10.2}$ & $    94.2 \pm      1.1{}_{-    15.5}^{+    15.5} {}_{-     3.0}^{+     2.9}$ \\ 
$  2.0 -   2.5$ & $   240.5 \pm      2.4{}_{-    20.3}^{+    20.3} {}_{-    13.2}^{+    11.3}$ & $    55.2 \pm      0.8{}_{-    14.2}^{+    14.2} {}_{-     1.9}^{+     1.9}$ \\ 
\hline
\end{tabular}
\caption{
Measured differential $ep$ cross-section $\seta$ for inclusive jet
photoproduction with $\etjet>17$ GeV. The jets have been selected
according to the combination of the jet shape and subjet multiplicity
into broad and narrow jets. The statistical
and systematic uncertainties $-$not associated with the absolute
energy scale of the jets$-$ are also indicated. The systematic
uncertainties associated to the absolute energy scale of the jets are
quoted separately.}
\label{tabfourteen}
\end{center}
\end{table}

\begin{table}[p]
\begin{center}
\begin{tabular}{|c||c||c|}                \hline
 & \multicolumn{2}{c|}{$\seta \pm\ {\rm stat.}\ \pm\ {\rm
    syst.}\ \pm\ \etjet-{\rm scale}$ (pb)} \\ \cline{2-3}
$\etajet$ & broad jets & narrow jets  \\ \hline
$ -1.0 -  -0.5$ & $     5.6 \pm      0.4{}_{-     0.9}^{+     0.9} {}_{-     0.4}^{+     0.4}$ & $    26.4 \pm      0.9{}_{-     6.1}^{+     6.1} {}_{-     1.5}^{+     1.7}$ \\ 
$ -0.5 -   0.0$ & $    21.1 \pm      0.7{}_{-     0.1}^{+     0.2} {}_{-     1.0}^{+     1.0}$ & $    95.4 \pm      1.5{}_{-     7.8}^{+     7.8} {}_{-     2.8}^{+     3.1}$ \\ 
$  0.0 -   0.5$ & $    41.4 \pm      1.0{}_{-     3.9}^{+     3.9} {}_{-     1.8}^{+     1.9}$ & $   181.6 \pm      2.1{}_{-    15.8}^{+    15.8} {}_{-     4.5}^{+     4.2}$ \\ 
$  0.5 -   1.0$ & $    51.4 \pm      1.1{}_{-     5.9}^{+     5.9} {}_{-     1.9}^{+     1.8}$ & $   236.1 \pm      2.5{}_{-    14.5}^{+    14.5} {}_{-     4.6}^{+     4.7}$ \\ 
$  1.0 -   1.5$ & $    57.3 \pm      1.3{}_{-     2.7}^{+     2.7} {}_{-     2.1}^{+     2.1}$ & $   252.3 \pm      2.7{}_{-     6.3}^{+     6.3} {}_{-     4.4}^{+     4.7}$ \\ 
$  1.5 -   2.0$ & $    57.5 \pm      1.1{}_{-     3.0}^{+     3.0} {}_{-     2.1}^{+     2.0}$ & $   256.3 \pm      2.6{}_{-     9.1}^{+     9.1} {}_{-     4.8}^{+     5.1}$ \\ 
$  2.0 -   2.5$ & $    49.4 \pm      1.0{}_{-     6.1}^{+     6.1} {}_{-     1.7}^{+     1.8}$ & $   211.2 \pm      2.4{}_{-     4.0}^{+     4.0} {}_{-     4.0}^{+     4.3}$ \\ 
\hline
\end{tabular}
\caption{
Measured differential $ep$ cross-section $\seta$ for inclusive jet
production in DIS with $\etjet>17$ GeV. The jets have been selected
according to their shape into broad and narrow
jets. The statistical
and systematic uncertainties $-$not associated with the absolute
energy scale of the jets$-$ are also indicated. The systematic
uncertainties associated to the absolute energy scale of the jets are
quoted separately.}
\label{tabfifteen}
\end{center}
\end{table}

\newpage
\clearpage
\begin{table}[p]
\begin{center}
\begin{tabular}{|c||c||c|}                \hline
 & \multicolumn{2}{c|}{$\set \pm\ {\rm stat.}\ \pm\ {\rm
    syst.}\ \pm\ \etjet-{\rm scale}$ (pb)} \\ \cline{2-3}
$\etjet$ (GeV) & broad jets & narrow jets  \\ \hline
$  17. -   21.$ & $   102.4 \pm      0.5{}_{-     3.7}^{+     3.7} {}_{-     5.4}^{+     5.1}$ & $   102.3 \pm      0.5{}_{-     6.8}^{+     6.8} {}_{-     3.5}^{+     3.3}$ \\ 
$  21. -   25.$ & $    27.2 \pm      0.3{}_{-     1.3}^{+     1.3} {}_{-     1.6}^{+     1.5}$ & $    42.7 \pm      0.3{}_{-     2.1}^{+     2.1} {}_{-     1.4}^{+     1.5}$ \\ 
$  25. -   29.$ & $    8.78 \pm     0.16{}_{-    0.37}^{+    0.37} {}_{-    0.53}^{+    0.45}$ & $    18.9 \pm      0.2{}_{-     1.5}^{+     1.5} {}_{-     0.8}^{+     0.7}$ \\ 
$  29. -   35.$ & $    3.02 \pm     0.08{}_{-    0.26}^{+    0.26} {}_{-    0.22}^{+    0.22}$ & $    7.82 \pm     0.12{}_{-    0.42}^{+    0.43} {}_{-    0.33}^{+    0.41}$ \\ 
$  35. -   41.$ & $   0.774 \pm    0.040{}_{-   0.062}^{+   0.061} {}_{-   0.045}^{+   0.042}$ & $    2.93 \pm     0.07{}_{-    0.18}^{+    0.18} {}_{-    0.17}^{+    0.13}$ \\ 
$  41. -   47.$ & $   0.231 \pm    0.022{}_{-   0.027}^{+   0.028} {}_{-   0.020}^{+   0.017}$ & $   1.094 \pm    0.046{}_{-   0.043}^{+   0.045} {}_{-   0.081}^{+   0.055}$ \\ 
$  47. -   55.$ & $   0.064 \pm    0.010{}_{-   0.010}^{+   0.010} {}_{-   0.003}^{+   0.005}$ & $   0.447 \pm    0.026{}_{-   0.014}^{+   0.014} {}_{-   0.033}^{+   0.035}$ \\ 
$  55. -   71.$ & $  0.0148 \pm   0.0036{}_{-  0.0016}^{+  0.0016} {}_{-  0.0003}^{+  0.0013}$ & $   0.068 \pm    0.006{}_{-   0.030}^{+   0.030} {}_{-   0.004}^{+   0.003}$ \\ 
$  71. -   95.$ & $ 0.00050{}_{- 0.00032}^{+ 0.00088}{}_{- 0.00001}^{+ 0.00098} {}_{- 0.00009}^{+ 0.00006}$ & $  0.0102 \pm   0.0027{}_{-  0.0039}^{+  0.0041} {}_{-  0.0003}^{+  0.0000}$ \\ 
\hline
\end{tabular}
\caption{
Measured differential $ep$ cross-section $\set$ for inclusive jet
photoproduction with $\etar$. The jets have been selected
according to their shape into broad and narrow jets. The statistical
and systematic uncertainties $-$not associated with the absolute
energy scale of the jets$-$ are also indicated. The systematic
uncertainties associated to the absolute energy scale of the jets are
quoted separately.}
\label{tabsixteen}
\end{center}
\end{table}

\begin{table}[p]
\begin{center}
\begin{tabular}{|c||c||c|}                \hline
 & \multicolumn{2}{c|}{$\set \pm\ {\rm stat.}\ \pm\ {\rm
    syst.}\ \pm\ \etjet-{\rm scale}$ (pb)} \\ \cline{2-3}
$\etjet$ (GeV) & broad jets & narrow jets  \\ \hline
$  17. -   21.$ & $    19.6 \pm      0.2{}_{-     1.6}^{+     1.6} {}_{-     0.8}^{+     0.7}$ & $    67.3 \pm      0.5{}_{-     3.2}^{+     3.2} {}_{-     1.2}^{+     1.0}$ \\ 
$  21. -   25.$ & $    7.76 \pm     0.15{}_{-    0.75}^{+    0.75} {}_{-    0.31}^{+    0.31}$ & $    35.2 \pm      0.3{}_{-     1.3}^{+     1.3} {}_{-     0.7}^{+     0.9}$ \\ 
$  25. -   29.$ & $    3.57 \pm     0.10{}_{-    0.30}^{+    0.30} {}_{-    0.13}^{+    0.16}$ & $    20.7 \pm      0.3{}_{-     1.1}^{+     1.1} {}_{-     0.3}^{+     0.4}$ \\ 
$  29. -   35.$ & $    1.70 \pm     0.06{}_{-    0.04}^{+    0.04} {}_{-    0.07}^{+    0.07}$ & $   10.81 \pm     0.15{}_{-    0.67}^{+    0.67} {}_{-    0.32}^{+    0.28}$ \\ 
$  35. -   41.$ & $   0.714 \pm    0.038{}_{-   0.045}^{+   0.045} {}_{-   0.008}^{+   0.049}$ & $    5.36 \pm     0.11{}_{-    0.27}^{+    0.27} {}_{-    0.11}^{+    0.16}$ \\ 
$  41. -   47.$ & $   0.312 \pm    0.026{}_{-   0.011}^{+   0.011} {}_{-   0.024}^{+   0.001}$ & $    3.02 \pm     0.08{}_{-    0.15}^{+    0.15} {}_{-    0.11}^{+    0.09}$ \\ 
$  47. -   55.$ & $   0.151 \pm    0.015{}_{-   0.015}^{+   0.015} {}_{-   0.003}^{+   0.008}$ & $   1.468 \pm    0.046{}_{-   0.035}^{+   0.035} {}_{-   0.052}^{+   0.073}$ \\ 
$  55. -   71.$ & $  0.0349 \pm   0.0052{}_{-  0.0095}^{+  0.0095} {}_{-  0.0020}^{+  0.0018}$ & $   0.572 \pm    0.021{}_{-   0.035}^{+   0.035} {}_{-   0.026}^{+   0.017}$ \\ 
$  71. -   95.$ & $  0.0099 \pm   0.0024{}_{-  0.0017}^{+  0.0017} {}_{-  0.0003}^{+  0.0012}$ & $  0.1185 \pm   0.0074{}_{-  0.0081}^{+  0.0081} {}_{-  0.0055}^{+  0.0089}$ \\ 
\hline
\end{tabular}
\caption{
Measured differential $ep$ cross-section $\set$ for inclusive jet
production in DIS with $\etar$. The jets have been selected
according to their shape into broad and narrow
jets. The statistical
and systematic uncertainties $-$not associated with the absolute
energy scale of the jets$-$ are also indicated. The systematic
uncertainties associated to the absolute energy scale of the jets are
quoted separately.}
\label{tabseventeen}
\end{center}
\end{table}

\newpage
\clearpage
\begin{table}[p]
\begin{center}
\begin{tabular}{|c||c||c|}                \hline
 & \multicolumn{2}{c|}{$\set \pm\ {\rm stat.}\ \pm\ {\rm
    syst.}\ \pm\ \etjet-{\rm scale}$ (pb)} \\ \cline{2-3}
$\etjet$ (GeV) & broad jets & narrow jets  \\ \hline
$  17. -   21.$ & $   116.6 \pm      0.6{}_{-     1.2}^{+     1.3} {}_{-     5.5}^{+     5.1}$ & $    44.7 \pm      0.3{}_{-    13.8}^{+    13.8} {}_{-     1.7}^{+     1.5}$ \\ 
$  21. -   25.$ & $    32.7 \pm      0.3{}_{-     0.6}^{+     0.6} {}_{-     1.7}^{+     1.7}$ & $    19.7 \pm      0.2{}_{-     4.6}^{+     4.6} {}_{-     0.7}^{+     0.8}$ \\ 
$  25. -   29.$ & $   10.91 \pm     0.18{}_{-    0.17}^{+    0.17} {}_{-    0.69}^{+    0.58}$ & $     9.2 \pm      0.1{}_{-     2.4}^{+     2.4} {}_{-     0.5}^{+     0.3}$ \\ 
$  29. -   35.$ & $    3.33 \pm     0.08{}_{-    0.16}^{+    0.16} {}_{-    0.22}^{+    0.21}$ & $    4.38 \pm     0.08{}_{-    0.64}^{+    0.65} {}_{-    0.20}^{+    0.22}$ \\ 
$  35. -   41.$ & $   0.868 \pm    0.040{}_{-   0.070}^{+   0.068} {}_{-   0.059}^{+   0.060}$ & $    1.74 \pm     0.05{}_{-    0.24}^{+    0.24} {}_{-    0.08}^{+    0.08}$ \\ 
$  41. -   47.$ & $   0.227 \pm    0.021{}_{-   0.030}^{+   0.030} {}_{-   0.018}^{+   0.016}$ & $    0.70 \pm     0.03{}_{-    0.16}^{+    0.16} {}_{-    0.05}^{+    0.03}$ \\ 
$  47. -   55.$ & $   0.084 \pm    0.010{}_{-   0.016}^{+   0.016} {}_{-   0.003}^{+   0.008}$ & $   0.327 \pm    0.021{}_{-   0.015}^{+   0.015} {}_{-   0.026}^{+   0.028}$ \\ 
$  55. -   71.$ & $  0.0133 \pm   0.0028{}_{-  0.0035}^{+  0.0035} {}_{-  0.0007}^{+  0.0016}$ & $   0.055 \pm    0.005{}_{-   0.035}^{+   0.035} {}_{-   0.003}^{+   0.002}$ \\ 
$  71. -   95.$ & $ 0.00016 {}_{- 0.00010}^{+ 0.00028}{}_{- 0.00004}^{+ 0.00017} {}_{- 0.00002}^{+ 0.00000}$ & $  0.0104 \pm   0.0028{}_{-  0.0031}^{+  0.0033} {}_{-  0.0004}^{+  0.0000}$ \\ 
\hline
\end{tabular}
\caption{
Measured differential $ep$ cross-section $\set$ for inclusive jet
photoproduction with $\etar$. The jets have been selected
according to their subjet multiplicity into broad and narrow
jets. The statistical
and systematic uncertainties $-$not associated with the absolute
energy scale of the jets$-$ are also indicated. The systematic
uncertainties associated to the absolute energy scale of the jets are
quoted separately.}
\label{tabeighteen}
\end{center}
\end{table}

\begin{table}[p]
\begin{center}
\begin{tabular}{|c||c||c|}                \hline
 & \multicolumn{2}{c|}{$\set \pm\ {\rm stat.}\ \pm\ {\rm
    syst.}\ \pm\ \etjet-{\rm scale}$ (pb)} \\ \cline{2-3}
$\etjet$ (GeV) & broad jets & narrow jets  \\ \hline
$  17. -   21.$ & $    82.7 \pm      0.5{}_{-     1.1}^{+     1.1} {}_{-     4.2}^{+     4.1}$ & $    40.5 \pm      0.3{}_{-    11.1}^{+    11.1} {}_{-     1.4}^{+     1.3}$ \\ 
$  21. -   25.$ & $    21.4 \pm      0.2{}_{-     0.9}^{+     0.9} {}_{-     1.3}^{+     1.2}$ & $    18.2 \pm      0.2{}_{-     4.0}^{+     4.0} {}_{-     0.6}^{+     0.7}$ \\ 
$  25. -   29.$ & $    6.72 \pm     0.14{}_{-    0.15}^{+    0.14} {}_{-    0.37}^{+    0.36}$ & $     8.7 \pm      0.1{}_{-     2.2}^{+     2.2} {}_{-     0.4}^{+     0.2}$ \\ 
$  29. -   35.$ & $    2.18 \pm     0.07{}_{-    0.10}^{+    0.10} {}_{-    0.17}^{+    0.16}$ & $    4.12 \pm     0.08{}_{-    0.57}^{+    0.57} {}_{-    0.18}^{+    0.21}$ \\ 
$  35. -   41.$ & $   0.505 \pm    0.032{}_{-   0.063}^{+   0.062} {}_{-   0.029}^{+   0.023}$ & $    1.66 \pm     0.05{}_{-    0.20}^{+    0.20} {}_{-    0.08}^{+    0.08}$ \\ 
$  41. -   47.$ & $   0.138 \pm    0.017{}_{-   0.011}^{+   0.011} {}_{-   0.010}^{+   0.011}$ & $    0.67 \pm     0.03{}_{-    0.12}^{+    0.12} {}_{-    0.05}^{+    0.02}$ \\ 
$  47. -   55.$ & $   0.044 \pm    0.008{}_{-   0.005}^{+   0.005} {}_{-   0.002}^{+   0.005}$ & $   0.317 \pm    0.021{}_{-   0.013}^{+   0.013} {}_{-   0.026}^{+   0.030}$ \\ 
$  55. -   71.$ & $  0.0071 \pm   0.0021{}_{-  0.0012}^{+  0.0012} {}_{-  0.0008}^{+  0.0010}$ & $   0.051 \pm    0.005{}_{-   0.030}^{+   0.030} {}_{-   0.003}^{+   0.002}$ \\ 
$  71. -   95.$ & & $  0.0091 \pm   0.0027{}_{-  0.0039}^{+  0.0041} {}_{-  0.0004}^{+  0.0000}$ \\ 
\hline
\end{tabular}
\caption{
Measured differential $ep$ cross-section $\set$ for inclusive jet
photoproduction with $\etar$. The jets have been selected
according to the combination of the jet shape and subjet multiplicity
into broad and narrow jets. The statistical
and systematic uncertainties $-$not associated with the absolute
energy scale of the jets$-$ are also indicated. The systematic
uncertainties associated to the absolute energy scale of the jets are
quoted separately.}
\label{tabnineteen}
\end{center}
\end{table}

\newpage
\clearpage
\begin{table}[p]
\begin{center}
\begin{tabular}{|c||c||c|}                \hline
 & \multicolumn{2}{c|}{$\scost \pm\ {\rm stat.}\ \pm\ {\rm
    syst.}\ \pm\ \etjet-{\rm scale}$ (pb)} \\ \cline{2-3}
$\cost$ & broad-broad dijets & narrow-narrow dijets  \\ \hline
$  0.0 -   0.2$ & $    3.93 \pm     0.49{}_{-    0.22}^{+    0.22} {}_{-    0.35}^{+    0.33}$ & $    43.9 \pm      1.6{}_{-     3.7}^{+     3.7} {}_{-     1.6}^{+     1.9}$ \\ 
$  0.2 -   0.4$ & $     7.3 \pm      0.7{}_{-     1.4}^{+     1.4} {}_{-     0.4}^{+     0.5}$ & $    49.0 \pm      1.7{}_{-     4.2}^{+     4.3} {}_{-     2.1}^{+     1.9}$ \\ 
$  0.4 -   0.6$ & $    10.4 \pm      0.8{}_{-     1.1}^{+     1.1} {}_{-     1.1}^{+     0.5}$ & $    67.9 \pm      2.0{}_{-     7.3}^{+     7.3} {}_{-     3.0}^{+     2.3}$ \\ 
$  0.6 -   0.8$ & $    29.9 \pm      1.2{}_{-     3.2}^{+     3.2} {}_{-     1.6}^{+     2.0}$ & $    86.3 \pm      2.3{}_{-     5.4}^{+     5.4} {}_{-     3.6}^{+     3.0}$ \\ 
\hline
\end{tabular}
\caption{
Measured differential $ep$ cross-section $\scost$ for dijet
photoproduction with $\mj>52$ GeV. The events have been selected
according to their shape into broad-broad and narrow-narrow dijets. 
The statistical
and systematic uncertainties $-$not associated with the absolute
energy scale of the jets$-$ are also indicated. The systematic
uncertainties associated to the absolute energy scale of the jets are
quoted separately.}
\label{tabtwenty}
\end{center}
\end{table}

\begin{table}[p]
\begin{center}
\begin{tabular}{|c||c|}                \hline
 & $\sccos_{\rm broad} \pm\ {\rm stat.}\ \pm\ {\rm
    syst.}\ \pm\ \etjet-{\rm scale}$ (pb) \\ \cline{2-2}
$\ccos_{\rm broad}$ & broad-narrow dijets \\ \hline
$ -0.8 -  -0.6$ & $    30.3 \pm      1.2{}_{-     4.2}^{+     4.2} {}_{-     1.5}^{+     1.4}$ \\ 
$ -0.6 -  -0.4$ & $    16.5 \pm      0.9{}_{-     2.6}^{+     2.6} {}_{-     0.9}^{+     0.8}$ \\ 
$ -0.4 -  -0.2$ & $    14.5 \pm      0.9{}_{-     2.0}^{+     2.0} {}_{-     0.4}^{+     0.6}$ \\ 
$ -0.2 -   0.0$ & $    10.3 \pm      0.7{}_{-     1.8}^{+     1.8} {}_{-     0.7}^{+     0.5}$ \\ 
$  0.0 -   0.2$ & $    12.8 \pm      0.8{}_{-     1.7}^{+     1.7} {}_{-     0.5}^{+     0.8}$ \\ 
$  0.2 -   0.4$ & $    17.0 \pm      1.0{}_{-     2.3}^{+     2.3} {}_{-     1.0}^{+     0.7}$ \\ 
$  0.4 -   0.6$ & $    31.7 \pm      1.4{}_{-     1.6}^{+     1.6} {}_{-     1.7}^{+     1.5}$ \\ 
$  0.6 -   0.8$ & $    66.4 \pm      1.9{}_{-     9.3}^{+     9.3} {}_{-     3.6}^{+     2.9}$ \\ 
\hline
\end{tabular}
\caption{
Measured differential $ep$ cross-section $\sccos_{\rm broad}$ for dijet
photoproduction with $\mj>52$ GeV. The events have been selected
according to their shape into broad-narrow dijets. 
The statistical
and systematic uncertainties $-$not associated with the absolute
energy scale of the jets$-$ are also indicated. The systematic
uncertainties associated to the absolute energy scale of the jets are
quoted separately.}
\label{tabtwentyone}
\end{center}
\end{table}

\newpage
\clearpage
\begin{table}[p]
\begin{center}
\begin{tabular}{|c||c||c|}                \hline
 & \multicolumn{2}{c|}{$\smj \pm\ {\rm stat.}\ \pm\ {\rm
    syst.}\ \pm\ \etjet-{\rm scale}$ (pb)} \\ \cline{2-3}
$\mj$ (GeV) & broad-broad dijets & narrow-narrow dijets  \\ \hline
$  52. -   57.$ & $   1.006 \pm    0.047{}_{-   0.036}^{+   0.038} {}_{-   0.057}^{+   0.060}$ & $    3.06 \pm     0.08{}_{-    0.47}^{+    0.47} {}_{-    0.15}^{+    0.11}$ \\ 
$  57. -   65.$ & $   0.387 \pm    0.022{}_{-   0.035}^{+   0.035} {}_{-   0.030}^{+   0.030}$ & $   1.842 \pm    0.053{}_{-   0.088}^{+   0.088} {}_{-   0.051}^{+   0.049}$ \\ 
$  65. -   83.$ & $   0.105 \pm    0.008{}_{-   0.004}^{+   0.004} {}_{-   0.006}^{+   0.004}$ & $   0.749 \pm    0.022{}_{-   0.078}^{+   0.078} {}_{-   0.031}^{+   0.032}$ \\ 
$  83. -  103.$ & $  0.0109 \pm   0.0023{}_{-  0.0015}^{+  0.0015} {}_{-  0.0011}^{+  0.0015}$ & $   0.210 \pm    0.011{}_{-   0.005}^{+   0.005} {}_{-   0.008}^{+   0.011}$ \\ 
$ 103. -  123.$ & $  0.0015 {}_{-  0.0010}^{+  0.0012}{}_{-  0.0003}^{+  0.0016} {}_{-  0.0001}^{+  0.0003}$ & $  0.0574 \pm   0.0059{}_{-  0.0053}^{+  0.0052} {}_{-  0.0074}^{+  0.0026}$ \\ 
\hline
\end{tabular}
\caption{
Measured differential $ep$ cross-section $\smj$ for dijet
photoproduction with $\cost<0.8$. The events have been selected
according to their shape into broad-broad and narrow-narrow dijets. 
The statistical
and systematic uncertainties $-$not associated with the absolute
energy scale of the jets$-$ are also indicated. The systematic
uncertainties associated to the absolute energy scale of the jets are
quoted separately.}
\label{tabtwentytwo}
\end{center}
\end{table}

\begin{table}[p]
\begin{center}
\begin{tabular}{|c||c||c|}                \hline
 & \multicolumn{2}{c|}{$\sxo \pm\ {\rm stat.}\ \pm\ {\rm
    syst.}\ \pm\ \etjet-{\rm scale}$ (pb)} \\ \cline{2-3}
$\xo$ & broad-broad dijets & narrow-narrow dijets  \\ \hline
$  0.0 -   0.2$ & $    31.6 \pm      1.5{}_{-     1.8}^{+     1.7} {}_{-     1.6}^{+     2.2}$ & $     9.0 \pm      0.8{}_{-     0.8}^{+     0.9} {}_{-     0.3}^{+     0.3}$ \\ 
$  0.2 -   0.4$ & $    79.2 \pm      2.3{}_{-     4.6}^{+     4.7} {}_{-     4.8}^{+     4.0}$ & $    48.7 \pm      1.7{}_{-     5.3}^{+     5.1} {}_{-     1.5}^{+     0.3}$ \\ 
$  0.4 -   0.6$ & $    66.0 \pm      2.0{}_{-     1.8}^{+     1.8} {}_{-     2.6}^{+     2.5}$ & $    78.1 \pm      2.2{}_{-     4.4}^{+     4.3} {}_{-     1.1}^{+     0.8}$ \\ 
$  0.6 -   0.8$ & $    71.9 \pm      2.0{}_{-    11.1}^{+    11.2} {}_{-     2.5}^{+     2.6}$ & $   131.1 \pm      2.8{}_{-    12.0}^{+    12.4} {}_{-     1.3}^{+     0.2}$ \\ 
$  0.8 -   1.0$ & $    79.0 \pm      1.9{}_{-    20.3}^{+    20.6} {}_{-     8.9}^{+     7.0}$ & $   459.6 \pm      5.4{}_{-    58.1}^{+    58.5} {}_{-    20.3}^{+    19.2}$ \\ 
\hline
\end{tabular}
\caption{
Measured differential $ep$ cross-section $\sxo$ for dijet
photoproduction. The events have been selected
according to their shape into broad-broad and narrow-narrow dijets. 
The statistical
and systematic uncertainties $-$not associated with the absolute
energy scale of the jets$-$ are also indicated. The systematic
uncertainties associated to the absolute energy scale of the jets are
quoted separately.}
\label{tabtwentythree}
\end{center}
\end{table}

\newpage
\clearpage
\begin{table}[p]
\begin{center}
\begin{tabular}{|c||c|}                \hline
 &  \\ 
$\langle\etjet\rangle$ (GeV) & $\asz \pm\ {\rm stat.}\ \pm\ {\rm syst.}\ \pm\ {\rm th.}$ \\
 &  \\ 
\hline\hline
 &  \\ 
$ 22.9$ & $  0.1177 \pm   0.0013{}_{-  0.0027}^{+  0.0005} {}_{-  0.0076}^{+  0.0096}$ \\ 
  &   \\ 
$ 26.8$ & $  0.1177 \pm   0.0018{}_{-  0.0036}^{+  0.0018} {}_{-  0.0073}^{+  0.0092}$ \\ 
  &   \\ 
$ 31.6$ & $  0.1179 \pm   0.0021{}_{-  0.0013}^{+  0.0021} {}_{-  0.0071}^{+  0.0089}$ \\ 
  &   \\ 
$ 37.7$ & $  0.1188 \pm   0.0032{}_{-  0.0044}^{+  0.0010} {}_{-  0.0074}^{+  0.0091}$ \\ 
  &   \\ 
$ 43.7$ & $  0.1134 \pm   0.0045{}_{-  0.0046}^{+  0.0067} {}_{-  0.0061}^{+  0.0077}$ \\ 
  &   \\ 
$ 50.5$ & $  0.1165 \pm   0.0058{}_{-  0.0061}^{+  0.0001} {}_{-  0.0070}^{+  0.0084}$ \\ 
  &   \\ 
$ 58.5$ & $  0.1126 \pm   0.0068{}_{-  0.0053}^{+  0.0165} {}_{-  0.0062}^{+  0.0075}$ \\ 
  &   \\ 
$ 79.3$ & $   0.120 \pm    0.014{}_{-   0.013}^{+   0.004} {}_{-   0.008}^{+   0.009}$ \\ 
  &   \\ 
 &  \\ 
\hline
\end{tabular}
\caption{
The $\asz$ values determined from the
QCD fit of the measured $\langle\psi(r=0.5)\rangle$ as a function of
$\etjet$ in DIS. The statistical, systematic and theoretical
uncertainties are also indicated.}
\label{tabtwentyfour}
\end{center}
\end{table}

\newpage
\clearpage
\begin{figure}[p]
\vfill
\setlength{\unitlength}{1.0cm}
%
\vspace{-1.5cm}
\caption
{\it 
Measured mean integrated jet shape corrected to the hadron level
(dots), $\langle\psi(r)\rangle$, for jets in photoproduction with
$\etjet>17$~GeV in different $\etajet$ regions. The error bars, which
are typically smaller than the dots, show the statistical and
systematic uncertainties added in quadrature. For comparison, the
predictions of {\sc Pythia} including resolved plus direct processes
for quark (dot-dashed lines), gluon (dashed lines) and all (solid
lines) jets are shown. The open circles show the fractional difference
of the data to the predictions of {\sc Pythia} for all jets.}
\label{fig1}
\vfill
\end{figure}

\newpage
\clearpage
\begin{figure}[p]
\vfill
\setlength{\unitlength}{1.0cm}
\begin{picture} (18.0,18.0)
\put (-1.0,0.5){\epsfig{figure=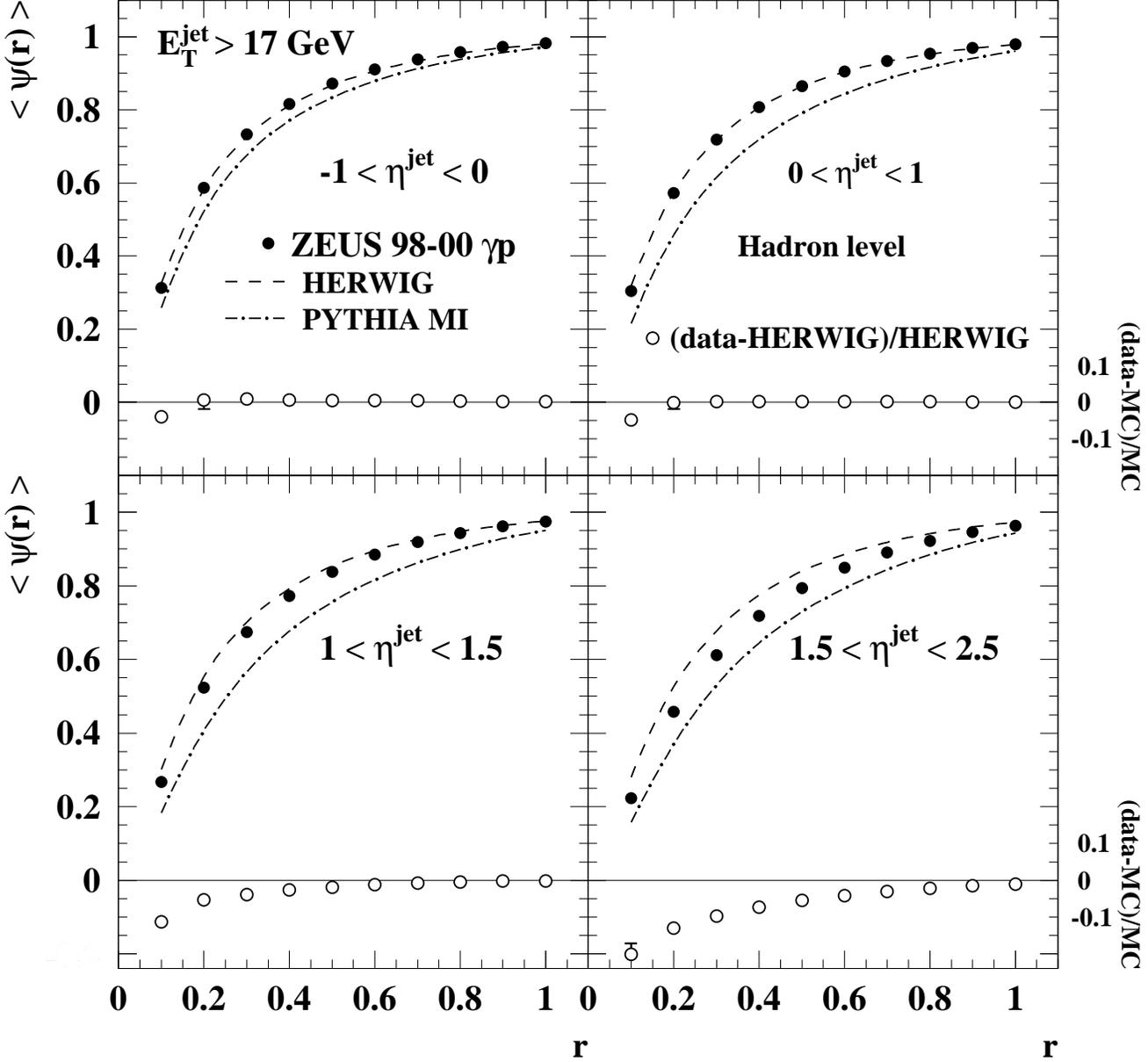,width=18cm}}
\end{picture}
\vspace{-1.5cm}
\caption
{\it 
Measured mean integrated jet shape corrected to the hadron level
(dots), $\langle\psi(r)\rangle$, for jets in photoproduction with
$\etjet>17$~GeV in different $\etajet$ regions. For comparison, the
predictions of {\sc Herwig} (dashed lines) and {\sc Pythia} MI
(dot-dashed lines) including resolved plus direct processes are
shown. The open circles show the fractional difference of the data to
the predictions of {\sc Herwig}. Other details are as in the caption to
Fig.~\protect\ref{fig1}.}
\label{fig2}
\vfill
\end{figure}

\newpage
\clearpage
\begin{figure}[p]
\vfill
\setlength{\unitlength}{1.0cm}
\begin{picture} (18.0,18.0)
\put (-1.0,0.5){\epsfig{figure=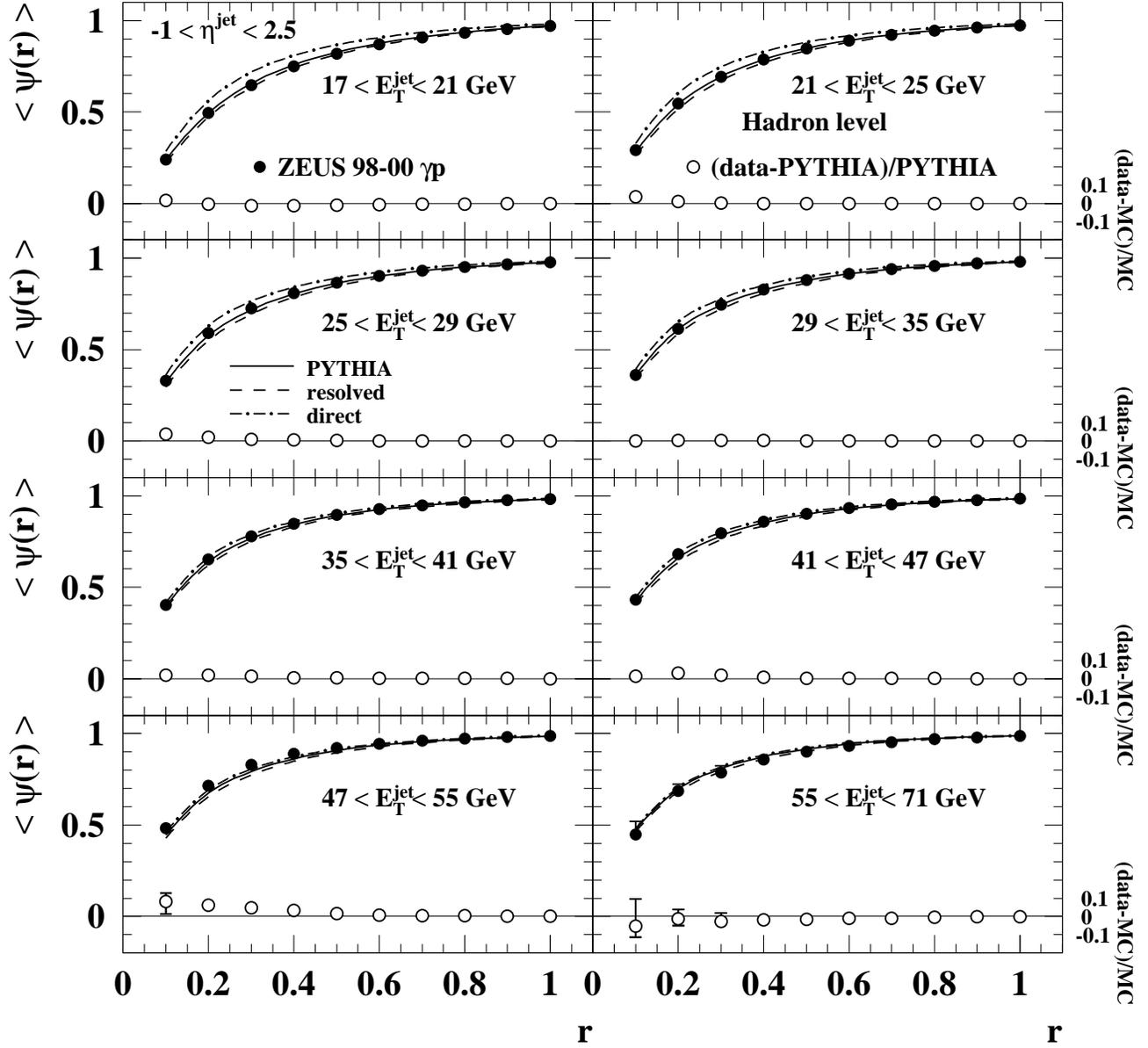,width=18cm}}
\end{picture}
\vspace{-1.5cm}
\caption
{\it 
Measured mean integrated jet shape corrected to the hadron level
(dots), $\langle\psi(r)\rangle$, for jets in photoproduction in the
range $\etar$ in different $\etjet$ regions. For comparison, the
predictions of {\sc Pythia} including resolved (dashed lines), direct
(dot-dashed lines) and resolved plus direct processes (solid lines) are
shown. Other details are as in the caption to Fig.~\protect\ref{fig1}.}
\label{fig3}
\vfill
\end{figure}

\newpage
\clearpage
\begin{figure}[p]
\vfill
\setlength{\unitlength}{1.0cm}
\begin{picture} (18.0,18.0)
\put (-1.0,0.5){\epsfig{figure=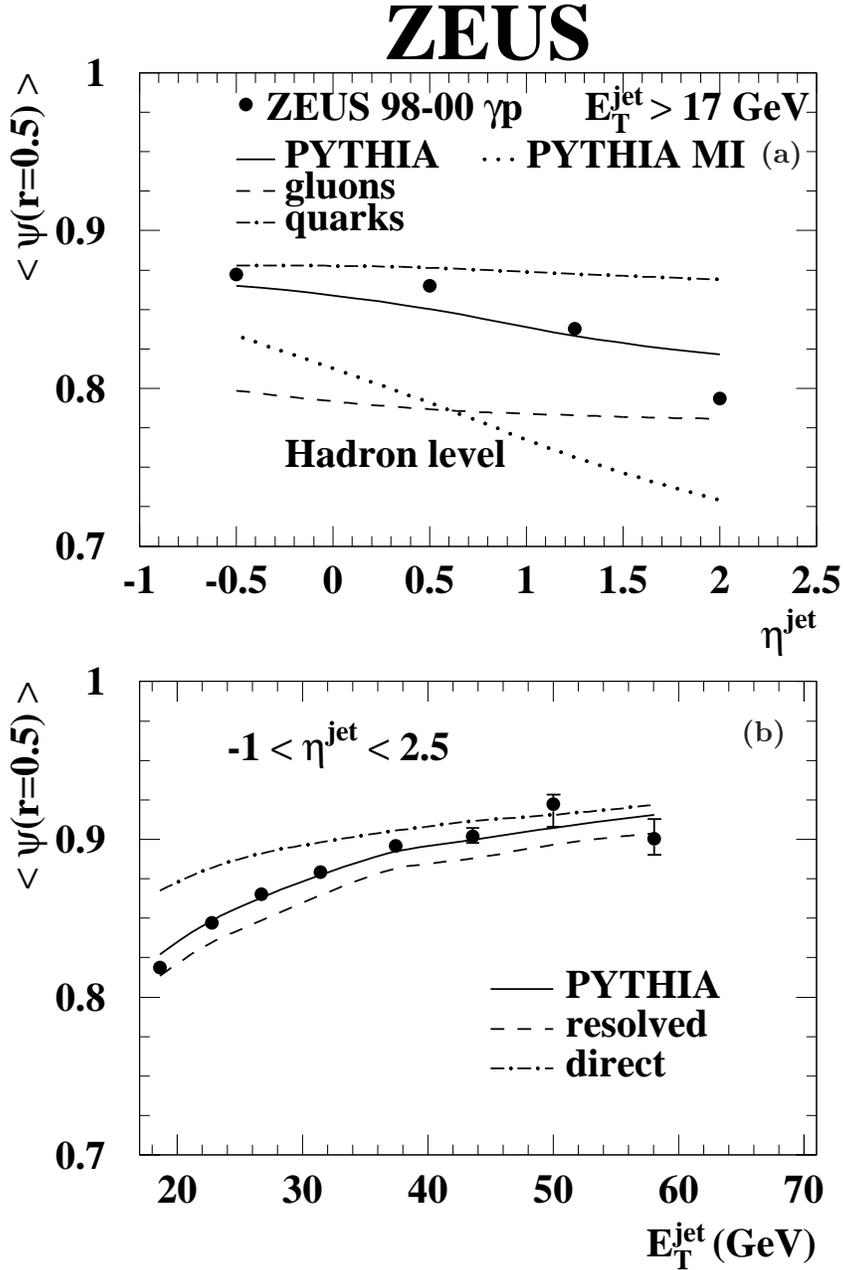,width=18cm}}
\put (11.75,15.5){\bf\small (a)}
\put (11.5,7.8){\bf\small (b)}
\end{picture}
\vspace{-1.5cm}
\caption
{\it 
Measured mean integrated jet shape in photoproduction corrected to the 
hadron level at a fixed value of $r=0.5$ (dots),
$\langle\psi(r=0.5)\rangle$, as a function of (a) $\etajet$ with
$\etjet>17$~GeV and (b) $\etjet$ with $\etar$. Other details are as in
the captions to Figs.~\protect\ref{fig1} and \protect\ref{fig3}.}
\label{fig4}
\vfill
\end{figure}

\newpage
\clearpage
\begin{figure}[p]
\vfill
\setlength{\unitlength}{1.0cm}
\begin{picture} (18.0,18.0)
\put (-1.0,0.5){\epsfig{figure=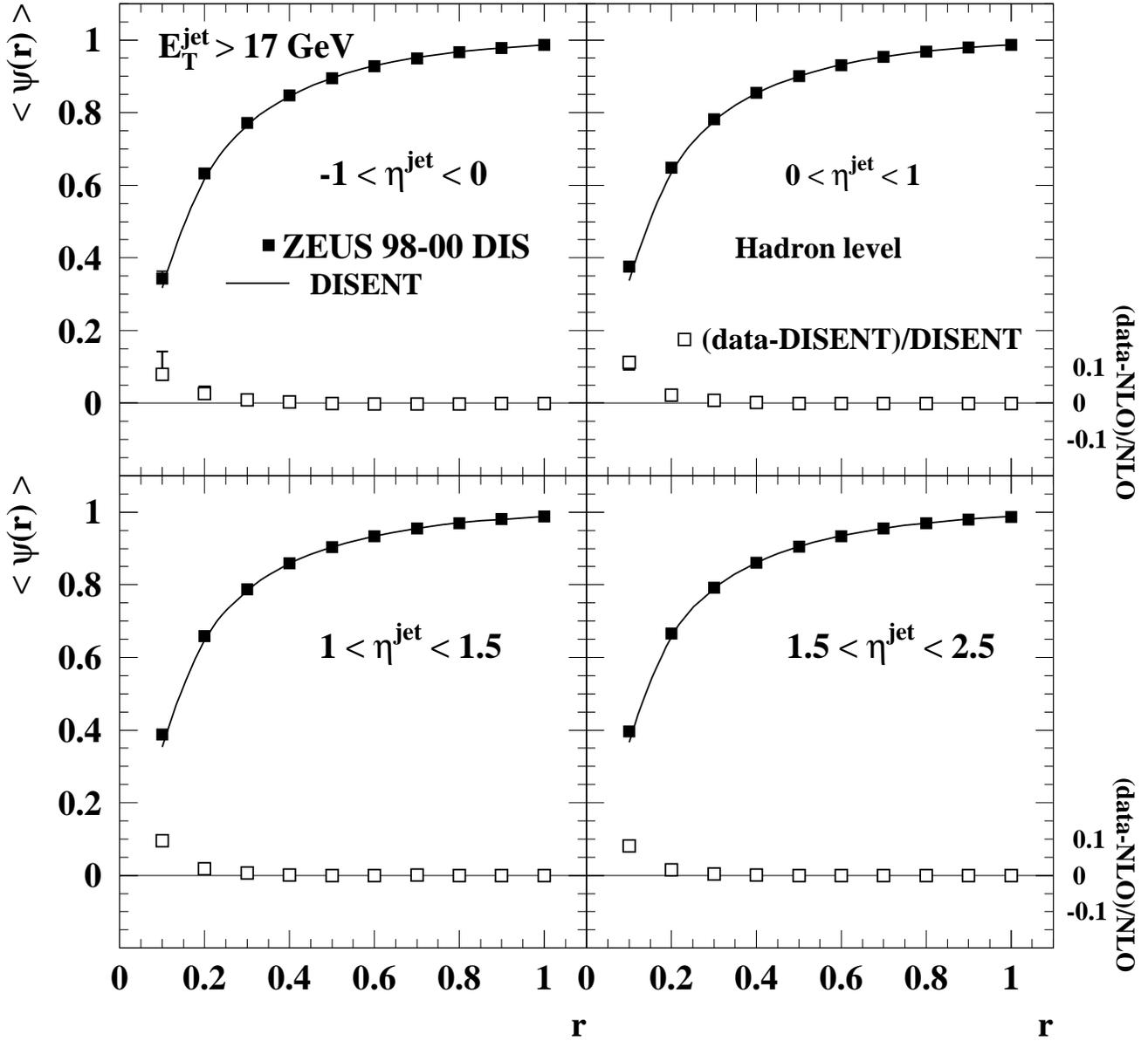,width=18cm}}
\end{picture}
\vspace{-1.5cm}
\caption
{\it 
Measured mean integrated jet shape corrected to the hadron level
and for electroweak radiative effects (squares), $\langle\psi(r)\rangle$,
for jets in DIS with $\etjet>17$~GeV in different $\etajet$
regions. For comparison, NLO predictions corrected for hadronisation and
$\z0$-exchange effects (solid lines) are shown. Other details are as in
the caption to Fig.~\protect\ref{fig1}.}
\label{fig5}
\vfill
\end{figure}

\newpage
\clearpage
\begin{figure}[p]
\vfill
\setlength{\unitlength}{1.0cm}
\begin{picture} (18.0,18.0)
\put (-1.0,0.5){\epsfig{figure=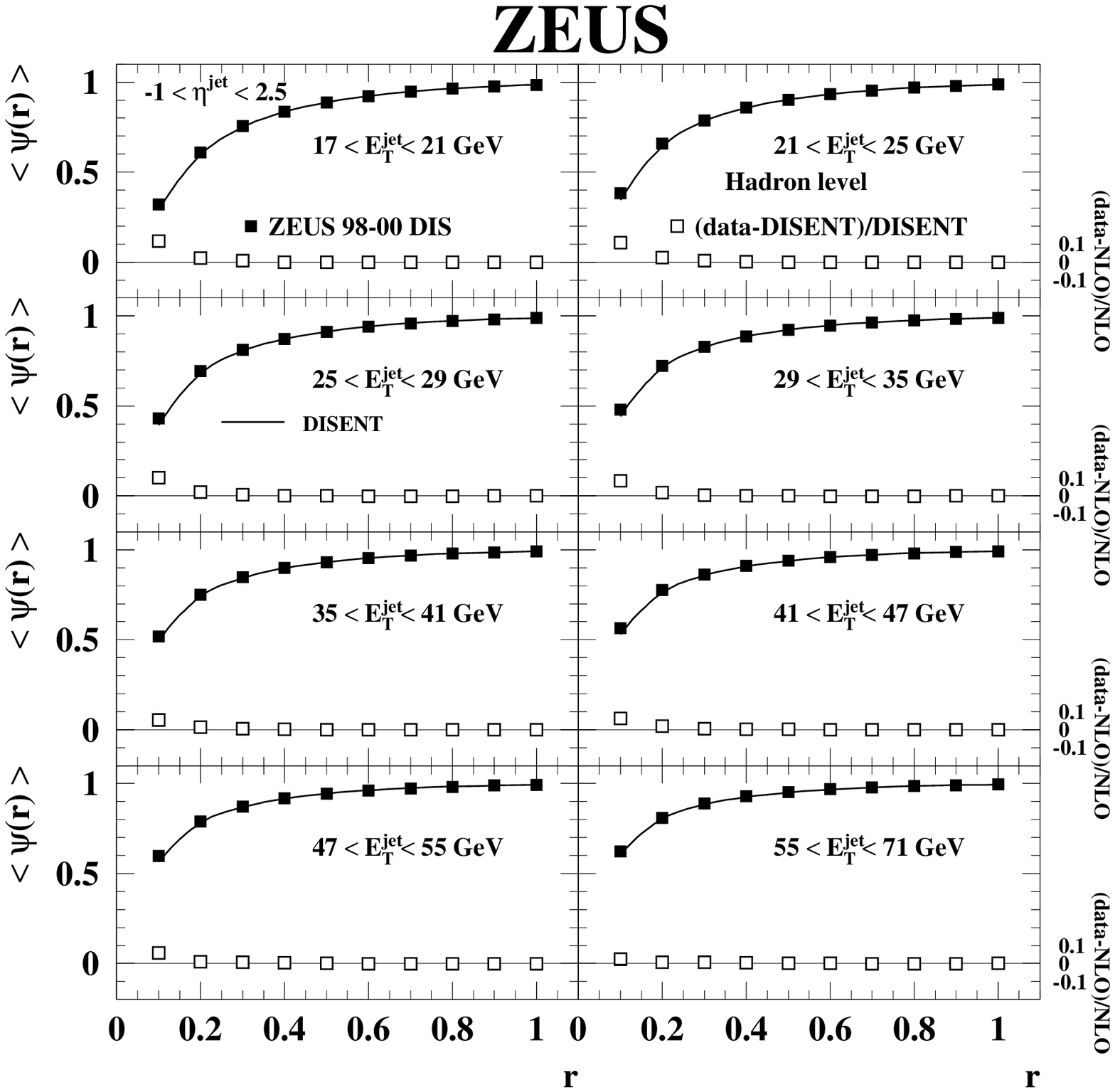,width=18cm}}
\end{picture}
\vspace{-1.5cm}
\caption
{\it 
Measured mean integrated jet shape corrected to the hadron level and
for electroweak radiative effects (squares), $\langle\psi(r)\rangle$,
for jets in DIS in the range $\etar$ in different $\etjet$
regions. Other details are as in the captions to Figs.~\protect\ref{fig1}
and \protect\ref{fig5}.}
\label{fig6}
\vfill
\end{figure}

\newpage
\clearpage
\begin{figure}[p]
\vfill
\setlength{\unitlength}{1.0cm}
\begin{picture} (18.0,18.0)
\put (-1.0,0.5){\epsfig{figure=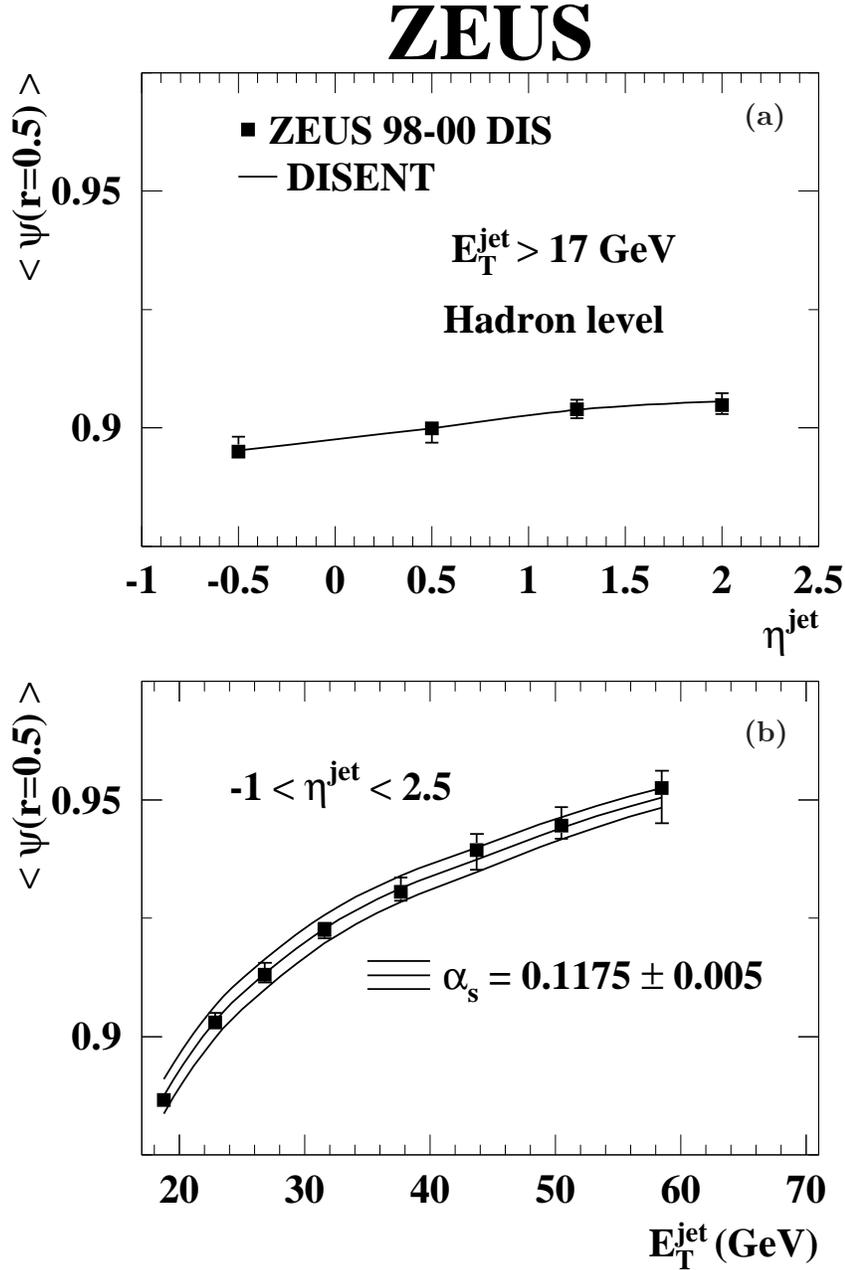,width=18cm}}
\put (11.5,16.0){\bf\small (a)}
\put (11.5,7.8){\bf\small (b)}
\end{picture}
\vspace{-1.5cm}
\caption
{\it 
Measured mean integrated jet shape in DIS corrected to the hadron level and
for electroweak radiative effects at a fixed value of $r=0.5$
(squares), $\langle\psi(r=0.5)\rangle$, as a function of (a) $\etajet$
with $\etjet>17$~GeV and (b) $\etjet$ with $\etar$. Other details are as
in the caption to Fig.~\protect\ref{fig5}.}
\label{fig7}
\vfill
\end{figure}

\newpage
\clearpage
\begin{figure}[p]
\vfill
\setlength{\unitlength}{1.0cm}
\begin{picture} (18.0,18.0)
\put (-1.0,0.5){\epsfig{figure=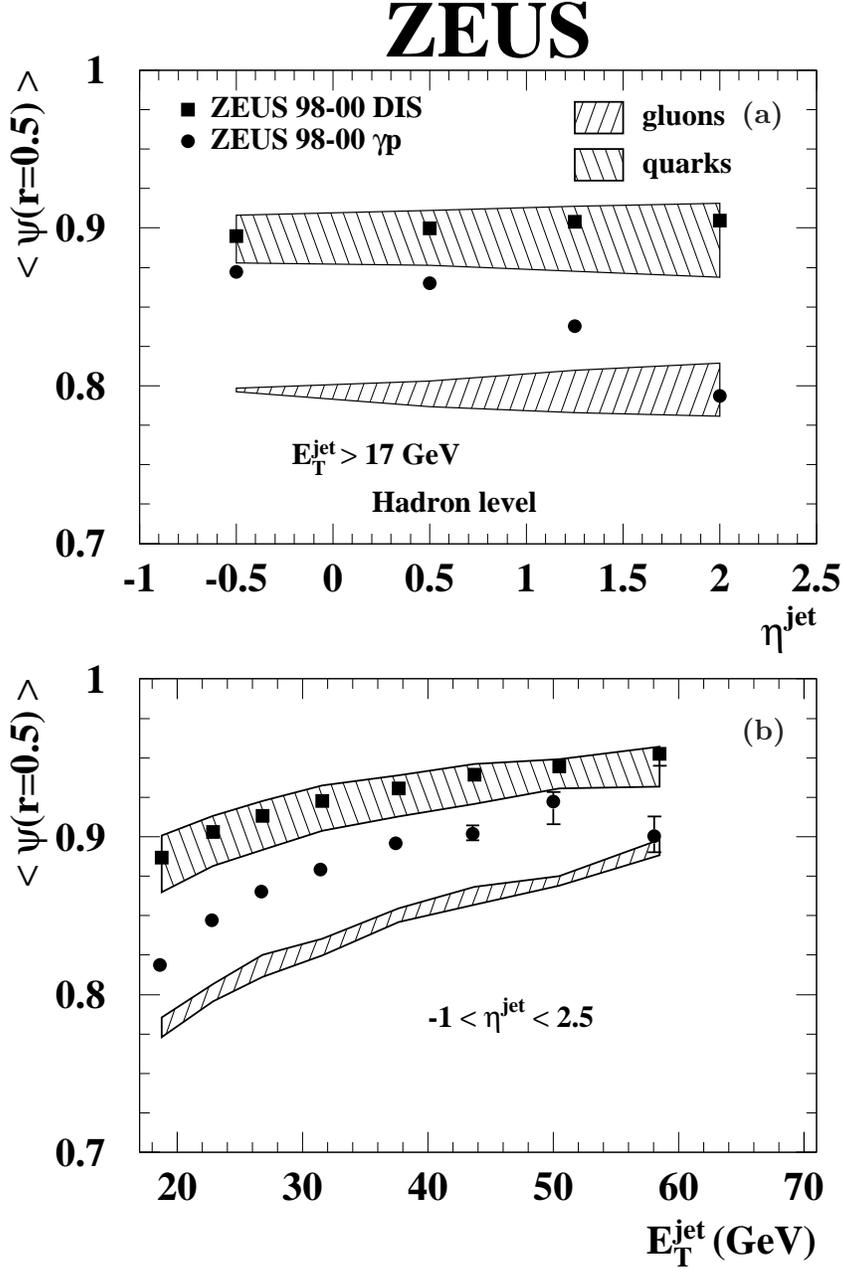,width=18cm}}
\put (11.5,16.0){\bf\small (a)}
\put (11.5,7.8){\bf\small (b)}
\end{picture}
\vspace{-1.5cm}
\caption
{\it 
Measured mean integrated jet shape corrected to the hadron level at a
fixed value of $r=0.5$ for DIS (squares) and photoproduction (dots), 
$\langle\psi(r=0.5)\rangle$, as a function of (a) $\etajet$ with
$\etjet>17$~GeV and (b) $\etjet$ with $\etar$. The predictions for
gluon-initiated (lower hatched areas) and quark-initiated (upper
hatched areas) jets are also shown. The bounds of each hatched area
are given by the predictions of CDM and {\sc Pythia}. Other details
are as in the caption to Fig.~\protect\ref{fig5}.}
\label{fig8}
\vfill
\end{figure}

\newpage
\clearpage
\begin{figure}[p]
\vfill
\setlength{\unitlength}{1.0cm}
\begin{picture} (18.0,18.0)
\put (-1.0,0.5){\epsfig{figure=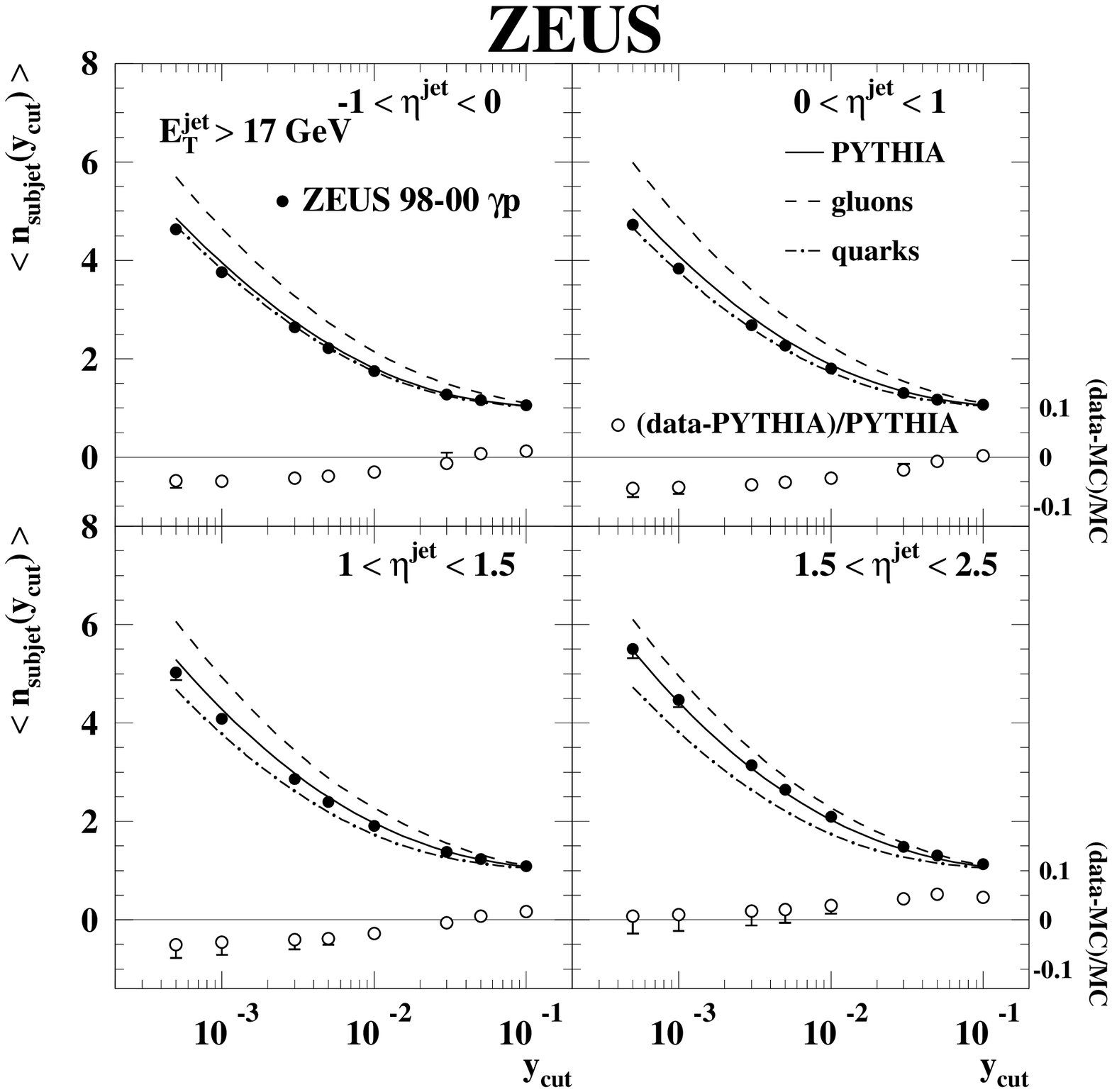,width=18cm}}
\end{picture}
\vspace{-1.5cm}
\caption
{\it 
Measured mean subjet multiplicity corrected to the hadron level
(dots), $\langle\ns(\yc)\rangle$, for jets in photoproduction with
$\etjet>17$~GeV in different $\etajet$ regions. Other details are as in
the caption to Fig.~\protect\ref{fig1}.} 
\label{fig9}
\vfill
\end{figure}

\newpage
\clearpage
\begin{figure}[p]
\vfill
\setlength{\unitlength}{1.0cm}
\begin{picture} (18.0,18.0)
\put (-1.0,0.5){\epsfig{figure=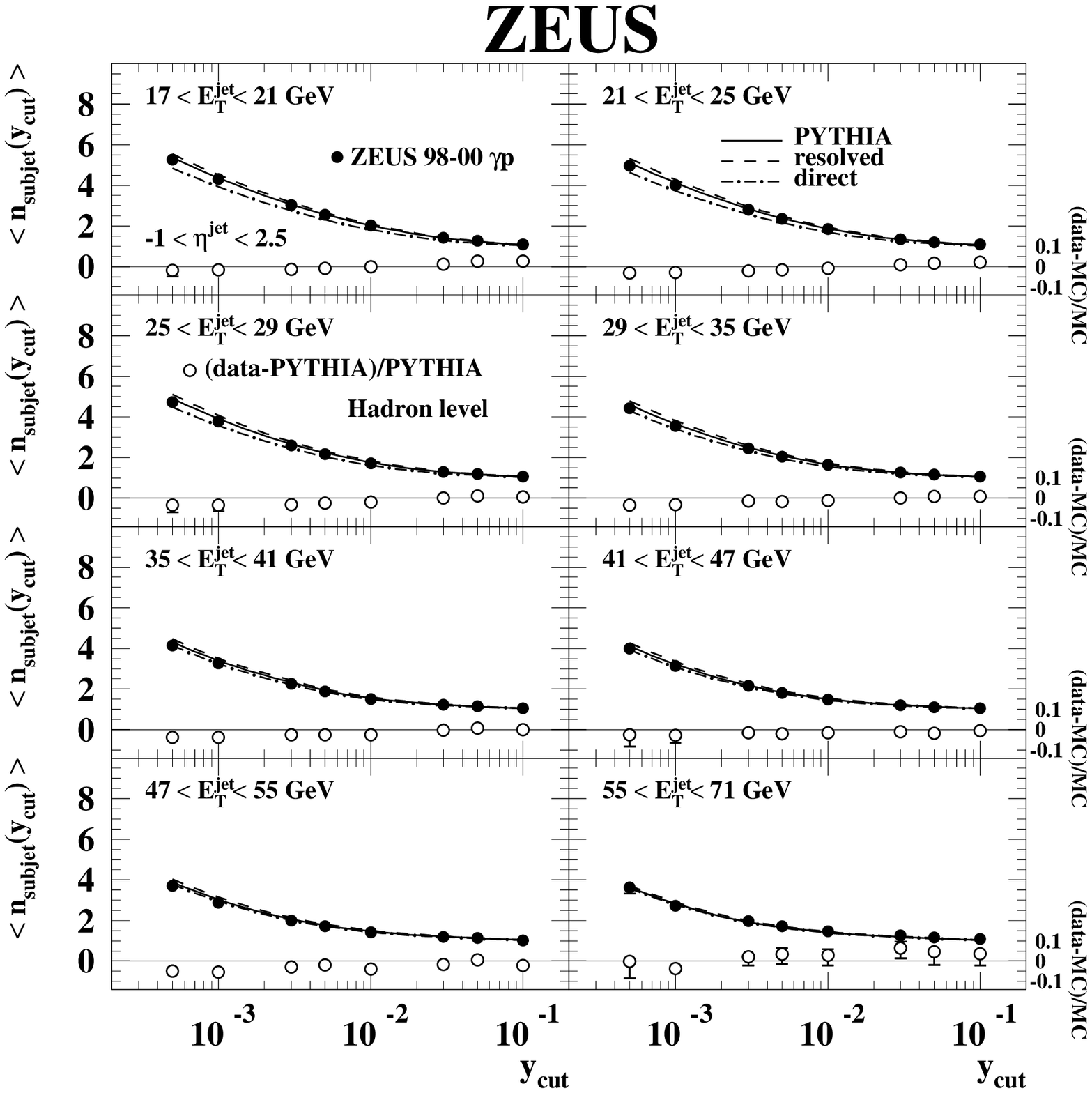,width=18cm}}
\end{picture}
\vspace{-1.5cm}
\caption
{\it 
Measured mean subjet multiplicity corrected to the hadron level
(dots), $\langle\ns(\yc)\rangle$, for jets in photoproduction in the
range $\etar$ in different $\etjet$ regions. Other details are as in the 
caption to Fig.~\protect\ref{fig3}.} 
\label{fig10}
\vfill
\end{figure}

\newpage
\clearpage
\begin{figure}[p]
\vfill
\setlength{\unitlength}{1.0cm}
\begin{picture} (18.0,18.0)
\put (-1.0,0.5){\epsfig{figure=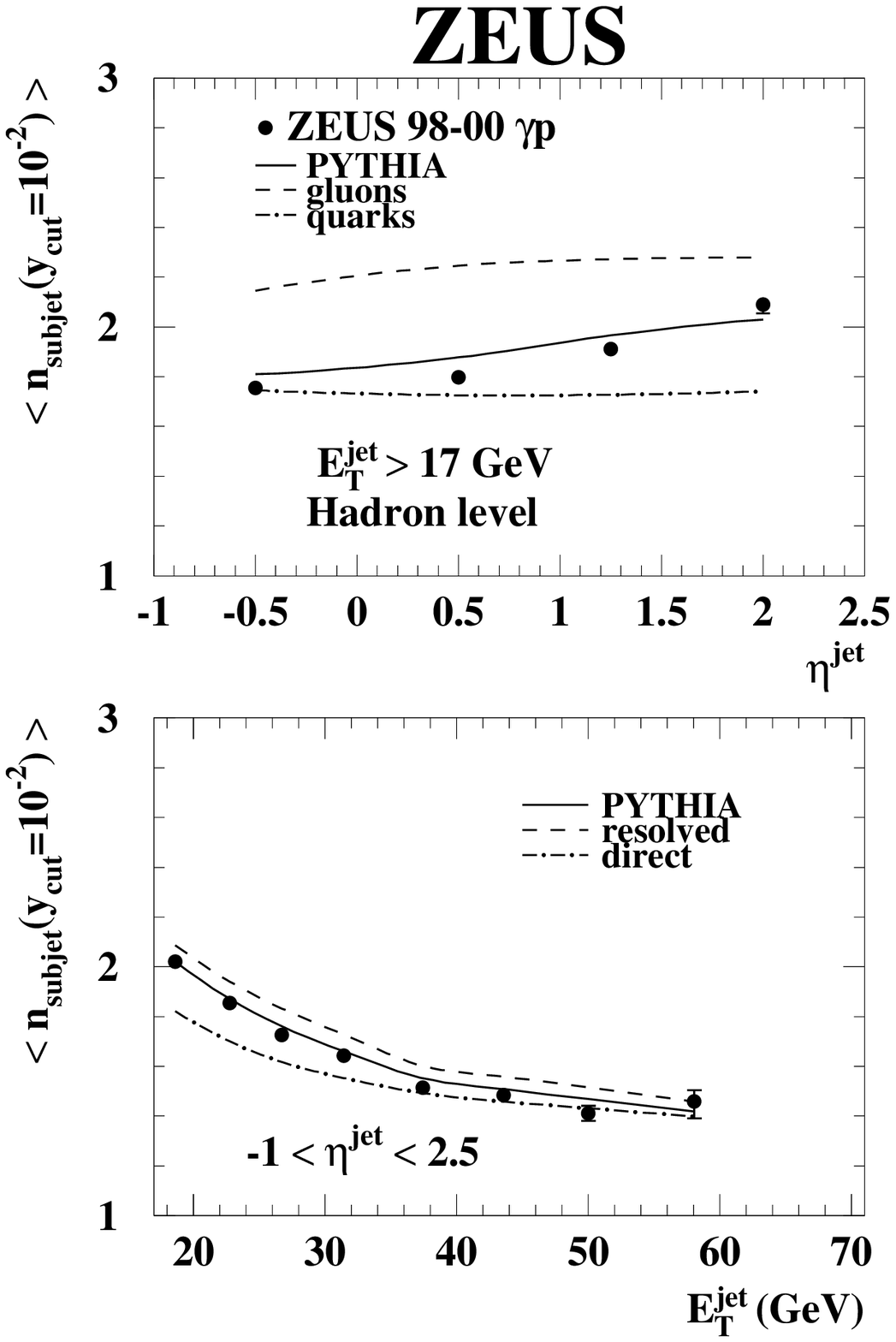,width=18cm}}
\put (11.5,16.0){\bf\small (a)}
\put (11.5,7.8){\bf\small (b)}
\end{picture}
\vspace{-1.5cm}
\caption
{\it 
Measured mean subjet multiplicity in photoproduction corrected to the 
hadron level at a fixed value of $\yc =10^{-2}$ (dots),
$\langle\ns(\yc=10^{-2})$, as a function of (a) $\etajet$ with
$\etjet>17$~GeV and (b) $\etjet$ with $\etar$. Other details are as in the
caption to Fig.~\protect\ref{fig4}.}
\label{fig11}
\vfill
\end{figure}

\newpage
\clearpage
\begin{figure}[p]
\vfill
\setlength{\unitlength}{1.0cm}
\begin{picture} (18.0,18.0)
\put (0.3,6.0){\epsfig{figure=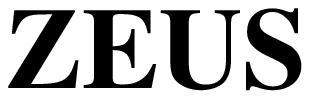,width=16cm}}
\put (-2.0,9.0){\epsfig{figure=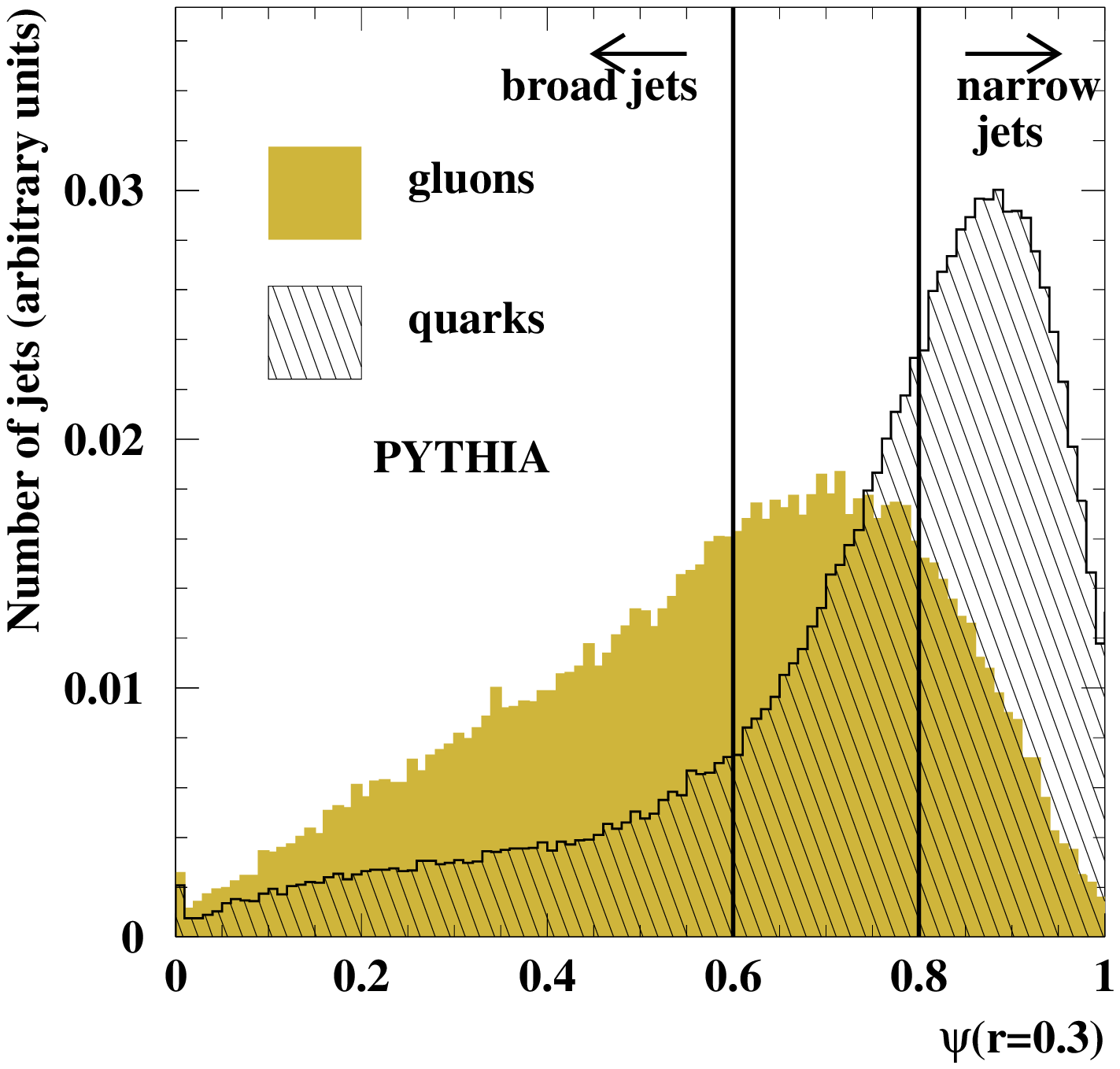,width=12cm}}
\put (7.0,9.0){\epsfig{figure=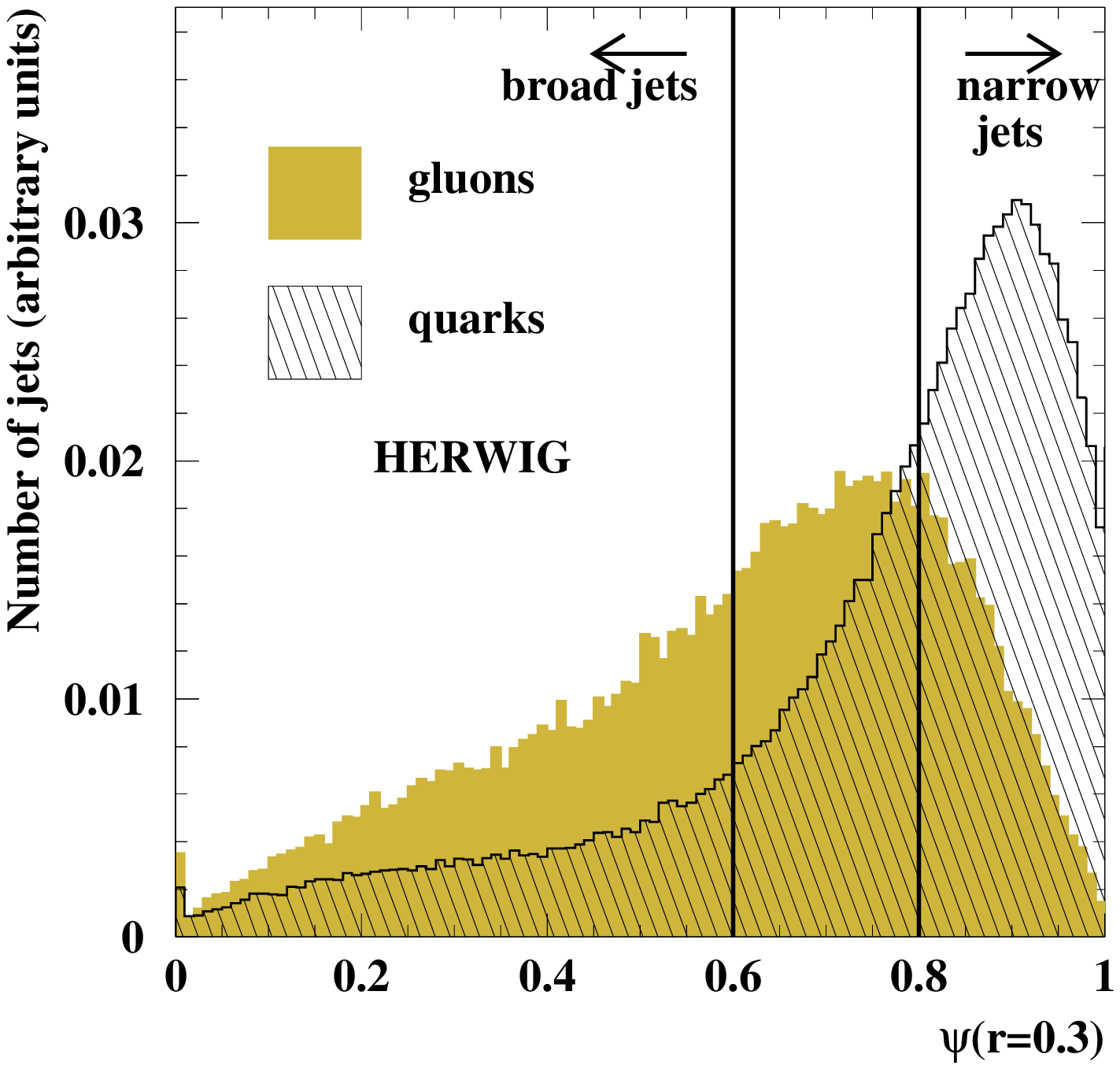,width=12cm}}
\put (-2.0,0.0){\epsfig{figure=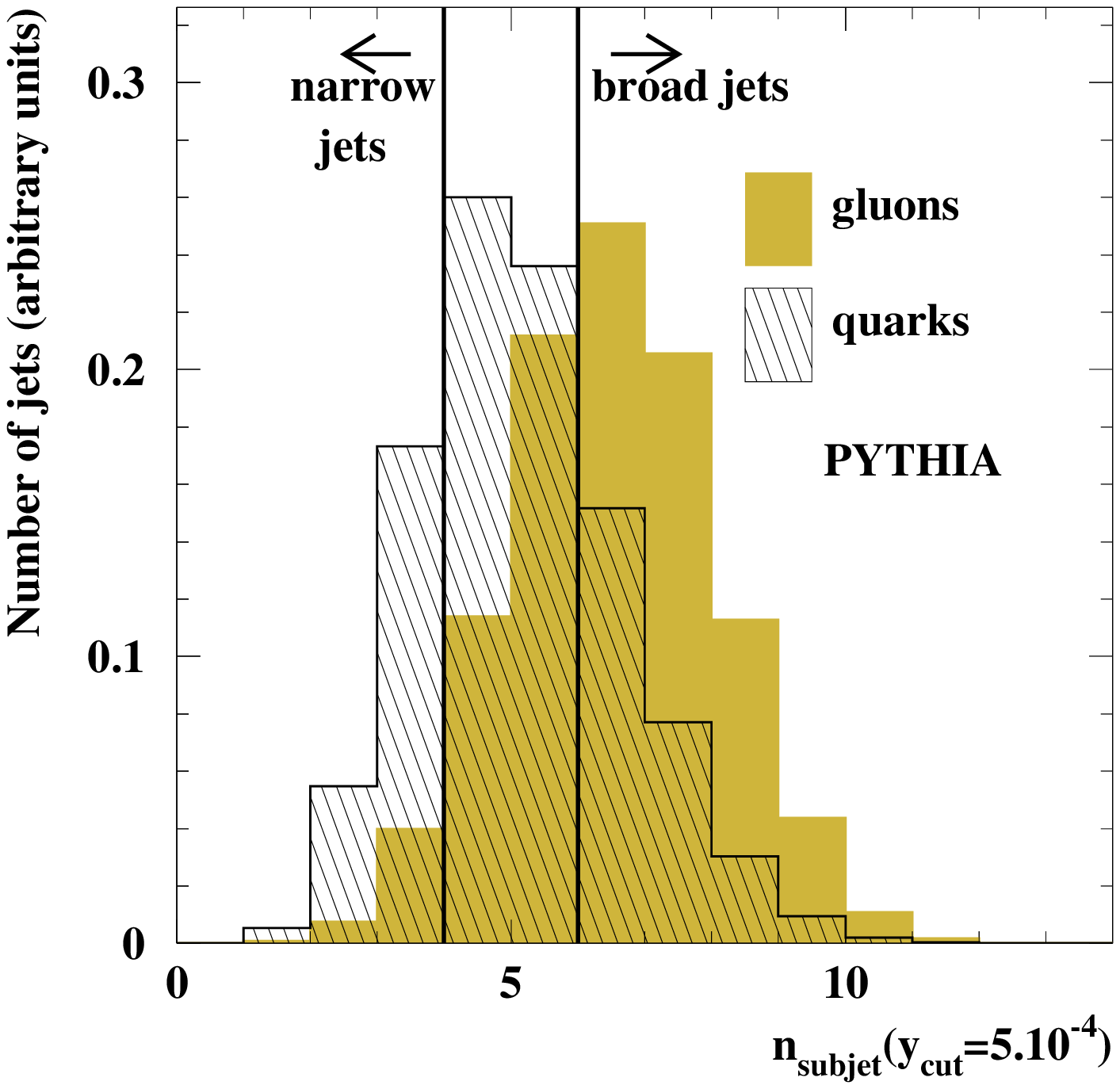,width=12cm}}
\put (7.0,0.0){\epsfig{figure=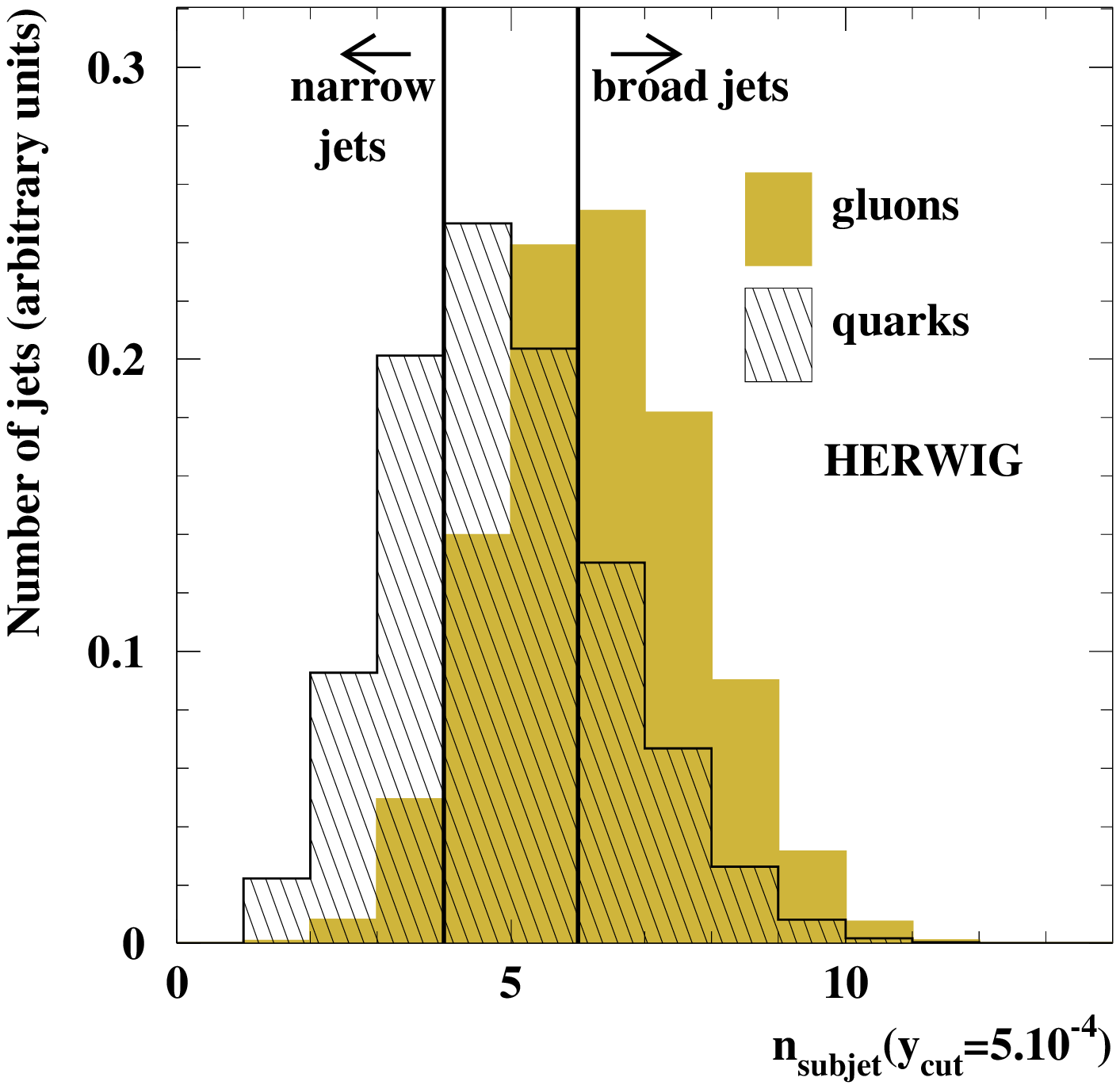,width=12cm}}
\put (0.5,18.0){\bf\small (a)}
\put (9.5,18.0){\bf\small (b)}
\put (0.5,9.0){\bf\small (c)}
\put (9.5,9.0){\bf\small (d)}
\end{picture}
\vspace{-1.5cm}
\caption
{\it 
(a) The predicted integrated jet shape distribution at $r=0.3$ and (c)
the predicted subjet multiplicity distribution at $\yc=5\cdot 10^{-4}$
at the hadron level for samples of gluon- (shaded histograms) and
quark-initiated (hatched histograms) jets simulated using the program
{\sc Pythia}; (b) and (d) show the same distributions for samples of
{\sc Herwig}.}
\label{fig12}
\vfill
\end{figure}

\newpage
\clearpage
\begin{figure}[p]
\vfill
\setlength{\unitlength}{1.0cm}
\begin{picture} (18.0,18.0)
\put (0.3,6.5){\epsfig{figure=DESY-04-072_0.eps,width=16cm}}
\put (-2.0,9.5){\epsfig{figure=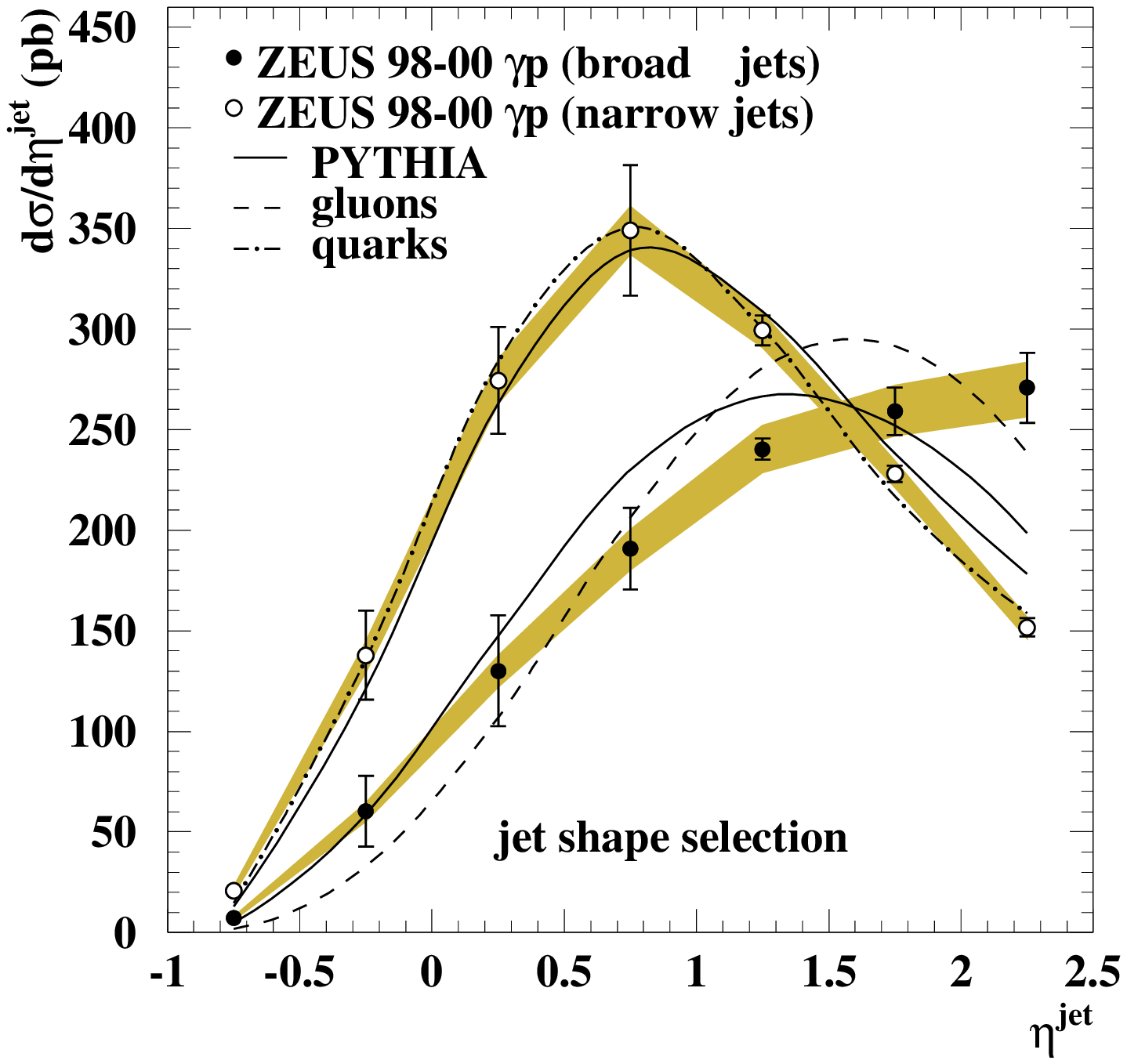,width=12cm}}
\put (7.0,9.5){\epsfig{figure=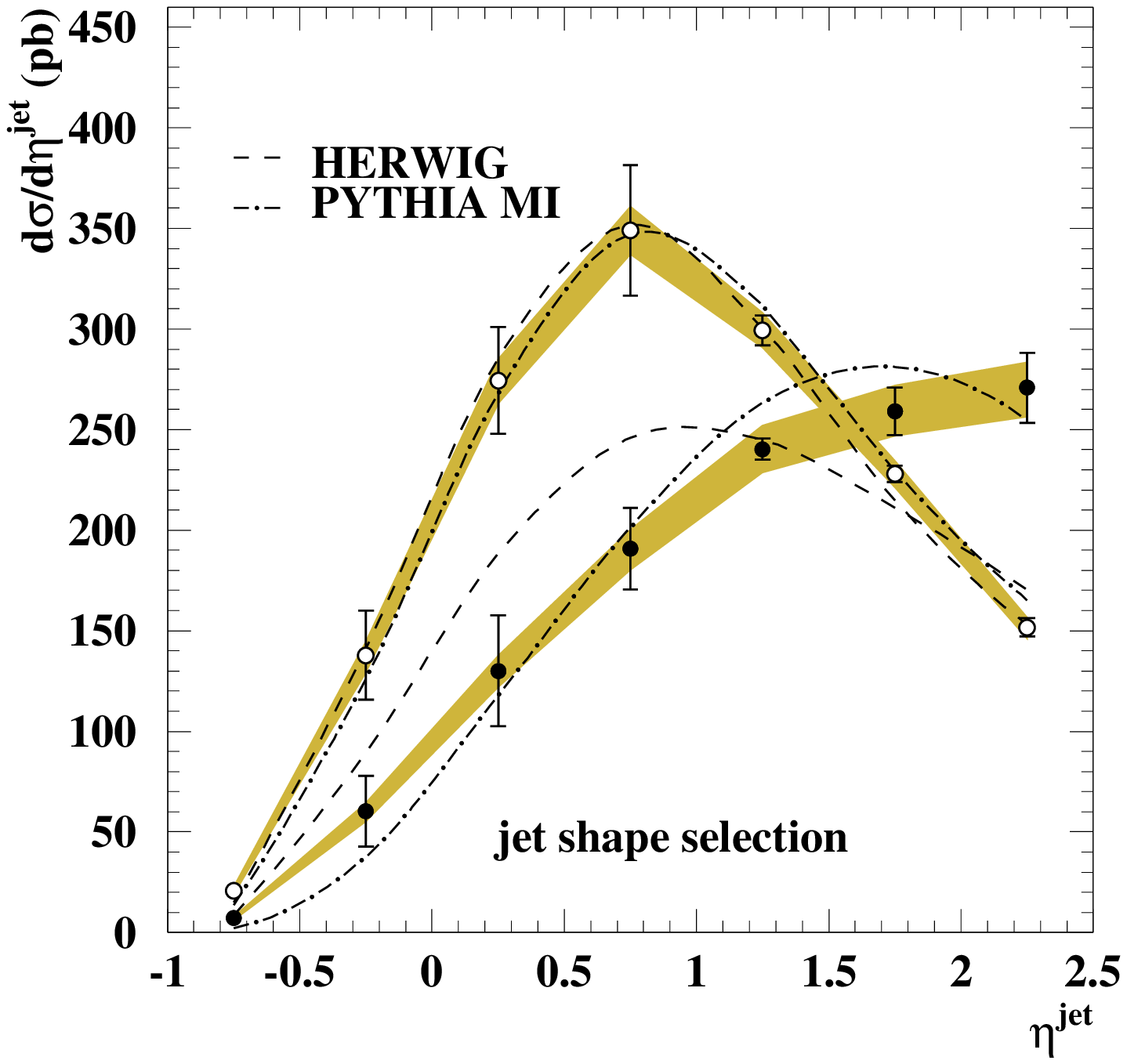,width=12cm}}
\put (-2.0,1.0){\epsfig{figure=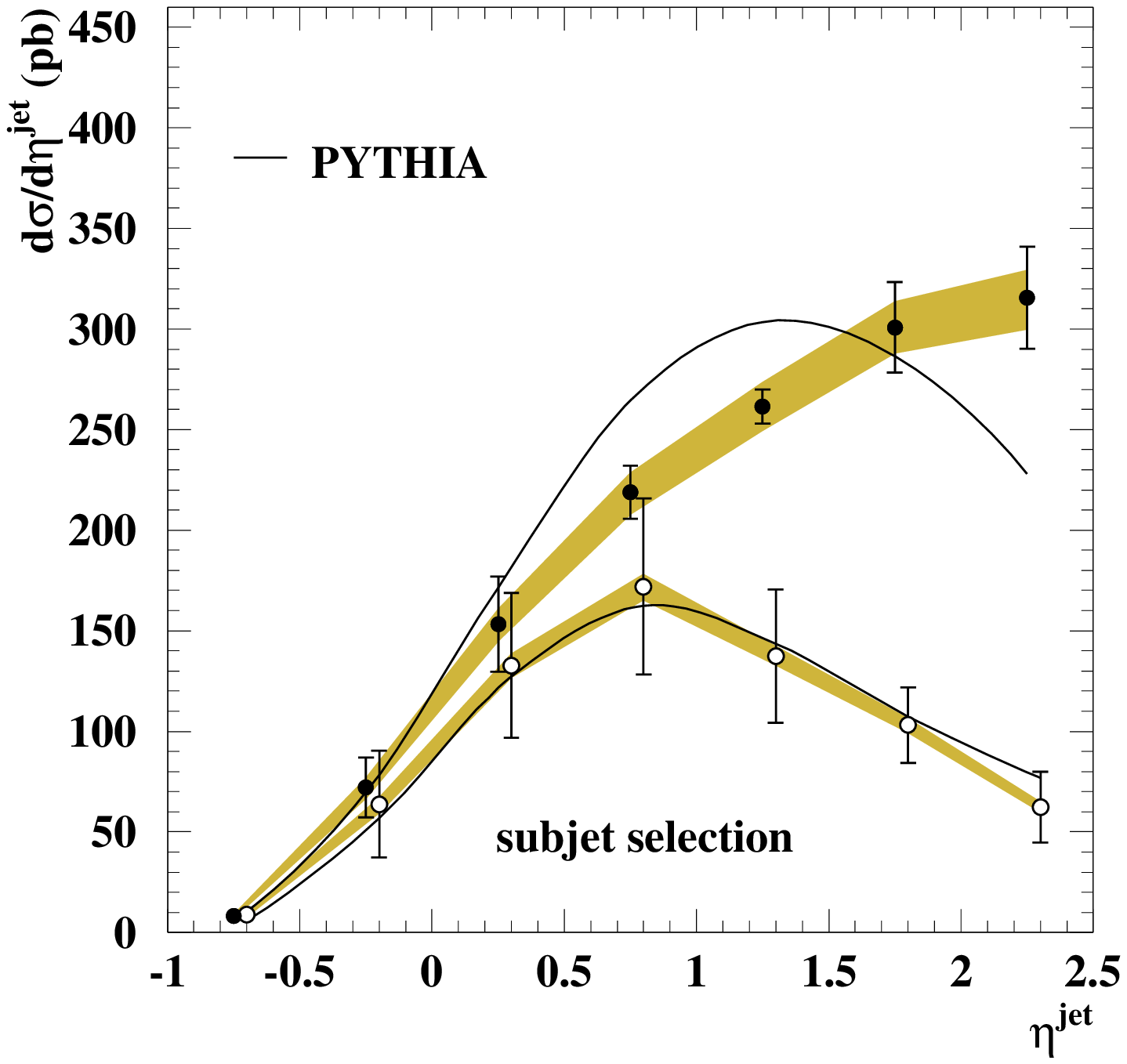,width=12cm}}
\put (7.0,1.0){\epsfig{figure=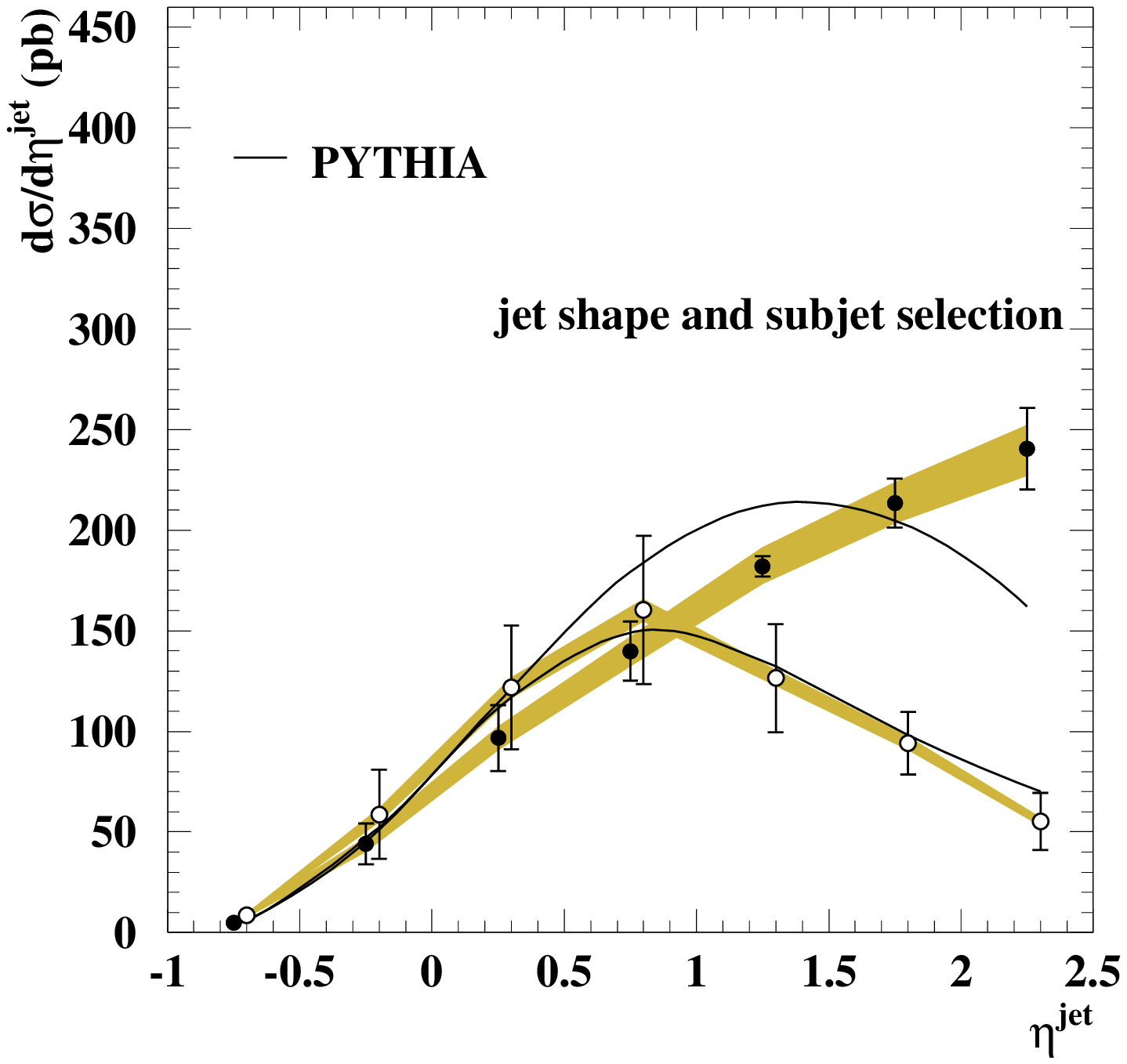,width=12cm}}
\put (6.5,18.5){\bf\small (a)}
\put (15.5,18.5){\bf\small (b)}
\put (6.5,10.0){\bf\small (c)}
\put (15.5,10.0){\bf\small (d)}
\end{picture}
\vspace{-3.5cm}
\caption
{\it 
Measured differential $ep$ cross-section $\seta$ for inclusive jet
photoproduction with $\etjet>17$ GeV in the kinematic region defined 
by $\q2<1$~\g2\ and \wrn. The jets have been selected according to
(a,b) their shape, (c) subjet multiplicity or (d) a combination of both
in broad jets (dots) and narrow jets (open circles). The thick error
bars (not visible) represent the statistical uncertainties of the data, and the thin
error bars show the statistical and systematic uncertainties $-$not
associated with the uncertainty in the absolute energy scale of the
jets, shown as a shaded band$-$ added in quadrature. The calculations
of {\sc Pythia} for resolved plus direct processes separated according
to the same criteria as in the data are included in (a,c,d) (solid
lines). In (a), the calculations of {\sc Pythia} for gluon (dashed
line) and quark (dot-dashed line) jets are also included. In (b), the
calculations of {\sc Herwig} (dashed lines) and {\sc Pythia} MI
(dot-dashed lines) with the same selection as in the data are
included. The MC calculations have been normalised to the total
measured cross section of each type. In (c) and (d), the measurements
and predictions for narrow jets have been plotted at $\etajet+0.05$
for clarity of presentation.}
\label{fig13}
\vfill
\end{figure}

\newpage
\clearpage
\begin{figure}[p]
\vfill
\setlength{\unitlength}{1.0cm}
\begin{picture} (18.0,10.0)
\put (0.3,0.0){\epsfig{figure=DESY-04-072_0.eps,width=16cm}}
\put (-2.0,1.0){\epsfig{figure=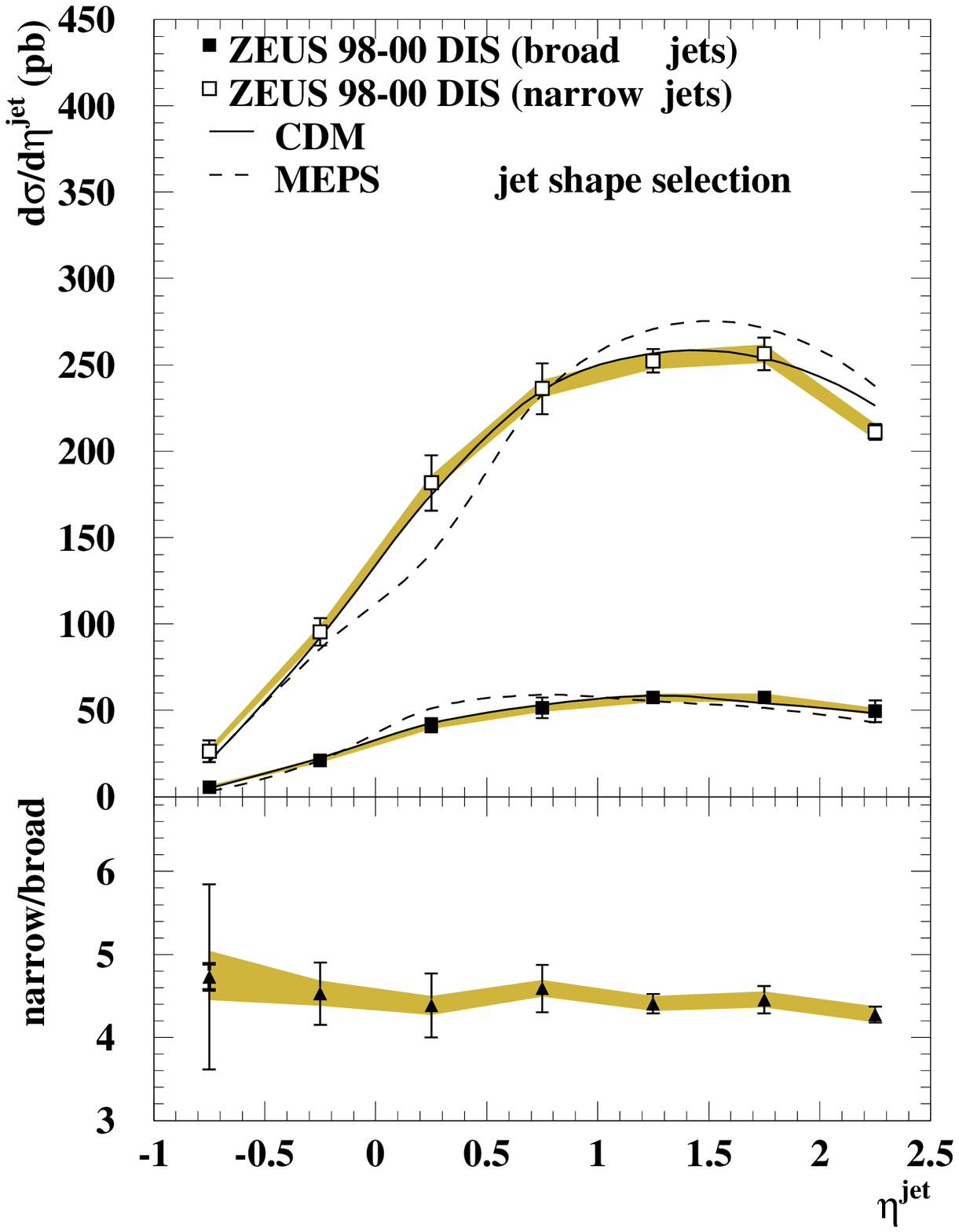,width=12cm}}
\put (7.0,2.65){\epsfig{figure=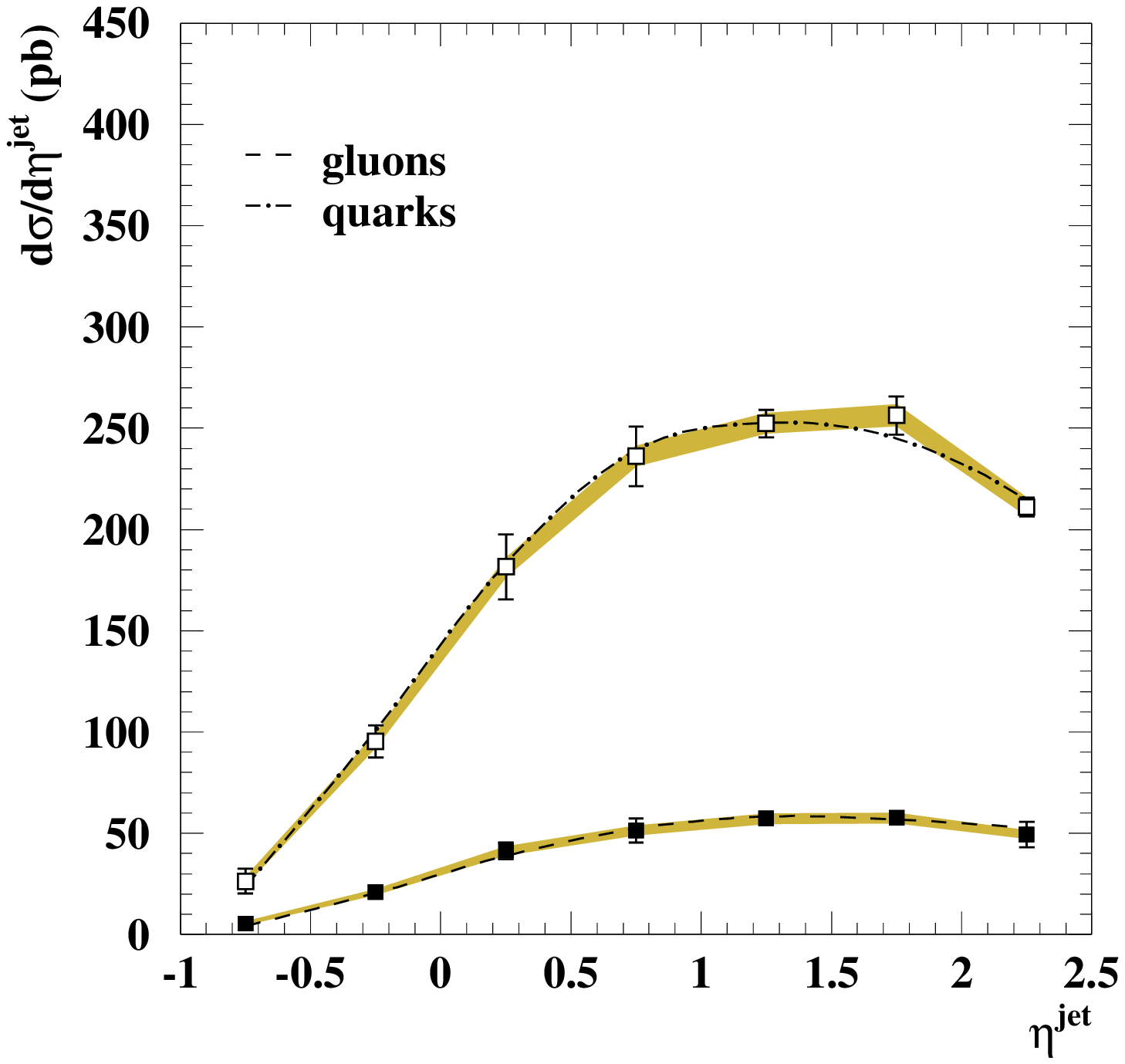,width=12cm}}
\put (6.5,11.5){\bf\small (a)}
\put (15.5,11.5){\bf\small (b)}
\end{picture}
\vspace{-1.5cm}
\caption
{\it
Measured differential $ep$ cross-section $\seta$ for inclusive jets in
DIS with $\etjet>17$ GeV in the kinematic region defined by 
$\q2>125$~\g2. The jets have been selected according to their shape as
broad jets (black squares) and narrow jets (white squares). The lower
part of (a) shows the ratio between the measured $\seta$ for the
narrow- and broad-jet samples (triangles). The calculations of CDM
(solid lines) and MEPS (dashed lines) are included in (a). In (b), the
calculations of CDM for gluon- (dashed line) and quark-initiated
(dot-dashed line) jets are included. Other details are as in the
caption to Fig.~\protect\ref{fig13}.}
\label{fig14}
\vfill
\end{figure}

\newpage
\clearpage
\begin{figure}[p]
\vfill
\setlength{\unitlength}{1.0cm}
\begin{picture} (18.0,18.0)
\put (0.3,6.0){\epsfig{figure=DESY-04-072_0.eps,width=16cm}}
\put (-2.0,9.0){\epsfig{figure=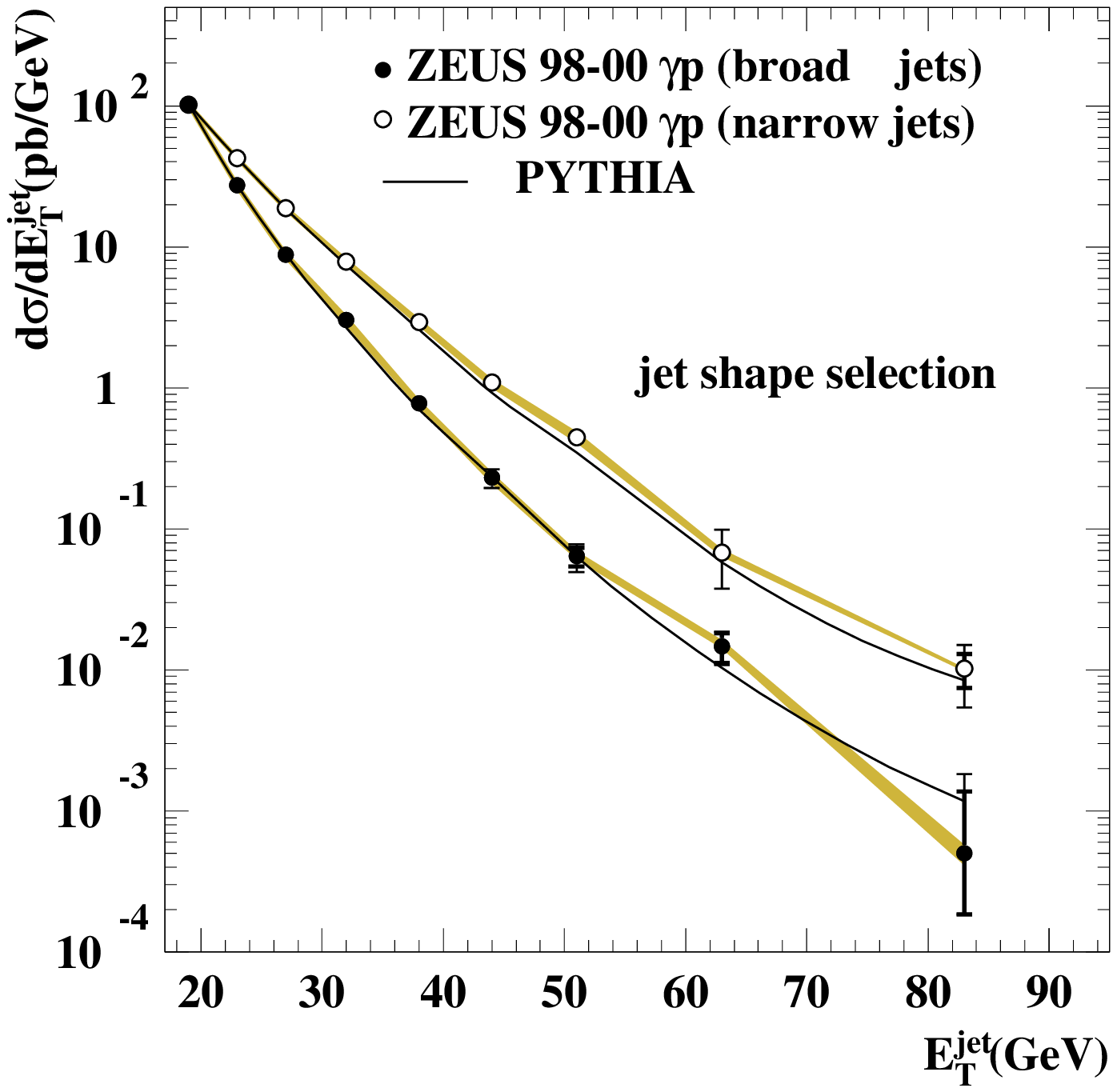,width=12cm}}
\put (7.0,9.0){\epsfig{figure=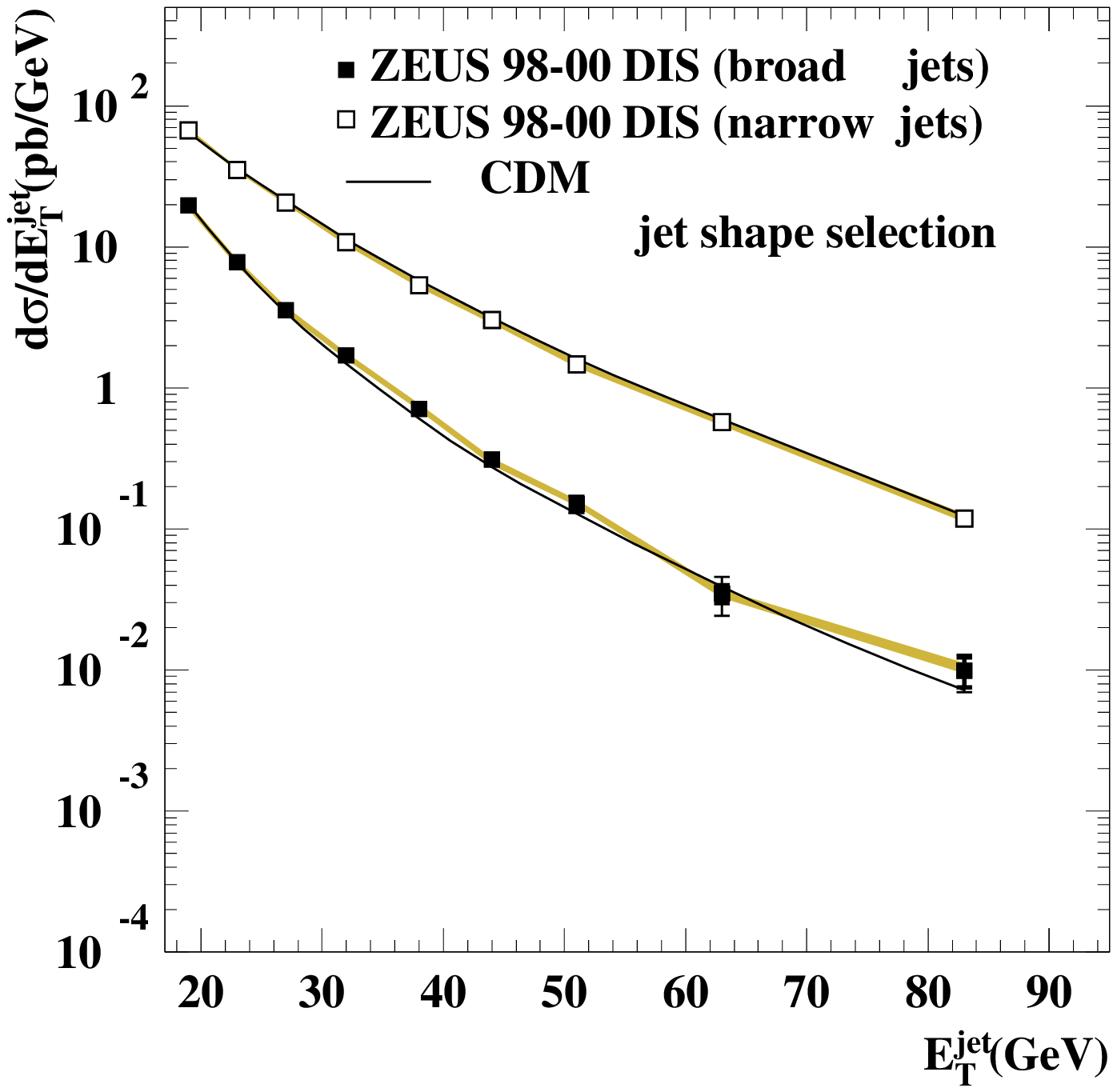,width=12cm}}
\put (-2.0,0.0){\epsfig{figure=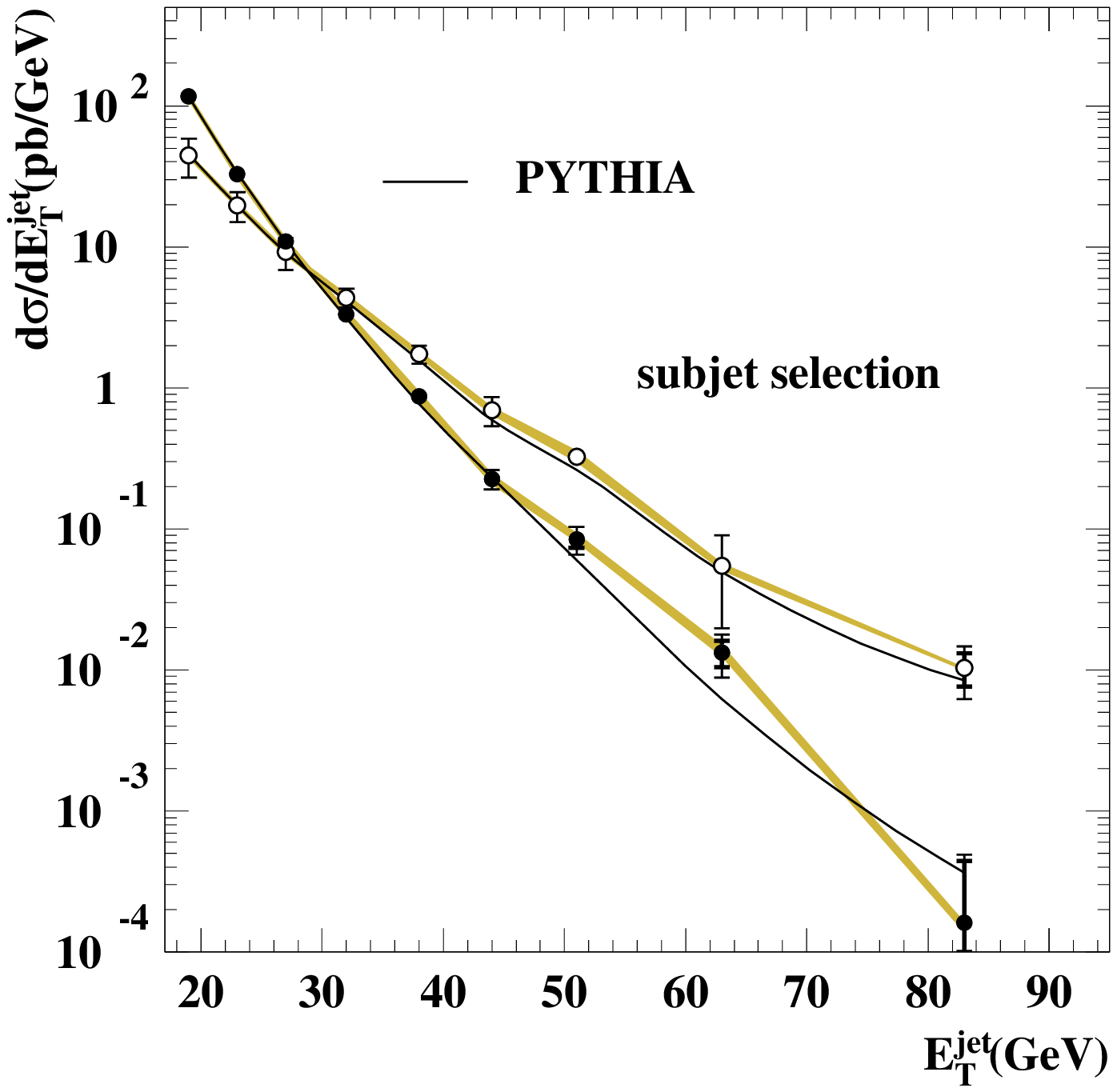,width=12cm}}
\put (7.0,0.0){\epsfig{figure=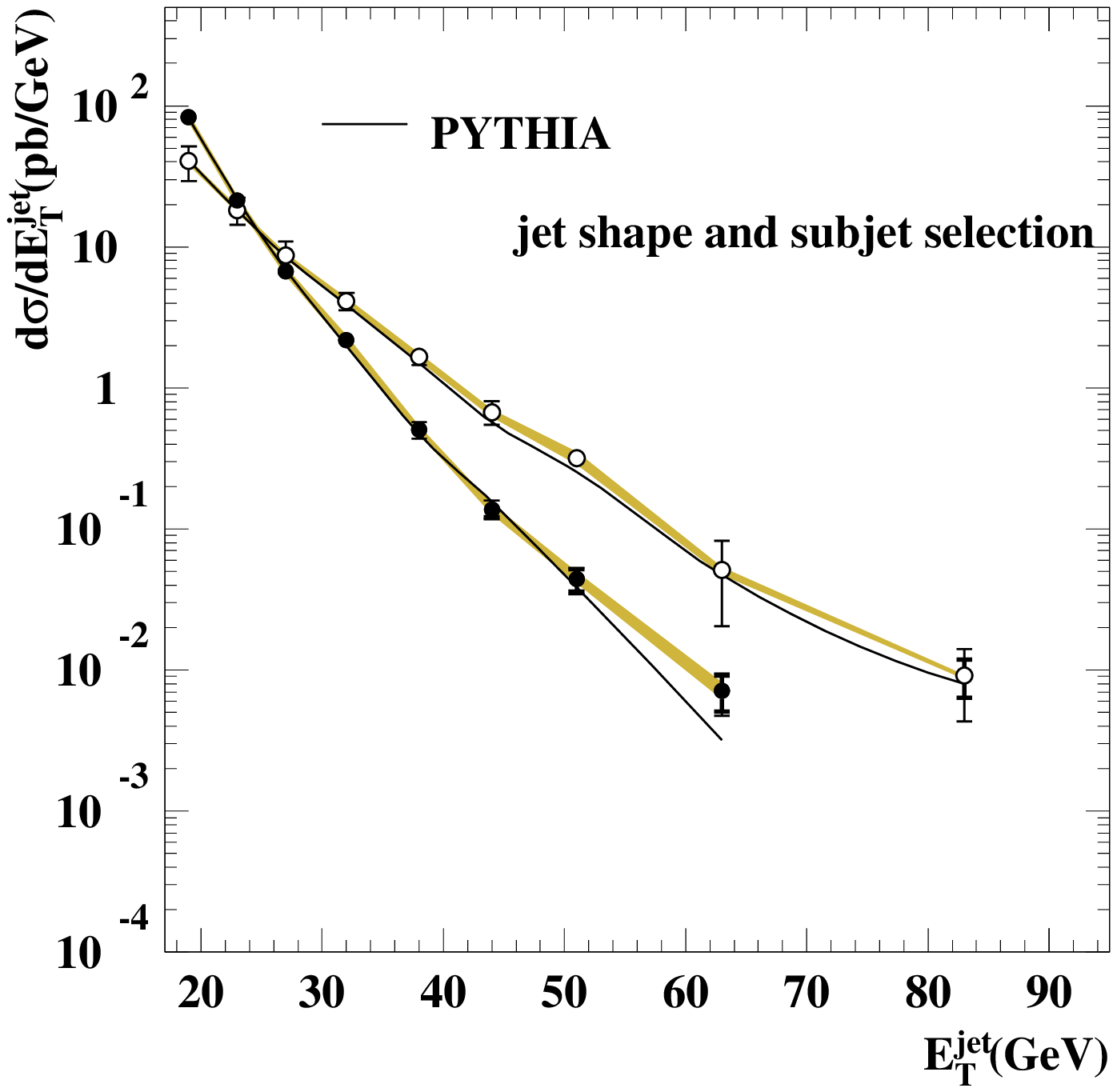,width=12cm}}
\put (6.8,18.0){\bf\small (a)}
\put (15.8,18.0){\bf\small (b)}
\put (6.5,9.0){\bf\small (c)}
\put (15.5,9.0){\bf\small (d)}
\end{picture}
\vspace{-1.5cm}
\caption
{\it 
(a) Measured differential $ep$ cross-section $\set$ for inclusive jet
photoproduction in the range $\etar$ in the kinematic region defined
by $\q2<1$~\g2\ and \wrn. (b) Measured differential $ep$ cross-section
$\set$ for inclusive jet DIS in the range $\etar$ in the kinematic
region defined by $\q2>125$~\g2. In (a) and (b), the jets have been
selected according to their shape. In (c) and (d), the photoproduced jets
have been selected according to the subjet multiplicity and the
combination of jet shape and subjet multiplicity, respectively. Other
details are as in the captions to Figs.~\protect\ref{fig13} and
\protect\ref{fig14}.}
\label{fig15}
\vfill
\end{figure}

\newpage
\clearpage
\begin{figure}[p]
\vfill
\setlength{\unitlength}{1.0cm}
\begin{picture} (18.0,18.0)
\put (0.3,6.0){\epsfig{figure=DESY-04-072_0.eps,width=16cm}}
\put (-2.0,9.0){\epsfig{figure=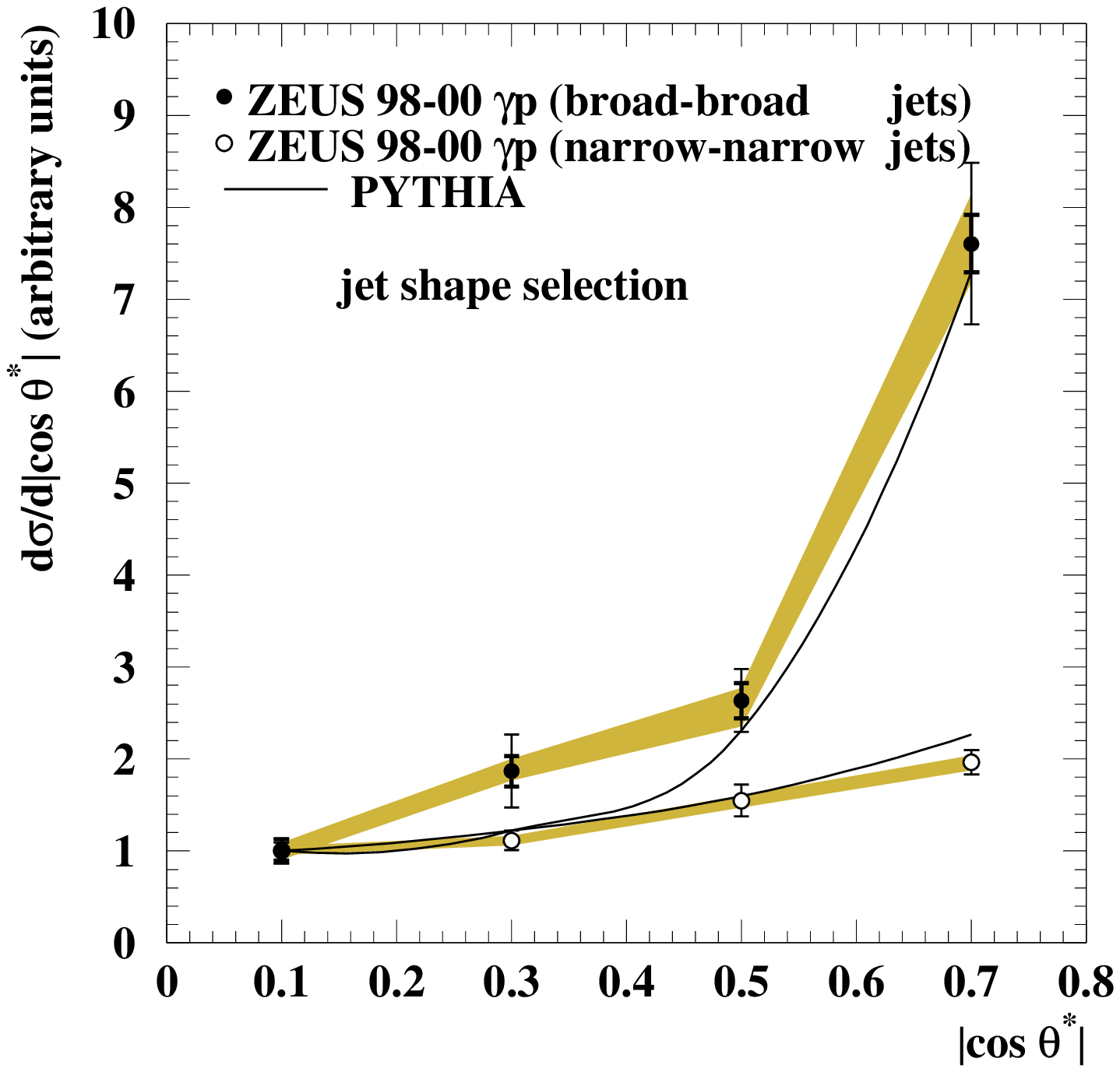,width=12cm}}
\put (7.0,9.0){\epsfig{figure=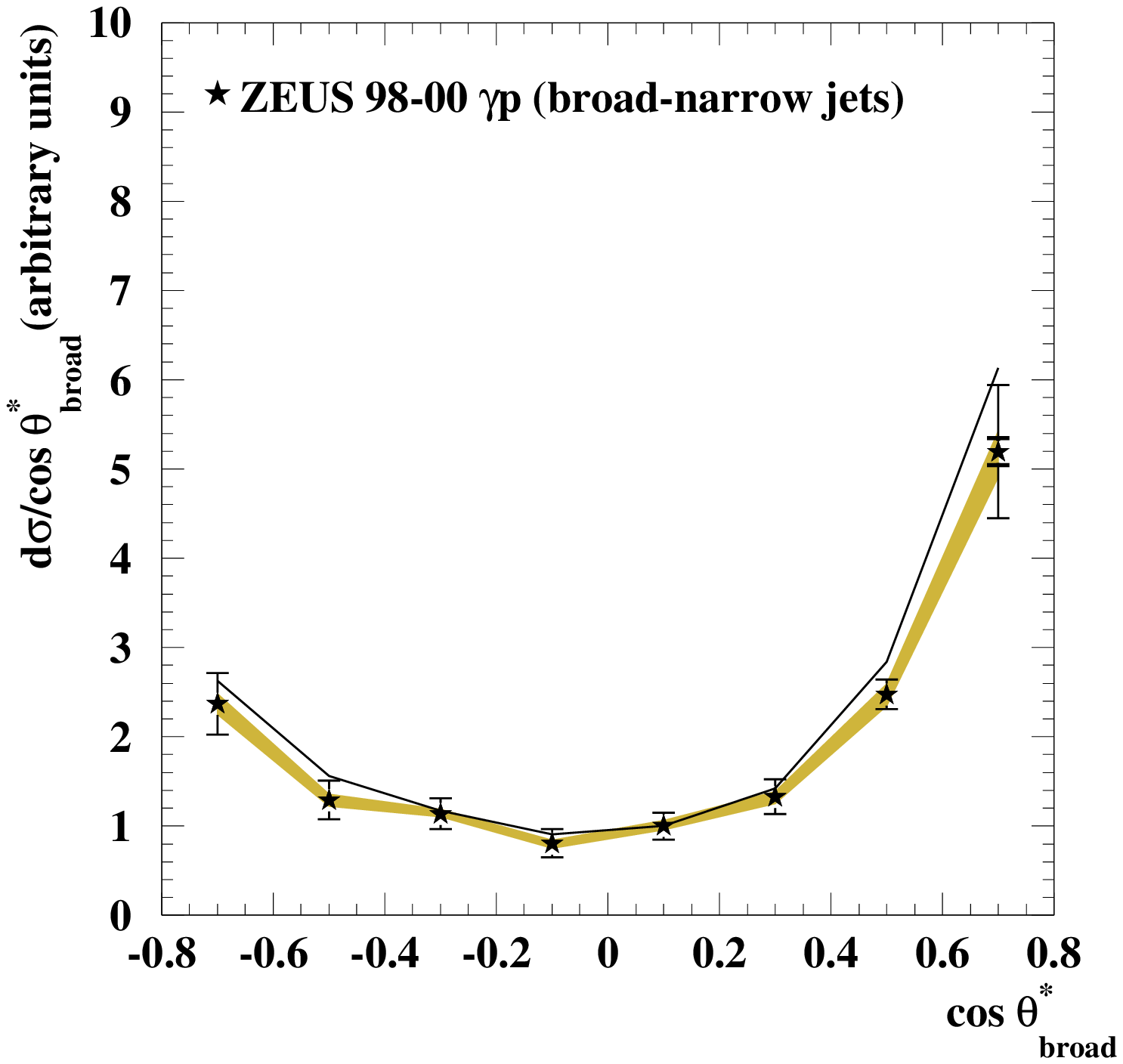,width=12cm}}
\put (-2.0,0.0){\epsfig{figure=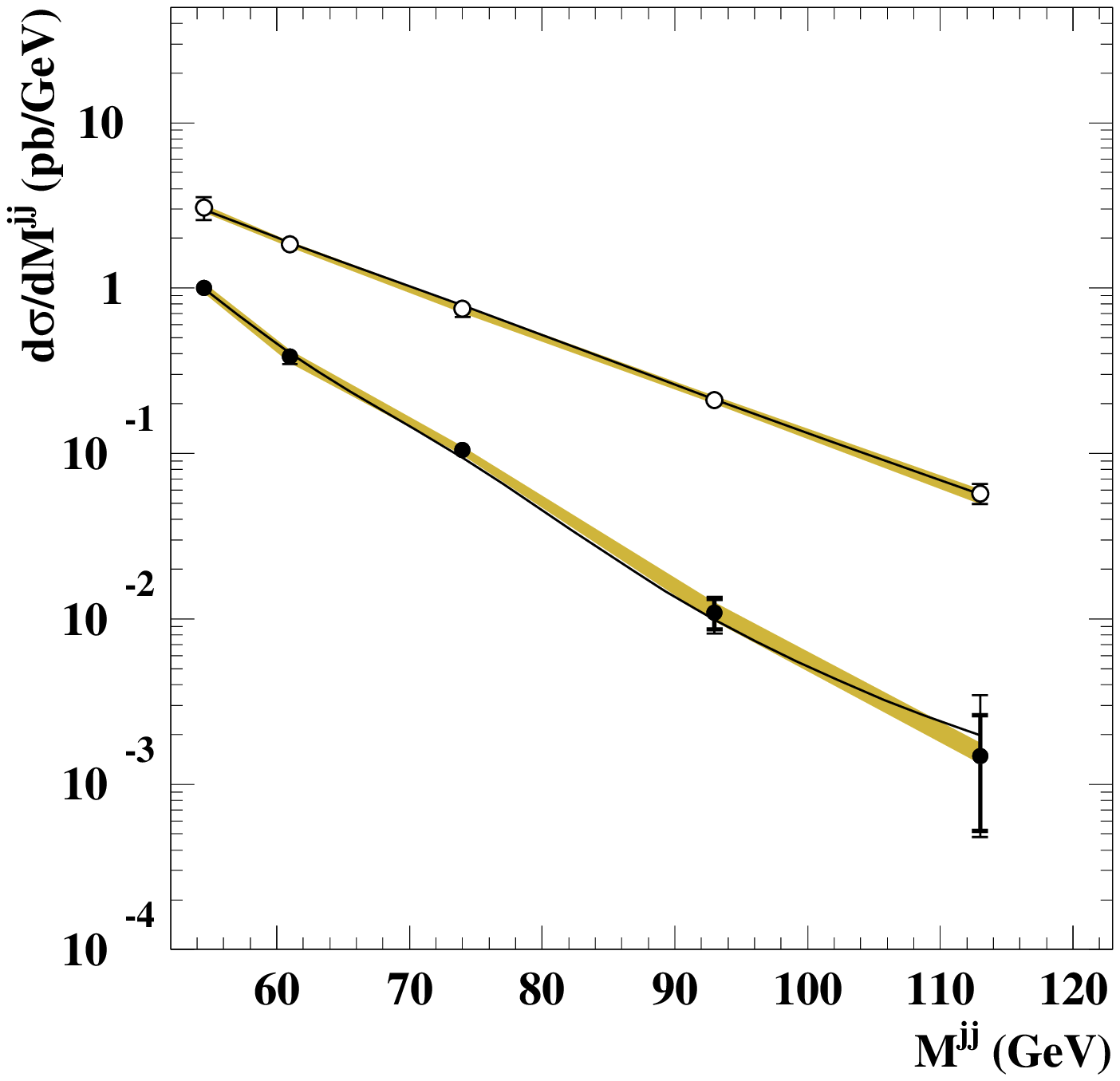,width=12cm}}
\put (7.0,0.0){\epsfig{figure=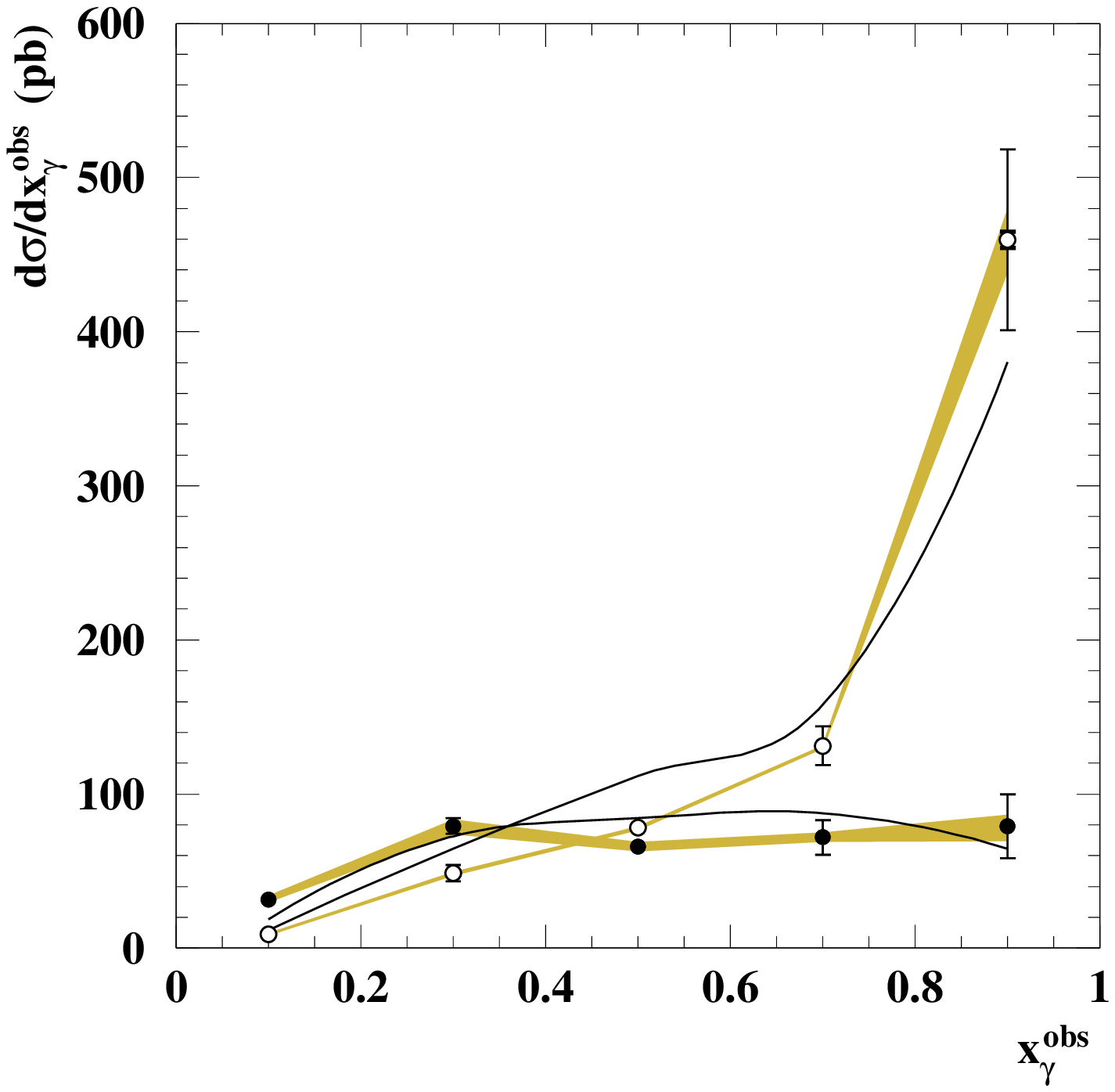,width=12cm}}
\put (6.8,18.0){\bf\small (a)}
\put (15.5,18.0){\bf\small (b)}
\put (6.5,9.0){\bf\small (c)}
\put (15.5,9.0){\bf\small (d)}
\end{picture}
\vspace{-1.5cm}
\caption
{\it 
Measured differential $ep$ cross sections for dijet photoproduction
with $\etj>17$ GeV, $\etjj>14$ GeV and $\etar$ in the kinematic region
defined by $\q2< 1$~\g2\ and \wrn\ as a function of (a)
$\cost$ and (b) $\ccos_{\rm broad}$ for $\mj>52$ GeV, (c) 
$\mj$ for $\cost<0.8$ and (d) $\xo$. The cross sections are for events
with broad-broad (dots), narrow-narrow (open circles) and
broad-narrow (stars) dijet configurations selected according to
their shape. Other details are as in the caption to
Fig.~\protect\ref{fig13}.} 
\label{fig16}
\vfill
\end{figure}

\newpage
\clearpage
\begin{figure}[p]
\vfill
\setlength{\unitlength}{1.0cm}
\begin{picture} (18.0,8.0)
\put (-1.0,-3.5){\epsfig{figure=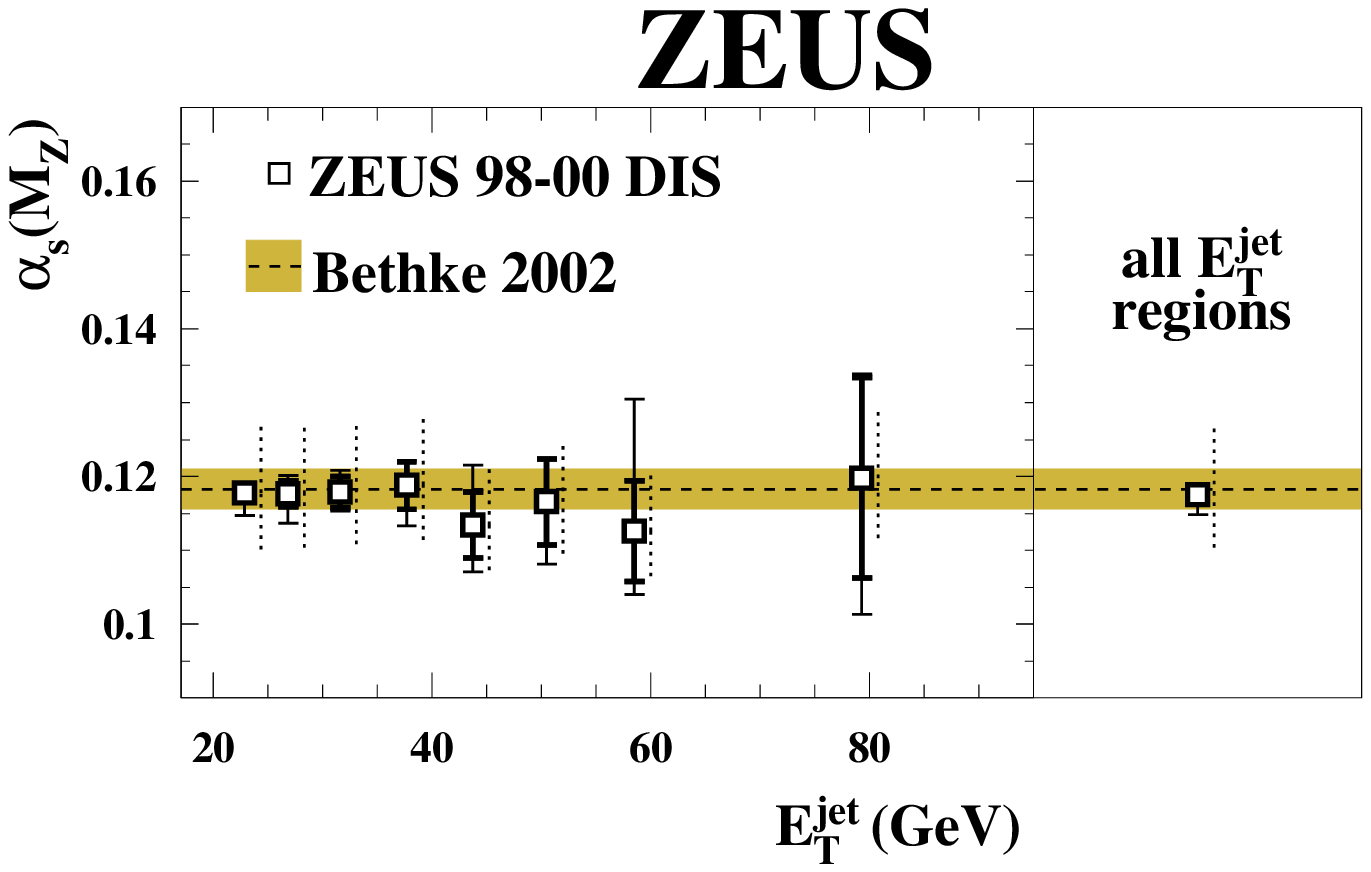,width=18cm}}
\put (9.5,7.5){\bf\small (a)}
\put (12.5,7.5){\bf\small (b)}
\end{picture}
\vspace{-1.5cm}
\caption
{\it 
(a) The $\as(\mz)$ values determined from the QCD fit of the measured
integrated jet shape $\langle \psi(r=0.5) \rangle$ in the different 
$\etjet$ regions (squares). (b) The combined value of $\as(\mz)$
obtained using all the $\etjet$ regions (square). In both plots, the
inner error bars represent the statistical uncertainties of the
data. The outer error bars show the statistical and systematic
uncertainties added in quadrature. The dotted vertical bars represent
the theoretical uncertainties.}
\label{fig17}
\vfill
\end{figure}


\begin{mcbibliography}{10}

\bibitem{pl:b507:70}
\colab{ZEUS}, J. Breitweg \etal,
\newblock Phys.\ Lett.{} B~507~(2001)~70\relax
\relax
\bibitem{pl:b531:9}
\colab{ZEUS}, S. Chekanov \etal,
\newblock Phys.\ Lett.{} B~531~(2002)~9\relax
\relax
\bibitem{epj:c23:615}
\colab{ZEUS}, S. Chekanov \etal,
\newblock Eur.\ Phys.\ J.{} C~23~(2002)~615\relax
\relax
\bibitem{pl:b547:164}
\colab{ZEUS}, S. Chekanov \etal,
\newblock Phys.\ Lett.{} B~547~(2002)~164\relax
\relax
\bibitem{pl:b560:7}
\colab{ZEUS}, S. Chekanov \etal,
\newblock Phys.\ Lett.{} B~560~(2003)~7\relax
\relax
\bibitem{epj:c31:149}
\colab{ZEUS}, J.~Chekanov \etal,
\newblock Eur.\ Phys.\ J.{} C~31~(2003)~149\relax
\relax
\bibitem{epj:c23:13}
\colab{ZEUS}, S. Chekanov \etal,
\newblock Eur.\ Phys.\ J.{} C~23~(2002)~13\relax
\relax
\bibitem{pl:b443:394}
\colab{ZEUS}, J.~Breitweg \etal,
\newblock Phys.\ Lett.{} B~443~(1998)~394\relax
\relax
\bibitem{epj:c19:289}
\colab{H1}, C. Adloff \etal,
\newblock Eur.\ Phys.\ J.{} C~19~(2001)~289\relax
\relax
\bibitem{epj:c25:13}
\colab{H1}, C.~Adloff \etal,
\newblock Eur.\ Phys.\ J.{} C~25~(2002)~13\relax
\relax
\bibitem{pl:b542:193}
\colab{H1}, C. Adloff \etal,
\newblock Phys.\ Lett.{} B~542~(2002)~193\relax
\relax
\bibitem{epj:c29:497}
\colab{H1}, C. Adloff \etal,
\newblock Eur.\ Phys.\ J.{} C~29~(2003)~497\relax
\relax
\bibitem{epj:c19:429}
\colab{H1}, C. Adloff \etal,
\newblock Eur.\ Phys.\ J.{} C~19~(2001)~429\relax
\relax
\bibitem{pl:b515:17}
\colab{H1}, C. Adloff \etal,
\newblock Phys.\ Lett.{} B~515~(2001)~17\relax
\relax
\bibitem{pl:b565:87}
\colab{ZEUS}, S. Chekanov \etal,
\newblock Phys.\ Lett.{} B~565~(2003)~87\relax
\relax
\bibitem{hep-ex-0312057}
\colab{ZEUS}, S. Chekanov \etal,
\newblock Preprint \mbox{DESY-03-212} (\mbox{hep-ex/0312057}), DESY, 2003\relax
\relax
\bibitem{pl:b79:83}
C.H. Llewellyn Smith,
\newblock Phys.\ Lett.{} B~79~(1978)~83\relax
\relax
\bibitem{np:b166:413}
I. Kang and C.H. Llewellyn Smith,
\newblock Nucl.\ Phys.{} B~166~(1980)~413\relax
\relax
\bibitem{pr:d21:54}
J.F. Owens,
\newblock Phys.\ Rev.{} D~21~(1980)~54\relax
\relax
\bibitem{zfp:c6:241}
M. Fontannaz, A. Mantrach and D. Schiff,
\newblock Z.\ Phys.{} C~6~(1980)~241\relax
\relax
\bibitem{proc:hera:1987:331}
W.J. Stirling and Z. Kunszt,
\newblock {\em Proc. HERA Workshop}, R.D. Peccei~(ed.), Vol.~2, p.~331.
\newblock DESY, Hamburg, Germany (1987)\relax
\relax
\bibitem{prl:61:275}
M. Drees and F. Halzen,
\newblock Phys.\ Rev.\ Lett.{} 61~(1988)~275\relax
\relax
\bibitem{prl:61:682}
M. Drees and R.M. Godbole,
\newblock Phys.\ Rev.\ Lett.{} 61~(1988)~682\relax
\relax
\bibitem{pr:d39:169}
M. Drees and R.M. Godbole,
\newblock Phys.\ Rev.{} D~39~(1989)~169\relax
\relax
\bibitem{zfp:c42:657}
H. Baer, J. Ohnemus and J.F. Owens,
\newblock Z.\ Phys.{} C~42~(1989)~657\relax
\relax
\bibitem{pr:d40:2844}
H. Baer, J. Ohnemus and J.F. Owens,
\newblock Phys.\ Rev.{} D~40~(1989)~2844\relax
\relax
\bibitem{prl:69:3615}
S.D. Ellis, Z. Kunszt and D.E. Soper,
\newblock Phys.\ Rev.\ Lett.{} 69~(1992)~3615\relax
\relax
\bibitem{np:b383:419}
S. Catani \etal,
\newblock Nucl.\ Phys.{} B~383~(1992)~419\relax
\relax
\bibitem{np:b421:545}
M.H. Seymour,
\newblock Nucl.\ Phys.{} B~421~(1994)~545\relax
\relax
\bibitem{pl:b378:279}
M.H. Seymour,
\newblock Phys.\ Lett.{} B~378~(1996)~279\relax
\relax
\bibitem{jhep:9909:009}
J.R. Forshaw and M.H. Seymour,
\newblock \JHEP{} 9909~(1999)~009\relax
\relax
\bibitem{pl:b384:401}
\colab{ZEUS}, M.~Derrick \etal,
\newblock Phys.\ Lett.{} B~384~(1996)~401\relax
\relax
\bibitem{np:b406:187}
S. Catani \etal,
\newblock Nucl.\ Phys.{} B~406~(1993)~187\relax
\relax
\bibitem{pr:d48:3160}
S.D. Ellis and D.E. Soper,
\newblock Phys.\ Rev.{} D~48~(1993)~3160\relax
\relax
\bibitem{pl:b293:465}
\colab{ZEUS}, M.~Derrick \etal,
\newblock Phys.\ Lett.{} B~293~(1992)~465\relax
\relax
\bibitem{zeus:1993:bluebook}
\colab{ZEUS}, U.~Holm~(ed.),
\newblock {\em The {ZEUS} Detector}.
\newblock Status Report (unpublished), DESY (1993),
\newblock available on
  \texttt{http://www-zeus.desy.de/bluebook/bluebook.html}\relax
\relax
\bibitem{nim:a279:290}
N.~Harnew \etal,
\newblock Nucl.\ Instr.\ Meth.{} A~279~(1989)~290\relax
\relax
\bibitem{npps:b32:181}
B.~Foster \etal,
\newblock Nucl.\ Phys.\ Proc.\ Suppl.{} B~32~(1993)~181\relax
\relax
\bibitem{nim:a338:254}
B.~Foster \etal,
\newblock Nucl.\ Instr.\ Meth.{} A~338~(1994)~254\relax
\relax
\bibitem{nim:a309:77}
M.~Derrick \etal,
\newblock Nucl.\ Instr.\ Meth.{} A~309~(1991)~77\relax
\relax
\bibitem{nim:a309:101}
A.~Andresen \etal,
\newblock Nucl.\ Instr.\ Meth.{} A~309~(1991)~101\relax
\relax
\bibitem{nim:a321:356}
A.~Caldwell \etal,
\newblock Nucl.\ Instr.\ Meth.{} A~321~(1992)~356\relax
\relax
\bibitem{nim:a336:23}
A.~Bernstein \etal,
\newblock Nucl.\ Instr.\ Meth.{} A~336~(1993)~23\relax
\relax
\bibitem{desy-92-066}
J.~Andruszk\'ow \etal,
\newblock Preprint \mbox{DESY-92-066}, DESY, 1992\relax
\relax
\bibitem{zfp:c63:391}
\colab{ZEUS}, M.~Derrick \etal,
\newblock Z.\ Phys.{} C~63~(1994)~391\relax
\relax
\bibitem{acpp:b32:2025}
J.~Andruszk\'ow \etal,
\newblock Acta Phys.\ Pol.{} B~32~(2001)~2025\relax
\relax
\bibitem{proc:chep:1992:222}
W.H.~Smith, K.~Tokushuku and L.W.~Wiggers,
\newblock {\em Proc.\ Computing in High-Energy Physics (CHEP), Annecy, France,
  Sept.~1992}, C.~Verkerk and W.~Wojcik~(eds.), p.~222.
\newblock CERN, Geneva, Switzerland (1992).
\newblock Also in preprint \mbox{DESY 92-150B}\relax
\relax
\bibitem{nim:a365:508}
H.~Abramowicz, A.~Caldwell and R.~Sinkus,
\newblock Nucl.\ Instr.\ Meth.{} A~365~(1995)~508\relax
\relax
\bibitem{nim:a391:360}
R.~Sinkus and T.~Voss,
\newblock Nucl.\ Instr.\ Meth.{} A~391~(1997)~360\relax
\relax
\bibitem{pl:b322:287}
\colab{ZEUS}, M.~Derrick \etal,
\newblock Phys.\ Lett.{} B~322~(1994)~287\relax
\relax
\bibitem{proc:epfacility:1979:391}
F.~Jacquet and A.~Blondel,
\newblock {\em Proc. of the Study for an $ep$ Facility for {Europe}},
  U.~Amaldi~(ed.), p.~391.
\newblock Hamburg, Germany (1979).
\newblock Also in preprint \mbox{DESY 79/48}\relax
\relax
\bibitem{pl:b558:41}
\colab{ZEUS}, S. Chekanov \etal,
\newblock Phys.\ Lett.{} B~558~(2003)~41\relax
\relax
\bibitem{proc:hera:1991:23}
S.~Bentvelsen, J.~Engelen and P.~Kooijman,
\newblock {\em Proc. of the Workshop on Physics at {HERA}}, W.~Buchm\"uller and
  G.~Ingelman~(eds.), Vol.~1, p.~23.
\newblock Hamburg, Germany, DESY (1992)\relax
\relax
\bibitem{proc:hera:1991:43}
{\em {\rm K.C.~H\"oger}}, ibid., p.~43\relax
\relax
\bibitem{proc:snowmass:1990:134}
J.E. Huth \etal,
\newblock {\em Research Directions for the Decade. Proc. of Summer Study on
  High Energy Physics, 1990}, E.L. Berger~(ed.), p.~134.
\newblock World Scientific (1992).
\newblock Also in preprint \mbox{FERMILAB-CONF-90-249-E}\relax
\relax
\bibitem{cpc:82:74}
T. Sj\"ostrand,
\newblock Comput.\ Phys.\ Comm.{} 82~(1994)~74\relax
\relax
\bibitem{cpc:135:238}
T. Sj\"ostrand \etal,
\newblock Comput.\ Phys.\ Comm.{} 135~(2001)~238\relax
\relax
\bibitem{cpc:67:465}
G. Marchesini \etal,
\newblock Comput.\ Phys.\ Comm.{} 67~(1992)~465\relax
\relax
\bibitem{jhep:0101:010}
G. Corcella \etal,
\newblock \JHEP{} 0101~(2001)~010\relax
\relax
\bibitem{pr:d45:3986}
M. Gl\"uck, E. Reya and A. Vogt,
\newblock Phys.\ Rev.{} D~45~(1992)~3986\relax
\relax
\bibitem{pr:d46:1973}
M. Gl\"uck, E. Reya and A. Vogt,
\newblock Phys.\ Rev.{} D~46~(1992)~1973\relax
\relax
\bibitem{pr:d55:1280}
H.L. Lai \etal,
\newblock Phys.\ Rev.{} D~55~(1997)~1280\relax
\relax
\bibitem{prep:97:31}
B. Andersson \etal,
\newblock Phys.\ Rep.{} 97~(1983)~31\relax
\relax
\bibitem{cpc:39:347}
T. Sj\"ostrand,
\newblock Comput.\ Phys.\ Comm.{} 39~(1986)~347\relax
\relax
\bibitem{cpc:43:367}
T. Sj\"ostrand and M. Bengtsson,
\newblock Comput.\ Phys.\ Comm.{} 43~(1987)~367\relax
\relax
\bibitem{np:b238:492}
B.R. Webber,
\newblock Nucl.\ Phys.{} B~238~(1984)~492\relax
\relax
\bibitem{pr:d36:2019}
T. Sj\"ostrand and M. van Zijl,
\newblock Phys.\ Rev.{} D~36~(1987)~2019\relax
\relax
\bibitem{epj:c2:61}
\colab{ZEUS}, J.~Breitweg \etal,
\newblock Eur.\ Phys.\ J.{} C~2~(1998)~61\relax
\relax
\bibitem{cpc:69:155}
A. Kwiatkowski, H. Spiesberger and H.-J. M\"ohring,
\newblock Comput.\ Phys.\ Comm.{} 69~(1992)~155\relax
\relax
\bibitem{spi:www:heracles}
H.~Spiesberger,
\newblock {\em An Event Generator for $ep$ Interactions at {HERA} Including
  Radiative Processes (Version 4.6)}, 1996,
\newblock available on\\ \texttt{http://www.desy.de/\til
  hspiesb/heracles.html}\relax
\relax
\bibitem{cpc:81:381}
K. Charcu\l a, G.A. Schuler and H. Spiesberger,
\newblock Comput.\ Phys.\ Comm.{} 81~(1994)~381\relax
\relax
\bibitem{spi:www:djangoh11}
H.~Spiesberger,
\newblock {\em {\sc heracles} and {\sc djangoh}: Event Generation for $ep$
  Interactions at {HERA} Including Radiative Processes}, 1998,
\newblock available on\\ \texttt{http://www.desy.de/\til
  hspiesb/djangoh.html}\relax
\relax
\bibitem{pl:b165:147}
Y. Azimov \etal,
\newblock Phys.\ Lett.{} B~165~(1985)~147\relax
\relax
\bibitem{pl:b175:453}
G. Gustafson,
\newblock Phys.\ Lett.{} B~175~(1986)~453\relax
\relax
\bibitem{np:b306:746}
G. Gustafson and U. Pettersson,
\newblock Nucl.\ Phys.{} B~306~(1988)~746\relax
\relax
\bibitem{zfp:c43:625}
B. Andersson \etal,
\newblock Z.\ Phys.{} C~43~(1989)~625\relax
\relax
\bibitem{cpc:71:15}
L. L\"onnblad,
\newblock Comput.\ Phys.\ Comm.{} 71~(1992)~15\relax
\relax
\bibitem{zfp:c65:285}
L. L\"onnblad,
\newblock Z.\ Phys.{} C~65~(1995)~285\relax
\relax
\bibitem{cpc:101:108}
G. Ingelman, A. Edin and J. Rathsman,
\newblock Comput.\ Phys.\ Comm.{} 101~(1997)~108\relax
\relax
\bibitem{epj:c12:375}
H.L.~Lai \etal,
\newblock Eur.\ Phys.\ J.{} C~12~(2000)~375\relax
\relax
\bibitem{tech:cern-dd-ee-84-1}
R.~Brun et al.,
\newblock {\em {\sc geant3}},
\newblock Technical Report CERN-DD/EE/84-1, CERN, 1987\relax
\relax
\bibitem{np:b485:291}
S. Catani and M.H. Seymour,
\newblock Nucl.\ Phys.{} B~485~(1997)~291.
\newblock Erratum in Nucl.\ Phys.{} B~510~(1998)~503.
\relax
\bibitem{np:b178:421}
R.K. Ellis, D.A. Ross and A.E. Terrano,
\newblock Nucl.\ Phys.{} B~178~(1981)~421\relax
\relax
\bibitem{epj:c4:463}
A.D. Martin \etal,
\newblock Eur.\ Phys.\ J.{} C~4~(1998)~463\relax
\relax
\bibitem{epj:c14:133}
A.D. Martin \etal,
\newblock Eur.\ Phys.\ J.{} C~14~(2000)~133\relax
\relax
\bibitem{disaster1}
D. Graudenz,
\newblock {\em Proc. of the Ringberg Workshop on New Trends in HERA physics},
  B.A. Kniehl, G. Kr\"amer and A. Wagner~(eds.).
\newblock World Scientific, Singapore (1998). Also in hep-ph/9708362
  (1997)\relax
\relax
\bibitem{disaster2}
D. Graudenz,
\newblock Preprint \mbox{hep-ph/9710244}, 1997\relax
\relax
\bibitem{proc:calor:2002:767}
M. Wing (on behalf of the \colab{ZEUS}),
\newblock {\em Proc. of the 10th International Conference on Calorimetry in
  High Energy Physics}, R. Zhu~(ed.), p.~767.
\newblock Pasadena, USA (2002).
\newblock Also in preprint \mbox{hep-ex/0206036}\relax
\relax
\bibitem{epj:c11:427}
\colab{ZEUS}, J. Breitweg \etal,
\newblock Eur.\ Phys.\ J.{} C~11~(1999)~427\relax
\relax
\bibitem{epj:c8:367}
\colab{ZEUS}, J.~Breitweg \etal,
\newblock Eur.\ Phys.\ J.{} C~8~(1999)~367\relax
\relax
\bibitem{jhep:0207:012}
J. Pumplin \etal,
\newblock \JHEP{} 0207~(2002)~012\relax
\relax
\bibitem{jhep:0310:046}
D. Stump \etal,
\newblock \JHEP{} 0310~(2003)~046\relax
\relax
\bibitem{jp:g26:r27}
S. Bethke,
\newblock J.\ Phys.{} G~26~(2000)~R27.
\newblock Updated in Preprint hep-ex/0211012, 2002\relax
\relax
\end{mcbibliography}
\end{document}